\pdfoutput=1
\pdfoutput=1\documentclass[final,3p,times]{elsarticle}

\usepackage[utf8]{inputenc}
\usepackage[T1]{fontenc}
\usepackage{booktabs}

\usepackage[font=small,labelfont=bf,labelsep=none]{caption}
\usepackage{longtable}
\usepackage{afterpage}
\usepackage{multirow}   
\usepackage{float}      
\usepackage{subfig}
\usepackage{amssymb}
\usepackage{algorithm}
\usepackage{algorithmic}

\usepackage{graphicx}
\usepackage{epstopdf}
\usepackage{ctable}
\usepackage{bm} 
\usepackage{amsmath}
\usepackage{chngcntr}
\usepackage{subcaption}
\usepackage{hyperref}
\biboptions{sort&compress}
\DeclareMathAlphabet{\mathcal}{OMS}{cmsy}{m}{n}
\DeclareSymbolFont{largesymbols}{OMX}{cmex}{m}{n}
\usepackage{array, threeparttable}
\usepackage{enumerate}
\usepackage{enumitem}
\hypersetup{
colorlinks=true,
linkcolor=blue,
anchorcolor=blue,
citecolor=blue}
\journal{Computer Methods in Applied Mechanics and Engineering}

\begin{document}
\captionsetup[figure]{labelfont={bf},labelformat={default},labelsep=period,name={Fig.}}

\begin{frontmatter}

\title{A hybrid framework integrating physics-informed neural operators with conventional methods for turbulence}

\author[inst1,inst2,inst3]{Sunan Zhao} 
\author[inst1,inst2,inst3]{Yunpeng Wang} 
\author[inst1,inst4]{Boyu Fan} 
\author[inst1,inst2,inst3]{Huiyu Yang} 
\author[inst1,inst2,inst3]{Zhihong Guo} 
\author[inst1,inst2,inst3]{Jianchun Wang\corref{cor1}} 
\cortext[cor1]{Corresponding author at: Department of Mechanics and Aerospace Engineering, Southern University of Science and Technology, Shenzhen 518055, China.\\
E-mail address: \href{wangjc@sustech.edu.cn}{wangjc@sustech.edu.cn} (J. Wang).
}  

\affiliation[inst1]{organization={Department of Mechanics and Aerospace Engineering},
            addressline={Southern University of Science and Technology}, 
            city={Shenzhen},
            postcode={518055},
            country={China}}
\affiliation[inst2]{organization={Shenzhen Key Laboratory of Complex Aerospace Flows},
            addressline={Southern University of Science and Technology}, 
            city={Shenzhen},
            postcode={518055}, 
            country={China}}
\affiliation[inst3]{organization={Guangdong Provincial Key Laboratory of Turbulence Research and Applications},
            addressline={Southern University of Science and Technology}, 
            city={Shenzhen},
            postcode={518055}, 
            country={China}}
\affiliation[inst4]{organization={Engineering Cluster},
            addressline={Singapore Institute of Technology}, 
            city={Singapore},
            postcode={828608}, 
            country={Singapore}}
\begin{abstract}
We develop large-eddy simulation nets (LESnets), a hybrid framework that integrates physics-informed neural operators with conventional computational methods for accurate and efficient simulation of wall-bounded turbulence. LESnets incorporates the large-eddy simulation (LES) equations into the training of the factorized Fourier neural operator (F-FNO), requiring no labeled training data and accommodating flexible time horizons during training. Boundary conditions are enforced as hard constraints and a wall model based on the law of the wall is incorporated to improve physical consistency and enable reliable coarse-grid simulations at high Reynolds numbers. Furthermore, the immersed boundary method (IBM) is embedded in the physics-informed loss function to accommodate complex geometries with boundaries that are not aligned with the computational grid. In addition, the subgrid-scale (SGS) model coefficient is optimized during training using a single flow snapshot. The framework is evaluated on turbulent channel flows at friction Reynolds numbers $Re_{\tau} \approx 180$, $590$, $1000$, and $2000$ and turbulent flows over periodic hills at Reynolds numbers $Re_h \approx 5600$ and $10595$. Numerical results show that LESnets achieves predictive accuracy and computational efficiency comparable to those of the implicit U-Net enhanced Fourier neural operator (IUFNO) and F-FNO, while attaining accuracy comparable to that of conventional LES with greater computational efficiency. These results support the integration of physics-informed neural operators with established computational methods as a promising approach to accurate and efficient simulation of wall-bounded turbulence.
\end{abstract}

\begin{highlights}
\item {LESnets enables large-eddy simulation of wall-bounded turbulence without labeled training data while enforcing boundary conditions as hard constraints.}
\item {A wall model and an immersed boundary method are incorporated into LESnets to improve coarse-grid predictions of wall-bounded turbulence at high Reynolds numbers.}
\item {The subgrid-scale model coefficient can be automatically optimized during training using a single flow snapshot as prior information.}
\end{highlights}



\end{frontmatter}


\section{Introduction}
\label{sec:introduction}


Turbulent flows are ubiquitous in meteorology, aerospace engineering, air pollution control, energy, and various industrial activities \cite{PopeTurbulence}. A large fraction of these flows occurs in wall-bounded configurations, including turbulent boundary layers over wings and turbine blades, channel and pipe flows in heat exchangers and pipelines, and separated flows over bluff bodies, where near-wall dynamics govern drag, heat transfer, and mixing. Computational fluid dynamics (CFD) has therefore become an indispensable tool for both fundamental studies of turbulence and engineering design. However, the wide range of flow scales makes direct numerical simulation (DNS) impractical at high Reynolds numbers \cite{ScaleI, StatP}. Consequently, the Reynolds-averaged Navier-Stokes (RANS) \cite{RANS} and large-eddy simulation (LES) \cite{LES} methods have been widely adopted to simulate turbulence at reduced computational cost. RANS solves for the mean flow and has proven efficient in many industrial applications \cite{Durbin18ARFM, Duraisamy19ARFM}, whereas LES captures large-scale motions that contain most of the energy, while the effects of small-scale fluctuations are represented through subgrid-scale (SGS) models, thereby
enabling more accurate predictions of flow structures \cite{Smagorinsky, Deardorff_1970, DSM}. Despite their superior efficiency compared with DNS, both methods remain computationally expensive when extensive parametric design spaces must be explored, as is commonly required in engineering design and optimization.

While conventional CFD methods continue to advance, machine-learning (ML) methods for flow prediction have attracted widespread research attention in recent years \cite{ling2016reynolds, wang2018investigations, xie2019artificial, yang2019predictive, yuan2020deconvolutional, Brunton20ARFM, Wang_Chu_2025, PLOGMANN2026118421}. ML-based turbulence models include, but are not limited to, data-driven RANS models with embedded invariance properties \cite{LING201622}, field inversion and machine learning (FIML) for improved closure models \cite{PARISH2016758}, blind deconvolution neural networks for reconstructing fine-scale flow fields from coarse-grid data \cite{Maulik_San_2017}, data-driven models for learning unclosed terms in LES \cite{BECK2019108910}, a metric to assess RANS equation conditioning for data-driven turbulence models \cite{Wu_Xiao_Sun_Wang_2019}, and data-driven wall shear stress models \cite{zhou21PRF}. These efforts share a common motivation: to reduce the cost and uncertainty of turbulence prediction in engineering practice. Meanwhile, many studies have directly employed ML methods to solve partial differential equations (PDEs) and fluid-flow problems \cite{PINN, DeepONets, FNO, PINO}.

One popular framework is physics-informed neural networks (PINNs) \cite{PINN}, which solve forward and inverse PDE problems by constraining neural networks with physical laws. This approach has been applied to a broad range of physical systems, including incompressible laminar flows \cite{RAO2020207}, vortex-induced vibrations \cite{raissi2019deep}, subdomain problems in three-dimensional (3D) turbulent channel flows \cite{NSFNets}, hemodynamics in the vasculature \cite{raissi2020hidden}, two-phase Darcy flows in heterogeneous porous media \cite{ZHANG2023111919}, high-resolution particulate matter transport \cite{feng2025physics}, and turbulent Rayleigh-Bénard convection \cite{Wu26_T2F}. However, the strong nonlinearity, high sensitivity to initial conditions, and multiscale dynamics of turbulence have caused frequent failures of PINNs training in turbulent flows \cite{wang2022and}, motivating numerous improvements \cite{gao2021phygeonet, shukla2021parallel, BPINN, List_Chen_Thuerey_2022, CHIU2022114909, MCCLENNY2023111722, Zhang2024filtered, Song2024VWPINN, zou2025uncertainty, FENN2025, CAO2025113494,WEI2025118139, WANG2026110220}. Nevertheless, applying PINNs to complex PDEs such as the 3D Navier-Stokes (NS) equations remains challenging \cite{wang2025simulating}.

Operator learning offers an alternative approach to efficiently solving PDEs. It involves learning a mapping from input conditions to PDEs solutions in a data-driven manner, enabling neural operators to solve a family of PDEs \cite{NeuralOperator}. DeepONet consists of a branch network for encoding the input function and a trunk network for encoding the spatial locations \cite{DeepONets}, whereas FNO applies the Fourier transform to map high-dimensional data into the frequency domain and approximates nonlinear operators by learning relationships in Fourier space \cite{FNO}. FNO models achieve exceptional accuracy in several one-dimensional (1D) and two-dimensional (2D) PDE problems \cite{guibas2021adaptive, FFNO, peng2022attention, meng2023fast, azizzadenesheli2024neural, zhou2024strategies, HUANG2025113601, CHEN2025114131, Gopakumar26}. Modeling 3D turbulent flows with neural operators is more difficult owing to the higher dimensionality of the data and the increased chaotic behavior of turbulence. Nevertheless, FNO and its variants now demonstrate outstanding performance across 3D homogeneous isotropic turbulence \cite{zhijie23TAML, peng2023linear}, turbulent mixing layers \cite{IUFNO}, compressible Rayleigh-Taylor turbulence \cite{luoFNO}, turbulent channel flows \cite{wang2024prediction}, and turbulent flows over periodic hills \cite{Wang26hills}. While well-trained operator-learning models are accurate and efficient for turbulence prediction, they still lack explicit physical principles, and their outputs depend on the labeled data available at a specified temporal resolution.

The physics-informed neural operator (PINO) improves operator-learning models by penalizing deviations from governing equations. Such new machine-learning models include, but are not limited to, the physics-informed DeepONets \cite{PI-DeepONets}, physics-informed Transformer \cite{zhao2023pinnsformer}, and physics-informed Fourier neural operator \cite{PINO}. PINO has demonstrated the ability to produce highly accurate results across various classical linear and nonlinear PDEs \cite{rosofsky2023applications, Goswami2023, Ehlers25POF, zhang2025omnifluids, Wang26NSR, LIU2026118668}. Most recently, Jiao et al. \cite{jiao2024solving} evaluated the effectiveness of physics-informed DeepONets in solving both forward and inverse problems of PDEs on unknown manifolds. Chen et al. \cite{chen2024physics} presented a novel physics-enhanced neural operator (PENO) that is improved by a self-augmentation mechanism to minimize the accumulated error in long-term simulations. Wang et al. \cite{wang2024beyond} proposed an alternative end-to-end learning approach using PINO, and trained the PINO model on data from a coarse-grid solver and then fine-tuned it with a small number of fully-resolved simulations and physics-based losses on a fine grid. Taassob et al. \cite{TAASSOB2024117163} developed a PINN-DeepONet framework for extracting turbulent combustion closure from multiscalar measurements. Zhao et al. \cite{LESnets} developed large-eddy simulation nets (LESnets) by integrating the resolved LES equations within two well-known FNO models for isotropic turbulence and turbulent mixing layers. Eshaghi et al. \cite{VINO} proposed a variational physics-informed neural operator (VINO), which efficiently solves the governing equations for physical losses by minimizing the energy formulation of PDEs. Roy et al. \cite{PIMRNO} introduced a resolution- independent neural operator (RINO) framework to handle arbitrarily (but sufficiently finely) discretized input functions. Guo et al. \cite{PITO} proposed a physics-informed Transformer operator (PITO) for predicting 3D turbulence, which was developed based on the vision Transformer (ViT) architecture. Recently, Zhang et al. \cite{ZHANG2026118667} developed a two-step DeepONet-based physics-informed neural operator (DeepOKAN) that maps pressure profiles to bubble radius responses. However, the PINO approach has not yet been developed for long-term predictions of 3D wall-bounded turbulent flows, a class central to many engineering applications.

Wall-bounded turbulence features solid walls, strong mean shear, anisotropic fluctuations, and distinct flow behavior along different spatial directions, and it underpins many flows of practical engineering interest, from aerodynamic drag and heat transfer to turbomachinery and internal cooling passages. Wall-modeled LES (WMLES) \cite{WMLES}, which represents wall effects through wall models or approximate boundary conditions on coarse grids, is a cost-effective alternative and an active research area \cite{Yang15POF, ARFM2018, CAI2023105893, LIU2025114029}. Combining wall models with machine-learning models offers a novel solution for wall-bounded turbulence at high Reynolds numbers \cite{BhaskaranIEEE, LEE2023108014, DUPUY2023112173}. In this work, we extend our recently proposed LESnets \cite{LESnets} into a general, self-supervised framework for predicting the temporal evolution of 3D wall-bounded turbulent flows. The factorized Fourier neural operator (F-FNO) serves as the neural operator module, capturing flow characteristics along the three spatial directions separately. Meanwhile, the near-wall treatment imposes hard boundary constraints to handle non-periodic boundary conditions and preserve physical consistency at solid walls, and incorporates a wall model into the loss function to correct the wall shear stress on coarse grids. Furthermore, the immersed boundary method (IBM) is integrated into the calculation of the physics-informed loss, extending the framework to complex, non-grid-aligned geometries such as turbulent flows over periodic hills. The proposed model is trained without using labeled data. To the best of our knowledge, the present PINO framework achieves, for the first time, long-term predictions of wall-bounded turbulent flows at high Reynolds numbers without using any labeled data.

The rest of the paper is organized as follows. Section \ref{sec:governing-equations-data} introduces the governing equations of the LES and the numerical framework used to generate the datasets, followed by a brief overview of data preprocessing. Section \ref{sec:methodology} provides a brief introduction to PINNs, FNO, and F-FNO, after which the LESnets framework for wall-bounded turbulent flows is presented. Section \ref{sec:results-and-discussion} showcases several numerical examples of turbulent channel flows and turbulent flows over periodic hills that demonstrate the capabilities of the proposed LESnets approach in modeling the long-term evolution of wall-bounded turbulent flows at high Reynolds numbers. We summarize our results in Section \ref{sec:conclusion}.

\section{Governing equations and numerical setup}
\label{sec:governing-equations-data}

In this section, we provide a concise overview of the governing equations for incompressible turbulence and the conventional LES method. We then introduce the wall model and the IBM commonly used in wall-bounded turbulence. Subsequently, we describe the numerical methods employed to solve these equations and summarize the DNS and LES parameters. Finally, we introduce the datasets used to train the machine-learning models in the subsequent section.

\subsection{Governing equations}
\label{subsec:governing-equations}

The conservation of mass and momentum for the incompressible Newtonian fluid is described by the 3D NS equations \cite{PopeTurbulence, ishihara2009study}, namely

\begin{equation}
    \frac{\partial u_i}{\partial x_i}=0,
\label{eq:ns-continuity}
\end{equation}

\begin{equation}
    \frac{\partial u_i}{\partial t}+\frac{\partial u_iu_j}{\partial x_j}=-\frac{\partial p}{\partial x_i}+\nu\frac{\partial^2u_i}{\partial x_j\partial x_j}+\mathcal{F}_i,
\label{eq:ns-momentum}    
\end{equation}
where ${t}$ is time, ${u_i}$ denotes the ${i}$th component of velocity, ${p}$ is the pressure divided by the constant density $\rho$, ${\nu}$ represents the kinematic viscosity, and ${\mathcal{F}_i}$ represents any forcing applied to the momentum of the fluid in the ${i}$th coordinate direction. This paper follows the summation notation convention unless stated otherwise. The three coordinate directions are denoted by $x$, $y$, and $z$ (corresponding to $x_1, x_2, x_3$), while the velocity components along these directions are denoted by $u$, $v$, and $w$ (corresponding to $u_1, u_2, u_3$), respectively. 

For turbulent channel flow, the friction Reynolds number is defined as \cite{PopeTurbulence}

\begin{equation}
    Re_{\tau}=\frac{u_{\tau}\delta}{\nu}=\frac{\delta}{\delta_\nu},
\label{eq:friction-reynolds-number}
\end{equation}
where $u_{\tau}=\sqrt{\tau_w/\rho}$ and $\delta$ are the friction velocity and channel half-width, respectively. $\delta_\nu=\nu\sqrt{\rho/\tau_w}=\nu/u_\tau$ is the viscous length scale. Here, the wall shear stress $\tau_w$ is calculated as

\begin{equation}
    {\tau_{w}} =\rho\nu \left\lvert \frac{\partial\langle{u}\rangle}{\partial y} \right\lvert_{y=\pm 1},
\label{eq:wall-shear-stress}
\end{equation}
where $\langle \cdot \rangle$ denotes a spatial average over the homogeneous streamwise and spanwise directions.

For turbulent flow over periodic hills, the Reynolds number is defined as 
\begin{equation}
    Re_{h}=\frac{U_{b}h}{\nu},
\label{eq:reynolds-number}
\end{equation}
where $U_b$ is the bulk velocity and $h$ denotes the hill height.

The LES directly solves the large-scale unsteady turbulent motions, whereas the effects of the subgrid-scale motions are modeled by subgrid-scale models \cite{PopeTurbulence}. A spatial filtering operation can be defined as
\begin{equation}
    \bar{f}(\mathbf{x})=\int_Mf(\mathbf{x}-\mathbf{r})G(\mathbf{r},\mathbf{x};\Delta)d\mathbf{r},
\label{eq:spatial-filter}
\end{equation}
where $f$ can be any physical quantity of interest, $G$ is the filter kernel, $\Delta$ is the filter width, and $M$ is the physical domain. By applying the spatial filter to Eq. \eqref{eq:ns-continuity} and Eq. \eqref{eq:ns-momentum}, respectively, the filtered incompressible NS equations can be expressed as \cite{PopeTurbulence}

\begin{equation}
    \frac{\partial \bar{u}_i}{\partial x_i}=0,
\label{eq:filtered-continuity}
\end{equation}
\begin{equation}
    \frac{\partial \bar{u}_i}{\partial t}+\frac{\partial \bar{u}_i\bar{u}_j}{\partial x_j}=-\frac{\partial \bar{p}}{\partial x_i}-\frac{\partial \tau_{ij}}{\partial x_j}+\nu\frac{\partial^2\bar{u}_i}{\partial x_j\partial x_j}+\bar{\mathcal{F}}_i.
\label{eq:filtered-momentum}
\end{equation}
Here, $\tau_{ij}$ is the unclosed SGS stress defined by $\tau_{ij}= \overline{u_iu_j}-\bar{u}_i\bar{u}_j$. To solve the LES equations, it is essential to model the SGS stress based on the filtered variables. 

For wall-bounded turbulent flows, a widely used SGS model is the wall-adapting local eddy-viscosity (WALE) model \cite{nicoud1999subgrid}, which can be written as

\begin{equation}
    \tau_{ij}^A=\tau_{ij}-\frac{\delta_{ij}}3\tau_{kk}=-2 \nu_t \bar{S}_{ij},
\label{eq:sgs-stress}
\end{equation}
with
\begin{equation}
    \nu_t=(C_w \Delta)^2\frac{(S^d_{ij}S^d_{ij})^{3/2}}{(\bar{S}_{ij}\bar{S}_{ij})^{5/2}+(S^d_{ij}S^d_{ij})^{5/4}},
\label{eq:wale-viscosity}
\end{equation}
where $\tau_{ij}^A$ denotes the anisotropic part of the SGS stress tensor, $\delta_{ij}$ represents the Kronecker symbol, $\bar{S}_{ij}=\frac{1}{2}(\partial\bar{u}_i/\partial x_j+\partial\bar{u}_j/\partial x_i)$ is the filtered strain rate tensor, $C_{w}$ is the WALE coefficient. Here, $S^d_{ij}S^d_{ij}=\frac{1}{6}(S^2S^2+\Omega^2\Omega^2)+\frac{2}{3}S^2\Omega^2+2IV_{S\Omega}$, where $S^2=\bar{S}_{ij}\bar{S}_{ij}$, $\Omega^2=\bar{\Omega}_{ij}\bar{\Omega}_{ij}$, and $IV_{S\Omega}=\bar{S}_{ik}\bar{S}_{kj}\bar{\Omega}_{jl}\bar{\Omega}_{li}$. $\bar{\Omega}$ is the filtered antisymmetric rate of rotation, given by

\begin{equation}    
    \bar{\Omega}_{ij}=\frac{1}{2}\left(\frac{\partial\bar{u}_i}{\partial x_j}-\frac{\partial\bar{u}_j}{\partial x_i}\right).
\label{eq:rotation-rate}
\end{equation}

\subsection{Wall model}
\label{subsec:Wall-model}

When LES is applied to wall-bounded turbulence, a characteristic length scale proportional to the wall-normal distance emerges. However, the computational cost of LES becomes comparable to that of DNS when the near-wall turbulent structures are fully resolved. WMLES is one of the most efficient strategies for reducing the computational cost associated with resolving near-wall turbulence \cite{CHUNG_PULLIN_2009}. Among such approaches, the wall-stress model is the most widely adopted type of WMLES method, as it ensures momentum conservation near the wall on relatively coarse grids, thereby enabling accurate predictions of the mean velocity distribution \cite{SCHUMANN1975376}. In this study, we employ the wall model proposed by Inagaki et al. \cite{Inagaki02} to approximate the log-law in a segmented manner, whose formulation is as follows:

\begin{equation}
U^+=
\begin{cases}
y^+, & y^+\leqslant y_{C_1}^+,\\
A_1(y^+)^{B_1}, & y_{C_1}^+<y^+\leqslant y_{C_2}^+,\\
A_2(y^+)^{B_2}, & y^+>y_{C_2}^+,
\end{cases}
\label{eq:wall-model}
\end{equation}
where $U^+=u/u_{\tau}$ and $y^+=y/\delta_\nu$ are the normalized velocity and normalized wall-normal distance, respectively. $A_1=2.7$, $A_2=8.6$, $B_1=1/2$, $B_2=1/7$, $y_{C_1}^+=A_1^{1/(1-B_1)}$, and $y_{C_2}^+=(A_2/A_1)^{1/(B_1-B_2)}$. Given the near-wall flow velocity and wall-normal distance, the friction velocity $u_{\tau}$ can be determined using the wall model. This subsequently enables the calculation of the modeled wall shear stress $\tau_w=\rho u_{\tau}^2$, which serves as the wall stress boundary condition.

\subsection{Immersed boundary method}
\label{subsec:IBM}
The IBM enforces wall boundary conditions on complex, non-grid-aligned geometries by directly modifying the discretized equations in the cells adjacent to the boundary, thereby avoiding the use of body-fitted meshes \cite{annurev_ibm}. It can be combined with the finite difference, finite volume, and lattice Boltzmann methods to solve a wide range of complex flow problems \cite{dbouk_df_ibmnscd_2016, fringand_stable_2024, YAN2025117898}. In this study, we adopt the direct forcing method to impose the boundary conditions accurately on rigid boundaries of complex shape. Specifically, the velocities at the grid nodes inside the solid body are reconstructed by 1D Lagrangian interpolation, with the interpolation performed separately along the three grid-line directions. Once the solid-node velocities are obtained, the derivatives along each grid-line direction can be computed from the velocities of all grid points, thereby enforcing the boundary conditions implicitly. The 1D Lagrangian interpolation reconstruction approach is illustrated in \ref{app:1D_lag}.

\subsection{Numerical framework and data preparation}
\label{subsec:numerical-framework-and-data-preparation}
We employ the open-source code Xcompact3D \cite{Xcompact3D} with a sixth-order finite difference (FD) solver to solve the DNS and LES equations numerically. All simulations are performed on an Intel Xeon Gold 6148 CPU at 2.40 GHz.

\begin{figure}[htbp]
\centering
\begin{minipage}{0.49\linewidth} 
\centerline{\includegraphics[width=\textwidth]{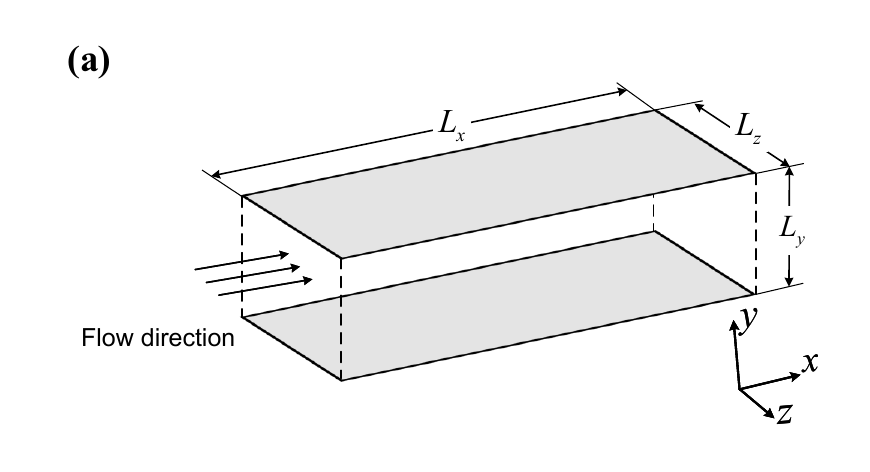}}
\end{minipage}
\hspace{-2pt} 
\begin{minipage}{0.49\linewidth} 
\centerline{\includegraphics[width=\textwidth]{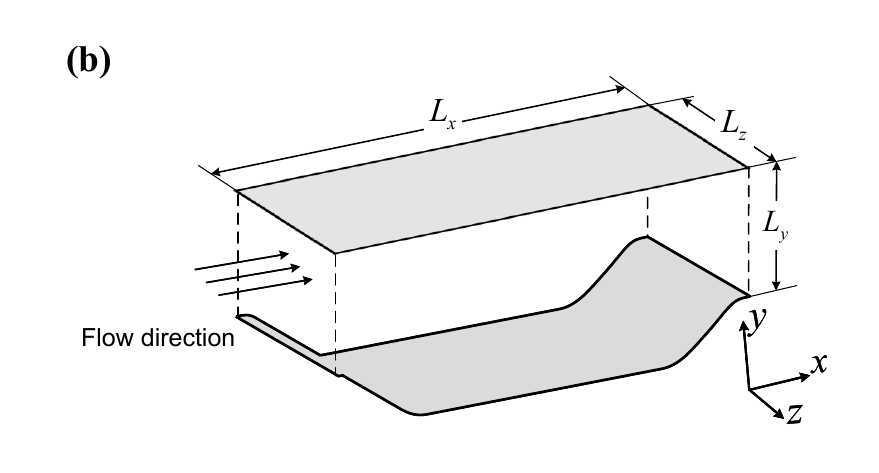}}
\end{minipage}
\caption{The computational domain of \textbf{(a)} the turbulent channel flows, and \textbf{(b)} turbulent flow over periodic hills.}
\label{fig:showcase}
\end{figure}

\subsubsection{Turbulent channel flows}
\label{subsubsec:turbulent-channel}
We first consider the turbulent channel flow, a canonical configuration for studying wall-bounded turbulence. The channel geometry is depicted in Fig. \ref{fig:showcase}(a). The computational domain is $L_x \times L_y \times L_z = [0,4\pi] \times [-1,1] \times [0,4\pi/3]$. The numbers of grid points in the streamwise ($x$), wall-normal ($y$), and spanwise ($z$) directions are $N_x$, $N_y$, and $N_z$, respectively, with a stretched mesh \cite{LAIZET20095989} used in the wall-normal direction. Here, $\Delta y_{w}^+$ denotes the wall-unit distance of the first grid point from the wall. No-slip boundary conditions are applied at the top and bottom walls of the channel ($y=\pm1$), with periodicity enforced along the streamwise and spanwise directions. The non-dimensional time step is chosen as large as possible while satisfying the CFL condition, giving $\Delta t=5.00\times10^{-3}$ for both DNS and LES.

In this work, four widely adopted benchmark cases at friction Reynolds numbers of $Re_{\tau} \approx 180$, $590$, $1000$, and $2000$ are driven by a constant pressure gradient $dp/dx=-u_{\tau}^2$, with different values of the kinematic viscosity $\nu$ and friction velocity $u_\tau$. In particular, the cases at $Re_{\tau} \approx 1000$ and $2000$ are simulated by wall-modeled LES (WMLES), in which the WALE model is still adopted as the SGS model, while the fourth grid point from the wall is used as input to the wall model to address the log-layer mismatch \cite{Log_mismatch}. For these cases, the DNS statistics of Lee et al. \cite{Lee_Moser_2015} are used as the benchmark data. Staggered meshes are employed to discretize the velocity and pressure fields \cite{AnnS1996}. An explicit second-order Adams-Bashforth scheme is used for time integration.

The training data are obtained from WALE and WMLES of three-dimensional turbulent channel flows at friction Reynolds numbers of $Re_{\tau} \approx 180$, $590$, $1000$, and $2000$. For the cases at $Re_{\tau} \approx 180$, $590$, and $1000$, the data are sampled over a duration of $200\Delta T=200$ with a time interval of $\Delta t_{\mathrm{train}}=\Delta T=200\Delta t=1.0$. For the wall-modeled cases at $Re_{\tau} \approx 2000$, the data are sampled over a duration of $100\Delta T=100$ with a time interval of $\Delta t_{\mathrm{train}}=\Delta T/2=100\Delta t=0.5$. Here, $\Delta T=1$ denotes the non-dimensional time unit. The DNS grids are $128 \times 129 \times 128$ and $384 \times 257 \times 192$ for $Re_{\tau} \approx 180$ and $590$, respectively; the LES grids with the WALE model are $32 \times 65 \times 32$ and $64 \times 65 \times 64$ for $Re_{\tau} \approx 180$ and $590$, respectively; and the WMLES grids are $32 \times 65 \times 32$ and $64 \times 65 \times 64$ for $Re_{\tau} \approx 1000$ and $2000$, respectively. The wall-unit distances of the first grid point from the wall are $\Delta y_w^+ = 1.0$ and $1.64$ for DNS at $Re_{\tau} \approx 180$ and $590$, respectively; $\Delta y_w^+ = 1.94$ and $6.14$ for WALE at $Re_{\tau} \approx 180$ and $590$, respectively; and $\Delta y_w^+ = 11.46$ and $11.11$ for WMLES at $Re_{\tau} \approx 1000$ and $2000$, respectively. The parameters of all DNS, WALE, and WMLES cases are listed in Table \ref{tab:channel-simulation-parameters}. Twenty groups of instantaneous velocity and pressure fields, generated from WALE or WMLES with different initial conditions, are used to construct the training datasets $\mathcal{C}_{\mathrm{train}}^{180}$, $\mathcal{C}_{\mathrm{train}}^{590}$, $\mathcal{C}_{\mathrm{train}}^{1000}$, and $\mathcal{C}_{\mathrm{train}}^{2000}$, where the superscript denotes the friction Reynolds number. Each dataset contains $T_{\mathrm{train}}=200$ training time steps for the four Reynolds-number cases. All machine-learning methods use the same training dataset for each Reynolds-number case to ensure a fair comparison.

\begin{table}[htbp]
\captionsetup{font=small,labelfont=bf, width=.885\textwidth}
\setlength{\abovecaptionskip}{0pt}
\setlength{\belowcaptionskip}{10pt}
\caption{Parameters for the DNS and LES of turbulent channel flows.}
\label{tab:channel-simulation-parameters}
\centering
\begin{tabular}{ccccccc}
\toprule
Method & $N_x \times N_y \times N_z$   & $Re_{\tau}$ & $\nu$ & $\Delta y_{w}^+$ & ${\Delta t}$ & $u_{\tau}$ \\
\midrule 
\multirow{2}{*}{DNS} & $128\times129\times128$ & $180$ & $1/4225$ & $1.00$ & $5.00\times10^{-3}$ & $4.26\times10^{-2}$ \\
                      & $384\times257\times192$ & $590$ & $1/16285$ & $1.64$ & $5.00\times10^{-3}$ & $3.62\times10^{-2}$ \\
\midrule 
\multirow{2}{*}{WALE} & $32\times65\times32$ & $180$ & $1/4225$ & $1.94$ & $5.00\times10^{-3}$ & $4.26\times10^{-2}$ \\
                       & $64\times65\times64$ & $590$ & $1/16285$ & $6.14$ & $5.00\times10^{-3}$ & $3.62\times10^{-2}$ \\
\midrule 
\multirow{2}{*}{WMLES} & $32\times65\times32$ & $1000$ & $1/30000$ & $11.46$ & $5.00\times10^{-3}$ & $3.42\times10^{-2}$ \\
                        & $64\times65\times64$ & $2000$ & $1/65475$ & $11.11$ & $5.00\times10^{-3}$ & $3.12\times10^{-2}$ \\
\bottomrule
\end{tabular}
\end{table}

\subsubsection{Turbulent flow over periodic hills}
\label{subsubsec:periodic-hill}

Next, we consider turbulent flow over periodic hills, a canonical configuration for studying flow separation and reattachment over a curved, non-grid-aligned boundary \cite{ALMEIDA199387}. The periodic hill geometry is depicted in Fig. \ref{fig:showcase}(b). The computational domain is $L_x \times L_y \times L_z = [0,9h] \times [0,3.036h] \times [0,4.5h]$, where $h=1$ denotes the hill height. The numbers of grid points in the streamwise ($x$), wall-normal ($y$), and spanwise ($z$) directions are $N_x$, $N_y$, and $N_z$, respectively, with a stretched mesh used in the wall-normal direction. The hill is a two-dimensional hill of height $h$ mounted on the lower wall, with its crest located at the streamwise center of the domain and the upper wall positioned $2.036h$ above the crest. Its surface profile is defined by a piecewise function that smoothly joins the flat bottom wall, following the periodic hill configuration of \cite{breuer_flow_2009}. Periodicity is enforced along the streamwise and spanwise directions, and no-slip boundary conditions are applied at the upper wall and at the hill surface. Since the curved hill surface is not aligned with the Cartesian grid, it is represented by the direct forcing IBM described in Section \ref{subsec:IBM}, in which the solid-node velocities are reconstructed by 1D Lagrangian interpolation along the three grid-line directions. The non-dimensional time step is chosen as large as possible while satisfying the CFL condition.

In this work, two widely adopted benchmark cases at Reynolds numbers $Re_h \approx 5600$ and $10595$ are considered. The case at $Re_h \approx 5600$ is simulated using DNS and WMLES, whereas the case at $Re_h \approx 10595$ is simulated using WRLES and WMLES, since DNS is prohibitively expensive at this Reynolds number. For the WMLES case at $Re_h \approx 5600$, the second grid point from the wall is used as input to the wall model. The same solver and time-integration scheme as in Section \ref{subsubsec:turbulent-channel} are adopted.

The training data are obtained from WMLES of turbulent flow over periodic hills at $Re_h \approx 5600$. The data are sampled over a duration of $160\Delta T=160$ with a time interval of $\Delta t_{\mathrm{train}}=\Delta T/5=100\Delta t=0.2$. The DNS grids are $768 \times 385 \times 128$ for $Re_h \approx 5600$; the WRLES grids with the WALE model are $256 \times 289 \times 128$ for $Re_h \approx 10595$; and the WMLES grids are $64 \times 73 \times 32$ for $Re_h \approx 5600$. The parameters of the periodic hill cases are listed in Table \ref{tab:hills-simulation-parameters}. Two groups of instantaneous velocity and pressure fields, generated from WMLES at $Re_h \approx 5600$ with different initial conditions, are used to construct the training dataset $\mathcal{H}_{\mathrm{train}}^{5600}$, where the superscript denotes the Reynolds number. The dataset contains $T_{\mathrm{train}} = 800$ training time steps. Accordingly, two models are trained independently on the same WMLES training dataset $\mathcal{H}_{\mathrm{train}}^{5600}$: the first is validated against the DNS at $Re_h \approx 5600$, while the second is validated against the WRLES at $Re_h \approx 10595$. This enables the framework to be examined at a Reynolds number nearly double that of the training case. All machine-learning methods use the same training strategy to ensure a fair comparison.

\begin{table}[htbp]
\captionsetup{font=small,labelfont=bf, width=.78\textwidth}
\setlength{\abovecaptionskip}{0pt}
\setlength{\belowcaptionskip}{10pt}
\caption{Parameters for the DNS and LES of turbulent flows over periodic
hills.}
\label{tab:hills-simulation-parameters}
\centering
\begin{tabular}{ccccc}
\toprule
Method & $N_x \times N_y \times N_z$  & $Re_h$  & $h$ & ${\Delta t}$ \\
\midrule
DNS & $768 \times 385 \times 128$ & $5600$ & $1.0$ & $2.00\times10^{-3}$ \\           
\midrule           
WRLES & $256 \times 289 \times 128$ & $10595$ & $1.0$ & $1.00\times10^{-3}$ \\           
\midrule  
WMLES & $64 \times 73 \times 32$ & $5600$ & $1.0$ & $2.00\times10^{-3}$ \\         
\bottomrule
\end{tabular}
\end{table}

\section{Methodology}
\label{sec:methodology}
In this section, we present a brief overview of the physics-informed machine-learning and operator-learning frameworks. Subsequently, we introduce our proposed LESnets in detail in the next subsections.

\subsection{Physics-informed neural networks}
\label{subsec:physics-informed-neural-networks}

Consider a general form of the dynamical system governed by the following PDEs:
\begin{equation}
\begin{aligned}
    \mathcal{R}(a,u;\Lambda)=0,\\
    \mathcal{B}(a,u)=0,\\
    \mathcal{I}(a,u)=0,
\label{eq:governing-pde-system}
\end{aligned}
\end{equation}
where $\mathcal{R}$, $\mathcal{B}$, and $\mathcal{I}$ are the nonlinear differential operator, boundary condition, and initial condition, respectively. $\Lambda$ denotes the parameter vector defining the coefficients of PDEs. $a\in\mathcal A$ and $u\in\mathcal U$ denote the vector-valued input and output functions, respectively, where $\mathcal A$ and $\mathcal U$ are Banach spaces. This formulation gives rise to the solution operator $\mathcal{G}:\mathcal{A}\to\mathcal{U}$ which maps the input space $\mathcal{A}$ to the output space $\mathcal{U}$. Prototypical examples include the NS equations introduced in Eq. \eqref{eq:ns-continuity} and Eq. \eqref{eq:ns-momentum}.

Given an instance ${a}^\dagger$ and a solution operator $\mathcal{G}^{\dagger}$, we denote $u^\dagger=\mathcal{G}^\dagger(a^\dagger)$ as the unique ground truth. The equation-solving task of PINNs is to approximate the true solution $u^\dagger$ using neural network-based solvers. Specifically, PINNs-type methods use a neural network $u_{\theta}$ with parameters ${\theta}$ to approximate the solution function $u^\dagger$. The parameters ${\theta}$ are obtained by minimizing the physics-informed loss, with exact derivatives computed using automatic differentiation (AD) \cite{autograd}. Typically, the loss function for a PINN is defined as the sum of the squared norms of the residuals on the left-hand sides of the equations in Eq. \eqref{eq:governing-pde-system}:

\begin{equation}
\label{eq:physics-informed-loss}
\begin{aligned}
\mathcal{L}_{\mathrm{PINN}}(a^\dagger,u_{\theta})& =\mathcal{L}_\mathrm{PDE}+\alpha\mathcal{L}_{\mathrm{bc}}+\beta\mathcal{L}_{\mathrm{ic}}=\left\|\mathcal{R}(a^\dagger,u_{\theta};\Lambda)\right\|_{L^{2}(D)}^{2}+\alpha\left\|\mathcal{B}(a^\dagger,u_{\theta})\right\|_{L^{2}(\partial D)}^{2}+\beta\left\|\mathcal{I}(a^\dagger,u_{\theta})\right\|_{L^{2}(D)}^{2}, \\
\end{aligned}
\end{equation}
where $\alpha,\beta$ are the hyper-parameters, and $D$ is the bounded domain. PINNs integrate physics into the learning process through the soft constraints. However, in practice, they often encounter some difficulties when solving PDEs for multiscale problems \cite{wang2022and}. In this study, we aim to alleviate the difficulties by applying physics-informed neural operators.
 
\begin{figure}[htbp]
\centering
\includegraphics [width=1.0\textwidth]{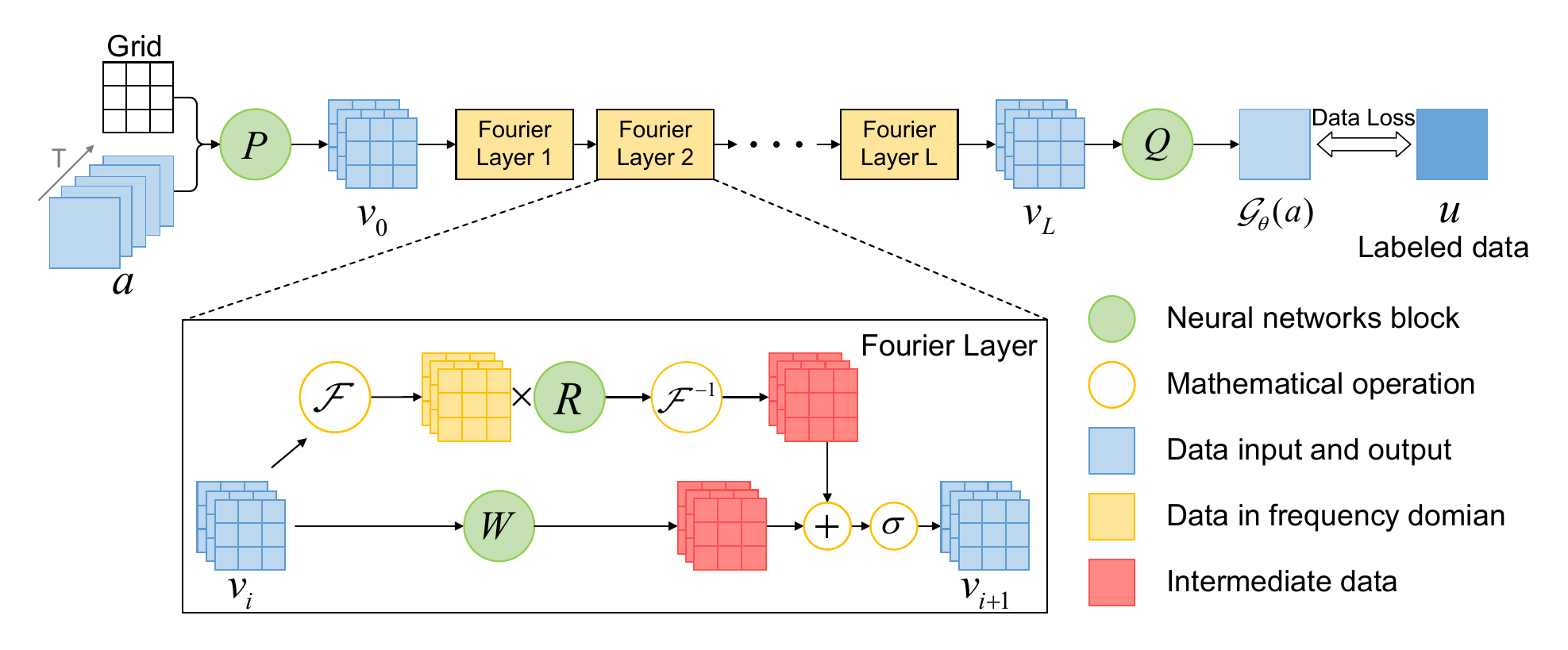}
\caption{{The architecture of the Fourier neural operator \cite{FNO}.}}
\label{fig:fno-architecture}
\end{figure}
\subsection{Neural operators}
\label{subsec:neural-operator}

Neural operators can provide an effective approximation for the solution operator $\mathcal{G}$, which is a nonlinear mapping between infinite-dimensional spaces \cite{NeuralOperator}. Specifically, given the PDEs described by Eq. \eqref{eq:governing-pde-system} and the corresponding solution operator $\mathcal{G}$, a neural operator $\mathcal{G}_{\theta}$ can be used as a surrogate model to approximate $\mathcal{G}$. Typically, we assume a dataset $\{a_j,u_j\}_{j=1}^N$ as a ground truth, where $\mathcal{G}(a_j)=u_j$ and each $a_j$ is drawn i.i.d. from a distribution $\mu$ supported on $\mathcal{A}$. Then, the neural operator can minimize the empirical error on a given pair of data:

\begin{equation}
\label{eq:data-loss}
    \mathcal{L}_{\mathrm{data}}(u^\dagger,\mathcal{G}_\theta(a^\dagger))=\|u^\dagger-\mathcal{G}_\theta(a^\dagger)\|_{L^{2}(\mathcal{U})}^{2}.
\end{equation}

\subsubsection{The Fourier neural operator (FNO)}
\label{subsubsec:fno}
Fourier neural operator (FNO) \cite{FNO} combines the kernel integral operator $\mathcal{K}$ and the point-wise linear operators $\mathcal{W}$ with a nonlinear activation function $\sigma$ to approximate highly nonlinear operators. The kernel integral operator $\mathcal{K}$ can be efficiently calculated in the frequency domain using the fast Fourier transform (FFT), whereas $\mathcal{W}$ corresponds to a convolution in the physical domain. In practice, the neural operator $\mathcal{G}_{\theta}$ can be formulated as an iterative architecture in the following form \cite{FNO}:

\begin{equation}
\label{eq:fno-operator}
    \mathcal{G}_\theta:=Q\circ(\mathcal{W}_L+\mathcal{K}_L)\circ\sigma(\mathcal{W}_{L-1}+\mathcal{K}_{L-1})\circ\cdots\circ\sigma(\mathcal{W}_1+\mathcal{K}_1)\circ{P},
\end{equation}
where ${P}$ and ${Q}$ are point-wise operators parameterized by neural networks. The integral kernel operators are linear transforms in Fourier space:
\begin{equation}
\label{eq:fno-kernel}
    \mathcal{K}_l(v_{l})=\mathcal{F}^{-1}(R_l\cdot\mathcal{F}(v_{l})),
\end{equation}
where $v_{l}$ is the input for each Fourier layer, $R_l$ are weight matrices, and $\mathcal{F},\mathcal{F}^{-1}$ denote the Fourier transform and its inverse. The FNO architecture is shown in Fig. \ref{fig:fno-architecture}. In particular, the point-wise operator $P$ maps the lower-dimensional input function $a$ to $v_0$ in a higher-dimensional channel space. In the Fourier layers, FFT is applied to map the spatial representation of the function to the frequency domain. After the spectral representation is modified using learnable weights defined on the retained Fourier modes ($|k|\le k_{max}$, where $k_{max}$ is the number of retained modes), an inverse fast Fourier transform (iFFT) is applied to map the modified function back into the spatial domain. Then the point-wise operator $Q$ projects $v_L$ onto a lower-dimensional output function space. For more information about the architecture of FNO, the reader is referred to \cite{FNO, NeuralOperator}. Temporally continuous inference of the flow field using FNO is achieved by appending the predicted velocity field to the preceding time history to form an extended input, which is then reintroduced into the FNO.

\subsubsection{Factorized Fourier neural operator (F-FNO)}
\label{subsubsec:ffno}

\begin{figure}[htbp]
\centering
\includegraphics [width=1.0\textwidth]{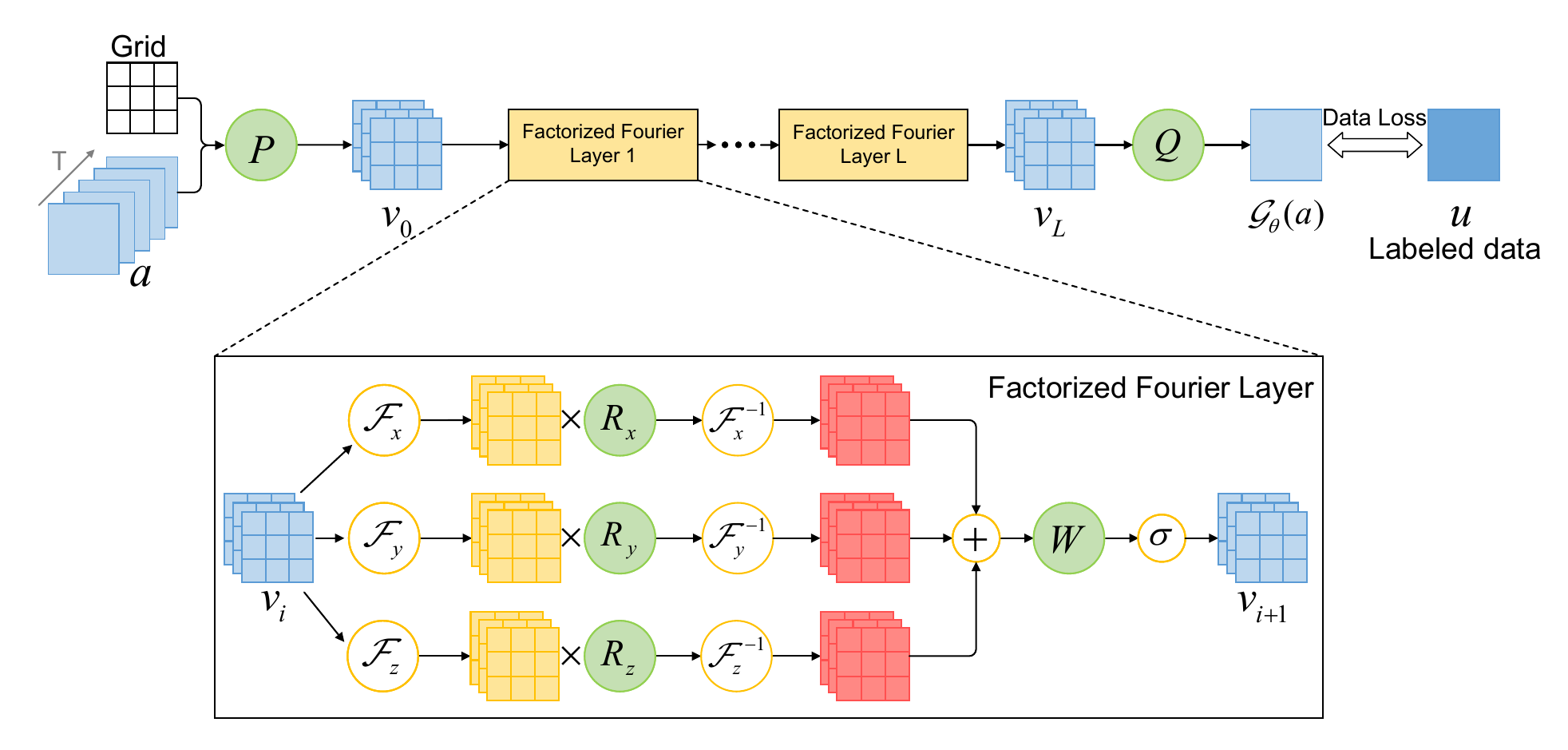}
\caption{{The architecture of the factorized Fourier neural operator \cite{FFNO}.}}
\label{fig:ffno-architecture}
\end{figure}

The factorized Fourier neural operator (F-FNO) \cite{FFNO} introduces an improved version of the Fourier layer, referred to as the factorized Fourier layer. The architecture of the F-FNO model is shown in Fig. \ref{fig:ffno-architecture}. The key modification compared to the original Fourier layer lies in the computation of the kernel integral operator $\mathcal{K}_l$. Instead of performing a 3D FFT, the factorized Fourier layer computes three 1D FFTs along each dimension and then sums the results:

\begin{equation}
\label{eq:ffno-kernel}
    \mathcal{K}_l(v_{l})=\mathcal{F}^{-1}_{(1)}(R_{l(1)}\cdot\mathcal{F}_{(1)}(v_l))+\mathcal{F}^{-1}_{(2)}(R_{l(2)}\cdot\mathcal{F}_{(2)}(v_l))+\mathcal{F}^{-1}_{(3)}(R_{l(3)}\cdot\mathcal{F}_{(3)}(v_l)),
\end{equation}
where the subscripts $(1)$, $(2)$, and $(3)$ denote the three spatial dimensions. This formulation drastically reduces the number of parameters per Fourier layer from $O(k_{{1}}k_{{2}}k_{{3}})$ to $O(k_{{1}}+k_{{2}}+k_{{3}})$, where $k_1$, $k_2$, and $k_3$ are the numbers of Fourier modes in the three directions. 

F-FNO has been used in 3D seismological applications and for solving implicit geometric PDEs \cite{elasticFFNO, MS-IUFFNO}, but to the best of our knowledge, it has not yet been applied to the prediction of 3D wall-bounded turbulent flows. Another open-source operator-learning framework, implicit U-Net enhanced Fourier neural operator (IUFNO), proposed by Li \cite{IUFNO}, has been used to simulate 3D turbulent channel flows \cite{wang2024prediction} (its architecture is detailed in \ref{app:iufno-architecture}). In this work, we employ two data-driven operator-learning models, IUFNO and F-FNO, for 3D wall-bounded turbulent flows. For IUFNO, the hyper-parameters and training strategy follow a previous study \cite{wang2024prediction}, whereas the F-FNO model is adapted and optimized for the present work: the coordinate inputs are removed, the two point-wise operators $P$ and $Q$ are replaced with convolutional encoder-decoder blocks, and the temporal dimensions of the input and output are adjusted such that the component and temporal dimensions are merged before the factorized Fourier layers. These modifications further reduce the number of parameters of F-FNO and significantly improve its accuracy in the present simulations; consequently, all F-FNO models used in the subsequent sections adopt this optimized version.

Data-driven operator learning requires a substantial amount of high-resolution data. However, obtaining such data is highly costly and time-consuming in computational fluid dynamics, thereby undermining its inherent computational efficiency. In the following subsection, we introduce the framework of LESnets and demonstrate its advantages in addressing various challenges.

\begin{figure}[htbp]
\centering
\includegraphics [width=0.9\textwidth]{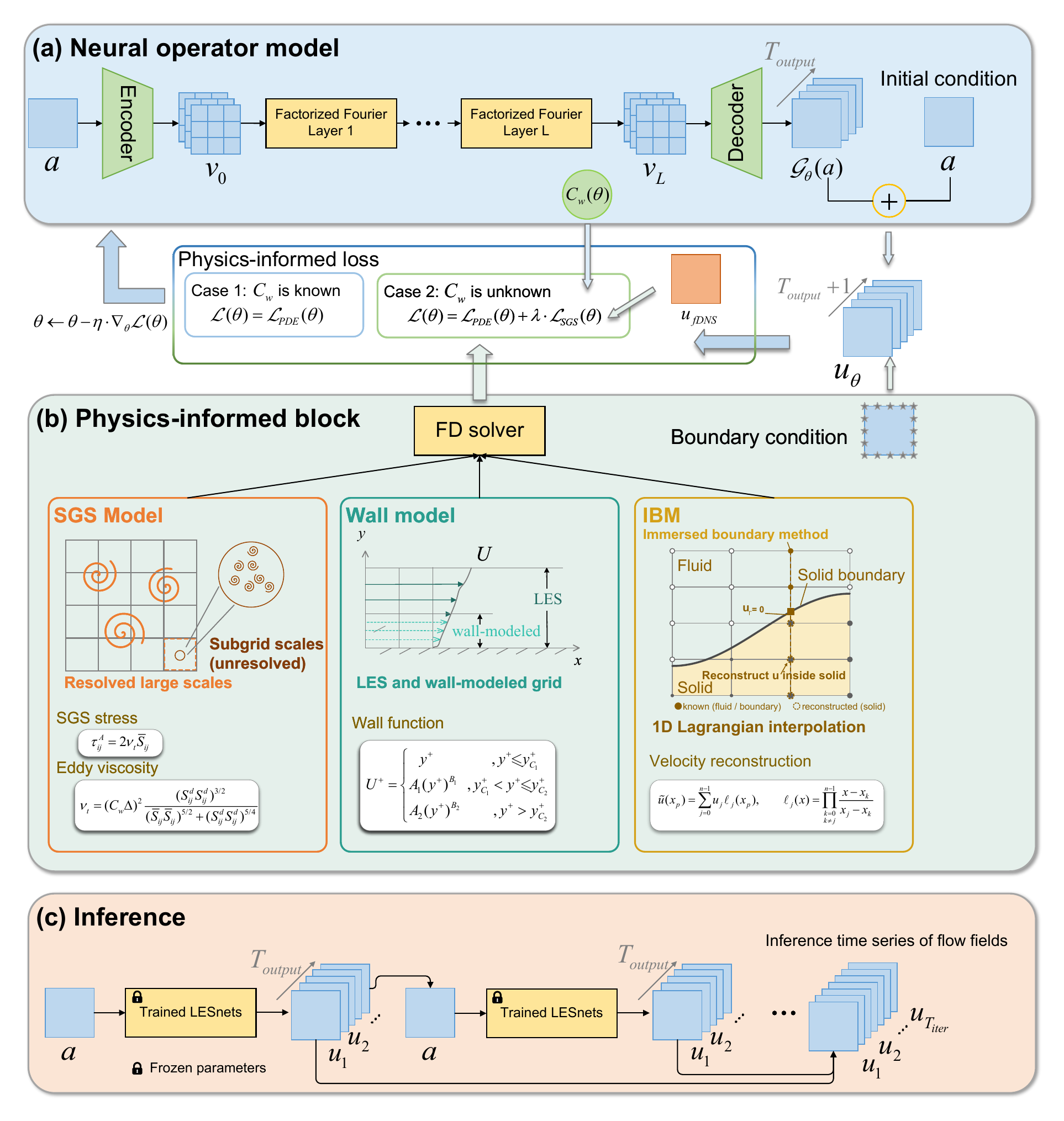}
\caption{{The architecture of LESnets. \textbf{(a)} The neural operator block for LESnets; \textbf{(b)} Physics-informed block of LESnets, comprising the subgrid-scale model, the wall model, and the immersed boundary method, together with hard-constrained boundary conditions and a finite difference solver used to calculate the physics-informed loss. Block \textbf{(a)} and \textbf{(b)} together constitute the training process of the LESnets model; \textbf{(c)} Inference process of LESnets.}}
\label{fig:lesnets-architecture}
\end{figure}

\subsection{Large-eddy simulation nets (LESnets)}
\label{subsec:lesnets}

To overcome the limitations of data-driven operator-learning methods that rely on labeled data, a physics-informed loss function is introduced to penalize deviations from the underlying physical laws. The physical constraints are then incorporated into the loss function of the neural operators. This approach is referred to as the physics-informed neural operator (PINO) \cite{PINO}. The loss function for the forward PDE problem can be written as follows:

\begin{equation}
\label{eq:pino-forward-loss}
    \mathcal{L}_\mathrm{forward}= \mathcal{L}_{\mathrm{PDE}}(a^\dagger,\mathcal{G}_\theta(a^\dagger))+\gamma\mathcal{L}_{\mathrm{data}}(u^\dagger,\mathcal{G}_\theta(a^\dagger)),
\end{equation}
where $\gamma$ is the hyper-parameter. Here, $\mathcal{L}_{\mathrm{PDE}}(a^\dagger,\mathcal{G}_\theta(a^\dagger))$ and $\mathcal{L}_{\mathrm{data}}(u^\dagger,\mathcal{G}_\theta(a^\dagger))$ are defined by Eq. \eqref{eq:physics-informed-loss} and Eq. \eqref{eq:data-loss}, respectively. $u_{\theta}$ is replaced by $\mathcal{G}_\theta(a)$ in operator learning. However, in the original PINO model \cite{PINO}, the loss term associated with the initial conditions may cause the iterative procedure to fail, unless the initial fields at later time steps are similar to those used during training. This issue has considerably hindered the further development of PINO models.

In our previous work \cite{LESnets}, we encoded the LES equations into neural operators for 3D decaying homogeneous isotropic turbulence and a temporally evolving turbulent mixing layer. In particular, initial conditions are combined directly with the subsequent output fields to evaluate the physics-informed loss, rather than relying on learnable conditions. LESnets predicted the temporal evolution of 3D turbulence by iteratively feeding the predicted velocity field back into the model as a new initial field. However, the previously proposed LESnets encounters the following challenges when applied to wall-bounded turbulent flows:
\begin{enumerate}[label=\roman*]
    \item Employing a standard FNO architecture as the neural operator can lead to reduced prediction accuracy for wall-bounded turbulent flows.
    \item The Fourier spectral methods adopted in our previous LESnets are not suitable for wall-bounded turbulence.
    \item The soft-constraint formulation of the physics-informed loss does not guarantee that the boundary conditions are satisfied exactly.
    \item The LES equations cannot accurately capture near-wall flow characteristics on coarse grids for wall-bounded turbulence at high Reynolds number without incorporating appropriate wall models.
    \item The FNO-type operator contains a large number of trainable parameters.
    \item The framework has not been extended to complex, non-grid-aligned geometries such as turbulent flow over periodic hills.
\end{enumerate}

Hence, to address these challenges in wall-bounded turbulent flows, we have made the following corresponding improvements to our previously proposed LESnets framework:

\begin{enumerate}[label=\roman*]
\item 
We enhance the neural operator architecture by integrating F-FNO, which is slightly modified and optimized for the present task, and whose factorized Fourier layers efficiently capture the dynamical features of turbulence.
\item  
We employ high-order finite difference schemes on non-uniform grids with non-periodic boundary conditions to calculate the physics-informed loss.
\item 
We enforce the boundary conditions as hard constraints in LESnets, which facilitates training and potentially improves generalization.
\item 
We incorporate a wall model into LESnets to enable efficient and accurate prediction of wall-bounded turbulence at high Reynolds number on coarse grids.
\item 
The point-wise operators $P$ and $Q$ are replaced by convolutional encoder-decoder blocks, and the coordinate inputs are removed, which reduces the number of trainable parameters.
\item 
We incorporate the immersed boundary method into the physics-informed loss so that complex, non-grid-aligned boundaries, such as the curved hill surface, can be handled on a Cartesian grid.

\end{enumerate}

\begin{algorithm}
    \caption{Training process of LESnets}
    \label{alg:lesnets-training}
    \begin{algorithmic}[1]
        \STATE \textbf{Input:} Training dataset ${\{{\bm{a}^{(i)}(0)}\}^{N}_{i=1}}, {\bm{a}^{(i)}(0)} = {\bm{a}^{(i)}(\bm{x},t=0)}$, \\ \quad\quad\quad Supplementary fDNS dataset ${\{{\bm{u}_{fDNS}^{(i)}}\}^{N_s}_{i=1}}, {\bm{u}_{fDNS}^{(i)}} = {\bm{u}_{fDNS}^{(i)}(\bm{x},t\in[1,...,T_{fDNS}])}$, \\ \quad\quad\quad Samples of training dataset $N$, samples of supplementary fDNS dataset $N_s$, number of epochs $E$, learning rate $\eta$, \\ \quad\quad\quad Batch size $N_{\mathrm{batch}}$, hyper-parameter $\lambda$, imposed boundary condition $P(\bm{y})$.
        \STATE $\theta_{0}\in\bm{\Theta}$ 
        \COMMENT{Initializing the model parameters}
        \FOR{$i=1$ to $E$}
            \STATE $\bm{u}^{pred,(i)}=\mathcal{G}_{\theta}\left({{\bm{a}^{(i)}(0)}}\right),\bm{u}^{pred,(i)}=\bm{u}^{pred,(i)}(\bm{x},t\in[1,...,T_{output}])$
            \COMMENT{Output prediction}
            \STATE $C_w(\theta)$ from $\theta$
            \COMMENT{WALE coefficient (if available)}
            \STATE ${\bm{u}^{(i)}_{\theta}}=\bm{u}^{pred,(i)}+\bm{a}^{(i)}(0),{\bm{u}^{(i)}_{\theta}}={\bm{u}^{(i)}_{\theta}}(\bm{x},t\in[1,...,T_{output}+1])$
            \COMMENT{Adding initial condition}
            \STATE ${\bm{u}^{(i)}_{\theta}(\bm{y})}=P(\bm{y}), \bm{y}\in\Omega_c$
            \COMMENT{Hard-constraining boundary condition}
            \STATE $\mathcal{L}_\mathrm{PDE}(\theta):= FD[\mathcal{R}(u^{(i)}_{\theta};C_w)]$
            \COMMENT{PDE loss}
            \STATE $\mathcal{L}_{\mathrm{SGS}}(\theta):= FD[\mathcal{R}(u^{(i)}_{fDNS};C_w(\theta))]$
            \COMMENT{SGS loss (if available)}
            \STATE $\mathcal{L}(\theta):= \mathcal{L}_\mathrm{PDE}(\theta)+\lambda\cdot\mathcal{L}_{\mathrm{SGS}}(\theta)$
            \COMMENT{Physics-informed loss}
            \STATE $\theta\leftarrow\theta-\eta\cdot\nabla_{\boldsymbol{\theta}}\mathcal{L}(\theta)$
            \COMMENT{Update neural operator parameters}
            \STATE Append $\theta$ to $\Theta$ 
            
        \ENDFOR
        \STATE Select proper $\theta$ from $\Theta$ based on physics-informed loss during the training process.
        \STATE \textbf{Output:} Trained model parameters $\theta$ and neural operator output $\mathcal{G}_{\theta}$.
    \end{algorithmic}  
\end{algorithm}

In this study, we compute the physics-informed loss function using a finite difference scheme \cite{LAIZET20095989}, with the hyper-parameters $\alpha = 0$ and $\beta = 0$ in Eq. \eqref{eq:physics-informed-loss}, while enforcing the boundary conditions as hard constraints, as detailed in Section \ref{subsubsec:hard-boundary-constraints}. Moreover, the proposed LESnets framework preserves the key advantage of not relying on any labeled data for supervision, i.e., $\gamma = 0$.

\subsubsection{The architecture of LESnets}
\label{subsubsec:lesnets-architecture}
Our LESnets architecture is shown in Fig. \ref{fig:lesnets-architecture}. Some specific features of the architecture are listed as follows:

\begin{enumerate}
\item  
The neural operator framework is similar to F-FNO, but we replace the point-wise operators $P$ and $Q$ with convolutional encoder-decoder blocks.
\item 
The input function $a$ is not a time sequence of flow fields but a single flow field. The input $a$ is lifted to a high-dimensional representation $v_0$ by an encoder block.
\item 
We remove the coordinate inputs that are commonly incorporated in the original FNO model to reduce the number of model parameters, thereby achieving better results compared with the original model.
\item 
Three 1D FFTs are performed on $v_0$ along each spatial dimension, with three independent neural network parameters $R_x$, $R_y$, and $R_z$.
\item  
By applying the iFFT to transform back to the physical space, we reconstruct the intermediate data, which are then passed through an activation function $\sigma$ to obtain $v_1$.
\item 
Finally, after passing through $L$ factorized Fourier layers, $v_L$ is projected to the desired output $\mathcal{G}_\theta(a)$, which includes $T_{output}$ output time steps using a decoder block.
\end{enumerate}

The physics-informed block is shown in Fig. \ref{fig:lesnets-architecture}(b). Here, we adopt the filtered NS equations with the WALE model as the governing equations. We use the finite difference schemes in the physical space to compute the necessary gradient operations of the PDE at specified collocation points. Additional details on the specific finite difference stencils used in this study are provided in \ref{app:finite-difference-loss}. Furthermore, we incorporate a wall model to represent the law of the wall at high Reynolds numbers, and an immersed boundary method (Section \ref{subsec:IBM}) to handle non-grid-aligned solid surfaces such as the curved hill of the turbulent flow over periodic hills. The parameters $\theta$, which include all the weights and biases of the neural networks, are optimized using the physics-informed loss function:

\begin{equation}
\label{eq:lesnets-pde-loss}
    \mathcal{L}(\theta)=\mathcal{L}_{\mathrm{PDE}}(\theta)= FD[\mathcal{R}(u_{\theta};C_w)].
\end{equation}
Here, $\mathcal{R}$ denotes the residual operator of the governing equations, i.e., Eqs. \eqref{eq:filtered-continuity} and \eqref{eq:filtered-momentum}, and FD denotes the finite difference solver.

When the coefficient of the WALE model $C_w$ is unknown, $C_w$ can be regarded as a learnable parameter in neural networks by incorporating fDNS data $u_{fDNS}$ in the training process and adding an additional term in the loss function, detailed in Section \ref{subsubsec:learning-les-coefficient}. Then the physics-informed loss function is given by

\begin{equation}
\label{eq:learnable-sgs-loss}
    \mathcal{L}(\theta)= \mathcal{L}_{\mathrm{PDE}}(\theta)+\lambda\cdot\mathcal{L}_{\mathrm{SGS}}(\theta)= FD[\mathcal{R}(u_{\theta};C_w)]+\lambda \cdot FD[\mathcal{R}(u_{fDNS};C_w(\theta))],
\end{equation}
where $\lambda$ is the loss weight hyper-parameter.

\begin{algorithm}
    \caption{Inference process of LESnets}
    \label{alg:lesnets-inference}
    \begin{algorithmic}[1]
        \STATE \textbf{Input:} Trained LESnets model $\mathcal{G}_{\theta}$, inference initial data ${\bm{a}^0}$, number of iterations $T_{iter}$.
        \STATE ${\bm{a}}={\bm{a}^0}$
        \COMMENT{Initializing the model input}
        \FOR{$k=1$ to $T_{iter}$}
            \STATE $\bm{u}^{pred}=\mathcal{G}_{\theta}\left({\bm{a},t\in[1,...,T_{output}]}\right)$
            \COMMENT{Apply the trained LESnets}
            \STATE Learned $C_w$
            \COMMENT{WALE coefficient (if available)}
            \STATE ${\bm{u}^{(k)}_{\theta}(\bm{y})}=P(\bm{y}), \bm{y}\in\Omega_c$
            \COMMENT{Hard-constraining boundary condition}
            \STATE $\bm{u}^{(k)}=(\bm{u}^{pred}_{t})_{t\in\mathcal{I}},\mathcal{I}\subseteq\{1,...,T_{output}\}$
            \COMMENT{Save the flow fields at specified time steps}
            \STATE ${\bm{a}}=\bm{u}^{pred}(t=T_{output})$ 
            \COMMENT{Set the new input data}

        \ENDFOR
        \STATE \textbf{Output:} The inference time series of flow fields $\{\bm{u}^{k}\}_{k=1}^{T_{iter}}$.
    \end{algorithmic}  
\end{algorithm}

\subsubsection{Training and inference}
\label{subsubsec:training-inference}
To ensure that physical constraints are properly integrated into the LESnets model, we develop the following training and prediction strategies. Before training, we select the hyper-parameters and impose additional constraints on the output fields. The complete training procedure is summarized in Algorithm \ref{alg:lesnets-training}, where $T_{fDNS}$ denotes the number of time steps of the supplementary fDNS fields $u_{fDNS}$. The training process is briefly summarized as follows:

\begin{enumerate}
\item
Input the prepared training data ${\bm{a}^{(i)}(0)}$ and output the prediction $\bm{u}^{pred,(i)}$ through the neural operator $\mathcal{G}_{\theta}$.
\item
Enforce the boundary conditions as hard constraints. 
\item
Calculate the PDE loss $\mathcal{L}_{\mathrm{PDE}}(\theta)$ and SGS loss $\mathcal{L}_{\mathrm{SGS}}(\theta)$ (if the coefficient of the SGS model is unknown).
\item
Optimize the model parameters and select the appropriate model parameters for the inference stage.
\end{enumerate}

For inference, we employ an autoregressive strategy, as shown in Fig. \ref{fig:lesnets-architecture}(c), iteratively feeding the predicted results back into the trained model to capture the long-term evolution of the corresponding turbulent flow structures. The autoregressive inference procedure is described in Algorithm \ref{alg:lesnets-inference}. The inference process is briefly summarized as follows:

\begin{enumerate}
\item
Input the initial data $\bm{a}^0$ as the model input to the trained LESnets model and begin iteration.
\item
Enforce the boundary conditions $P(\bm{y})$ as hard constraints on the output data ${\bm{u}^{(k)}_{\theta}(\bm{y})}$ at the $k$th iteration. 
\item
The resulting flow fields are used as input data (initial conditions) for the next iteration. 
\item
Collect the full inference time series of flow fields $\{\bm{u}^{k}\}_{k=1}^{T_{iter}}$, where ${T_{iter}}=T_{inf}/T_{output}$ is the number of iterations and $T_{inf}$ is the number of inference time steps.  

\end{enumerate}

\subsubsection{Hard constraints on boundary conditions}
\label{subsubsec:hard-boundary-constraints}

It is important to note that FNO-type models are particularly well-suited to problems with periodic boundary conditions, owing to their network architecture. This limitation can be alleviated by extracting the mean velocity fields \cite{wang2024prediction} or mapping the complex, non-periodic physical geometric domains into standard rectangular or cubic computational domains \cite{Geo_FNO}. Another problem is that the loss function presented in Eq. \eqref{eq:physics-informed-loss} adopts a soft-constraint formulation, in which the initial and boundary conditions are enforced through the loss function. As a result, the optimizer minimizes the deviation from the true initial and boundary conditions, but there is no strict guarantee that the resulting solution will satisfy them exactly. Consequently, designing the network architecture to explicitly contain exact physical constraints can accelerate the training process and enhance the model's generalization ability.

\begin{figure}[htbp]
\centering
\begin{minipage}{0.49\linewidth} 
\centerline{\includegraphics[width=\textwidth]{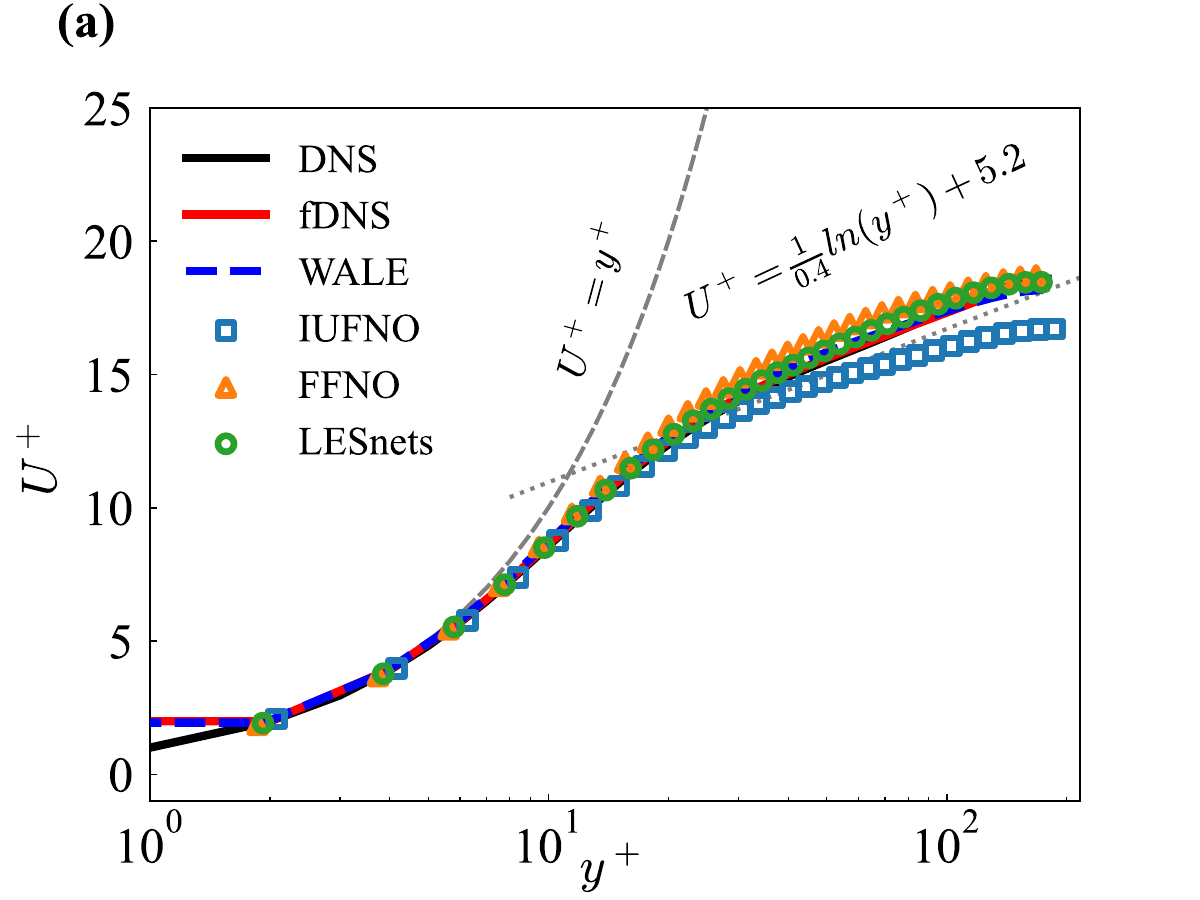}}
\end{minipage}
\hspace{-2pt} 
\begin{minipage}{0.49\linewidth} 
\centerline{\includegraphics[width=\textwidth]{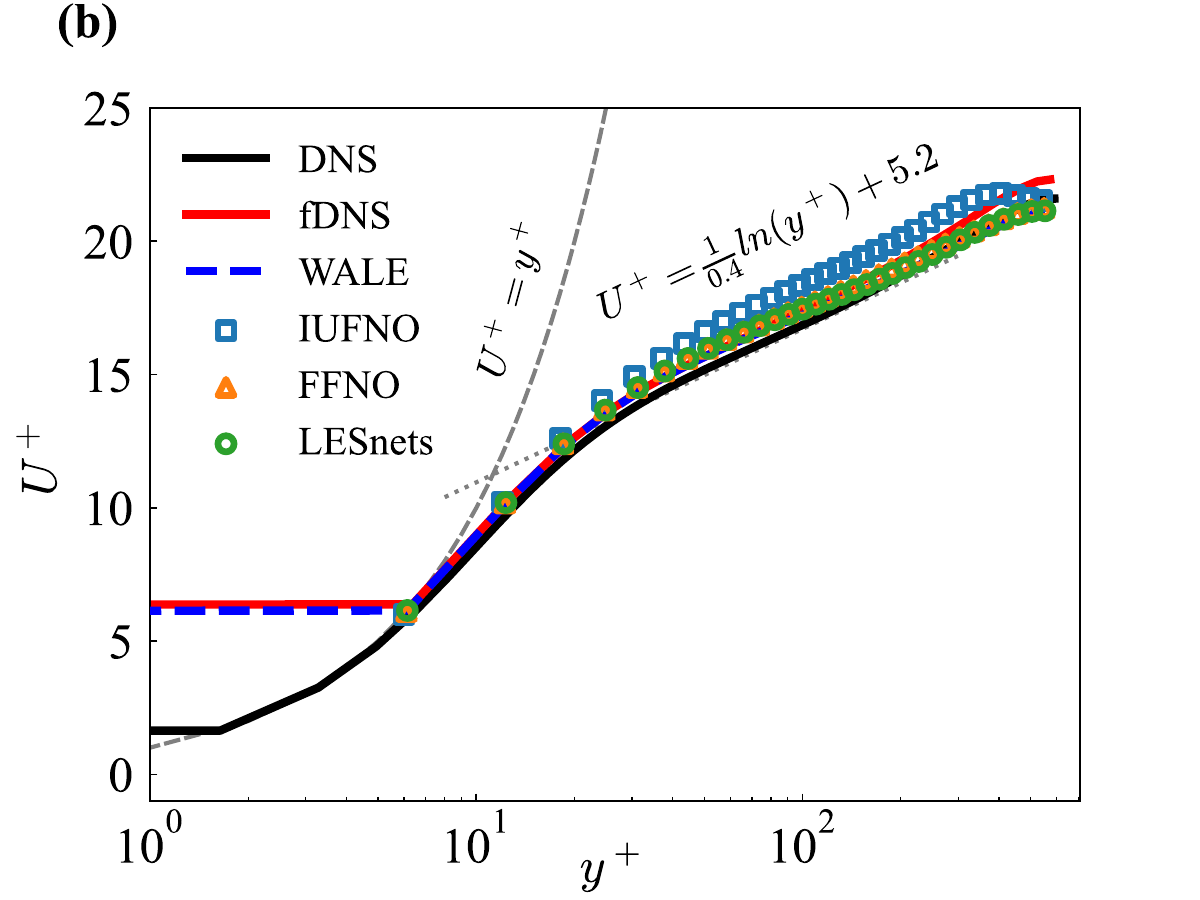}}
\end{minipage}

\begin{minipage}{0.49\linewidth} 
\centerline{\includegraphics[width=\textwidth]{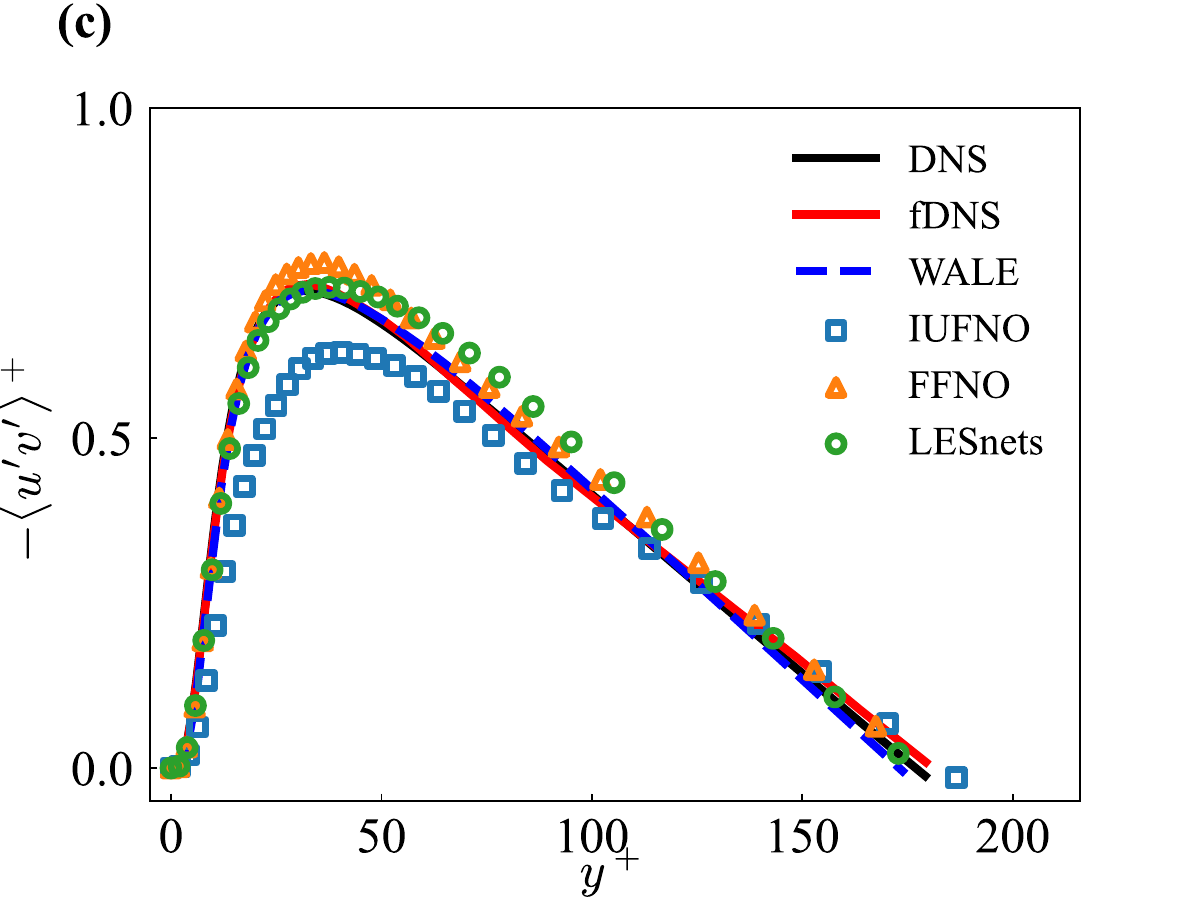}}
\end{minipage}
\hspace{-2pt} 
\begin{minipage}{0.49\linewidth} 
\centerline{\includegraphics[width=\textwidth]{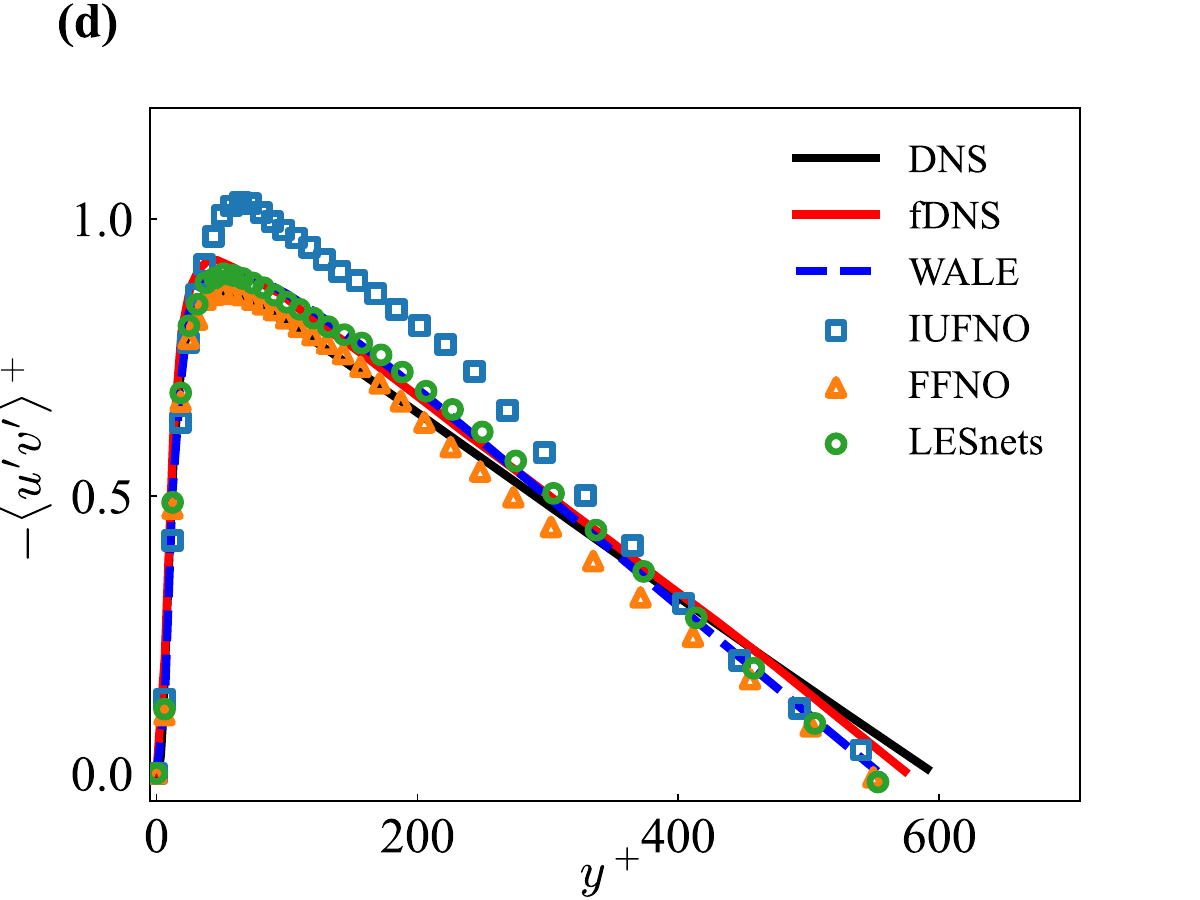}}
\end{minipage}
\caption{Comparison between DNS, fDNS, WALE, and three machine-learning models of turbulent channel flow at two friction Reynolds numbers: \textbf{(a)} $Re_{\tau}\approx180$, normalized mean streamwise velocity profiles, normalized by the respective friction velocity $U^+ = \langle \overline{u}\rangle/u_{\tau}$; \textbf{(b)} $Re_{\tau}\approx590$, normalized mean streamwise velocity profiles; \textbf{(c)} $Re_{\tau}\approx180$, normalized Reynolds shear stress $-\langle u'v'\rangle^+$, normalized by the square of respective friction velocity $u_{\tau}^2$; \textbf{(d)} $Re_{\tau}\approx590$, normalized Reynolds shear stress $-\langle u'v'\rangle^+$.}
\label{fig:mean-velocity-reynolds-stress}
\end{figure}

To address this issue, we enforce the boundary conditions as hard constraints, applied to the predicted velocity field after the model output and before the physics-informed loss is calculated. For the turbulent channel flow, the no-slip condition is imposed by setting the velocity to zero on the two wall planes, $y=\pm1$, while the periodic directions remain unconstrained. For the turbulent flow over periodic hills, whose curved surface is not aligned with the Cartesian grid, the solid region is first identified by a mask and the velocity at every grid point inside this mask is set to zero; the immersed boundary method described in Section \ref{subsec:IBM} then reconstructs the velocities at the grid nodes inside the solid body by 1D Lagrangian interpolation, so that the finite difference stencil near the curved boundary is well-posed. Formally,

\begin{equation}
    {\bm{u}(\bm{y})}=P(\bm{y}),\qquad \bm{y}\in\Omega_c,
\label{eq:no-slip-boundary}
\end{equation}
where $P(\bm{y})=0$, and $\Omega_c$ denotes the set of constrained grid points, i.e., the two wall planes for the channel flow, or the solid mask for the turbulent flow over periodic hills.

\begin{figure}[htbp]
\centering
\begin{minipage}{0.49\linewidth}
\centerline{\includegraphics[width=\textwidth]{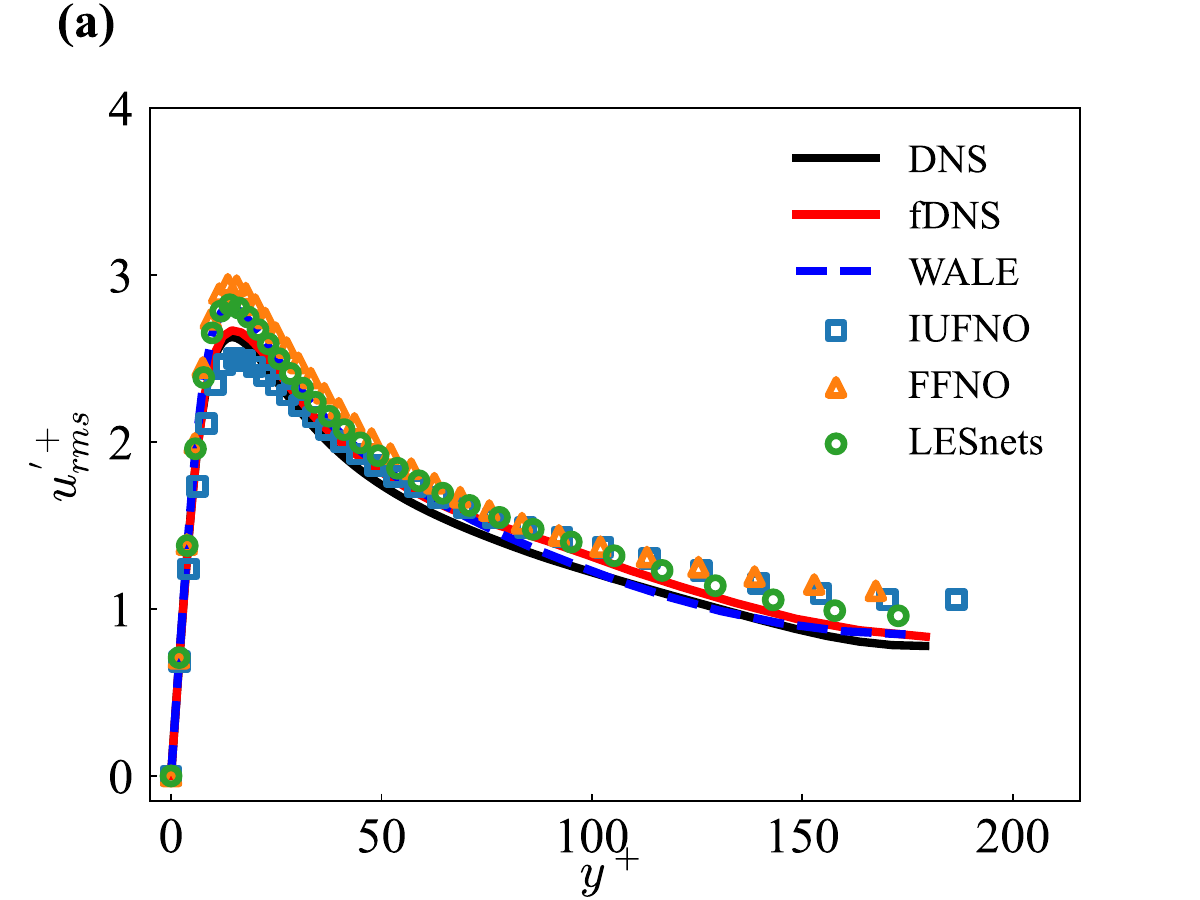}}
\end{minipage}
\hspace{-2pt}
\begin{minipage}{0.49\linewidth}
\centerline{\includegraphics[width=\textwidth]{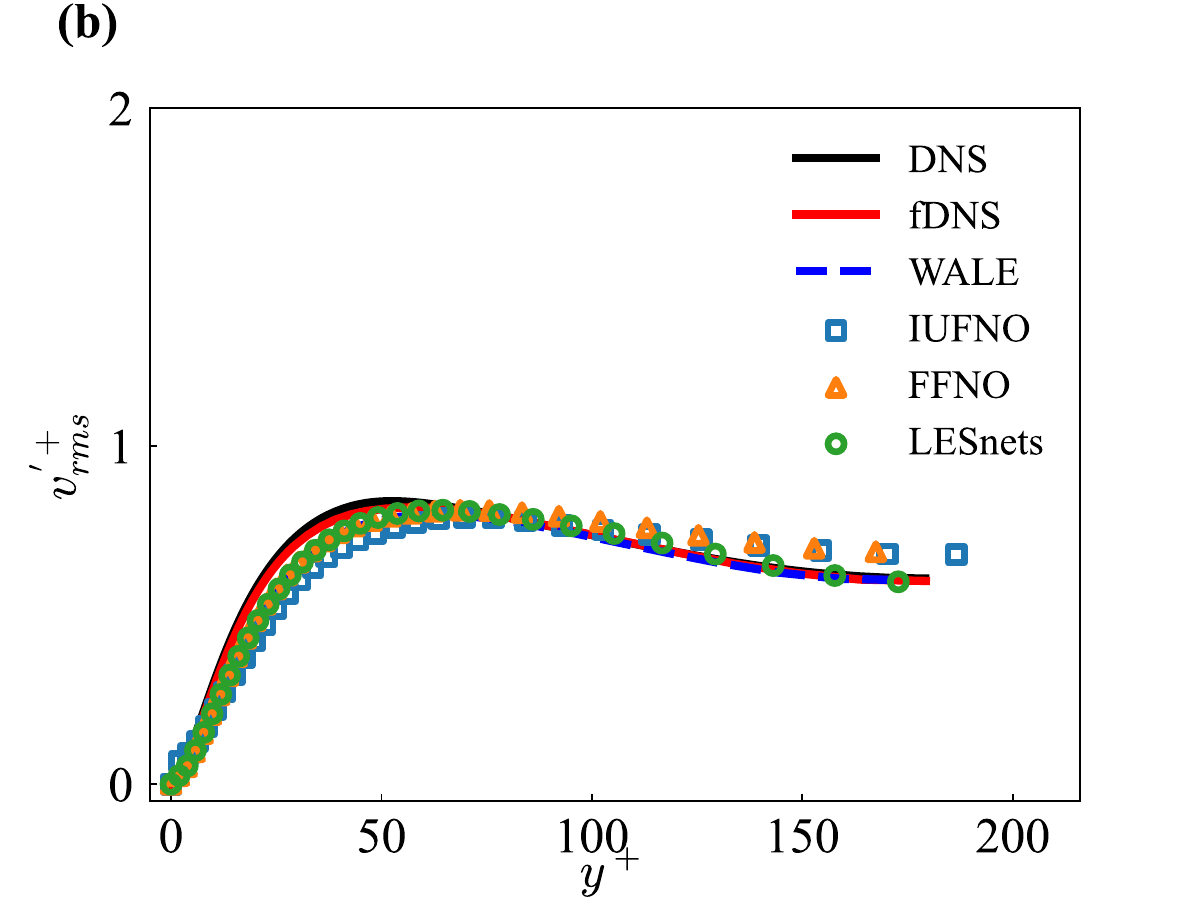}}
\end{minipage}

\begin{minipage}{0.49\linewidth}
\centerline{\includegraphics[width=\textwidth]{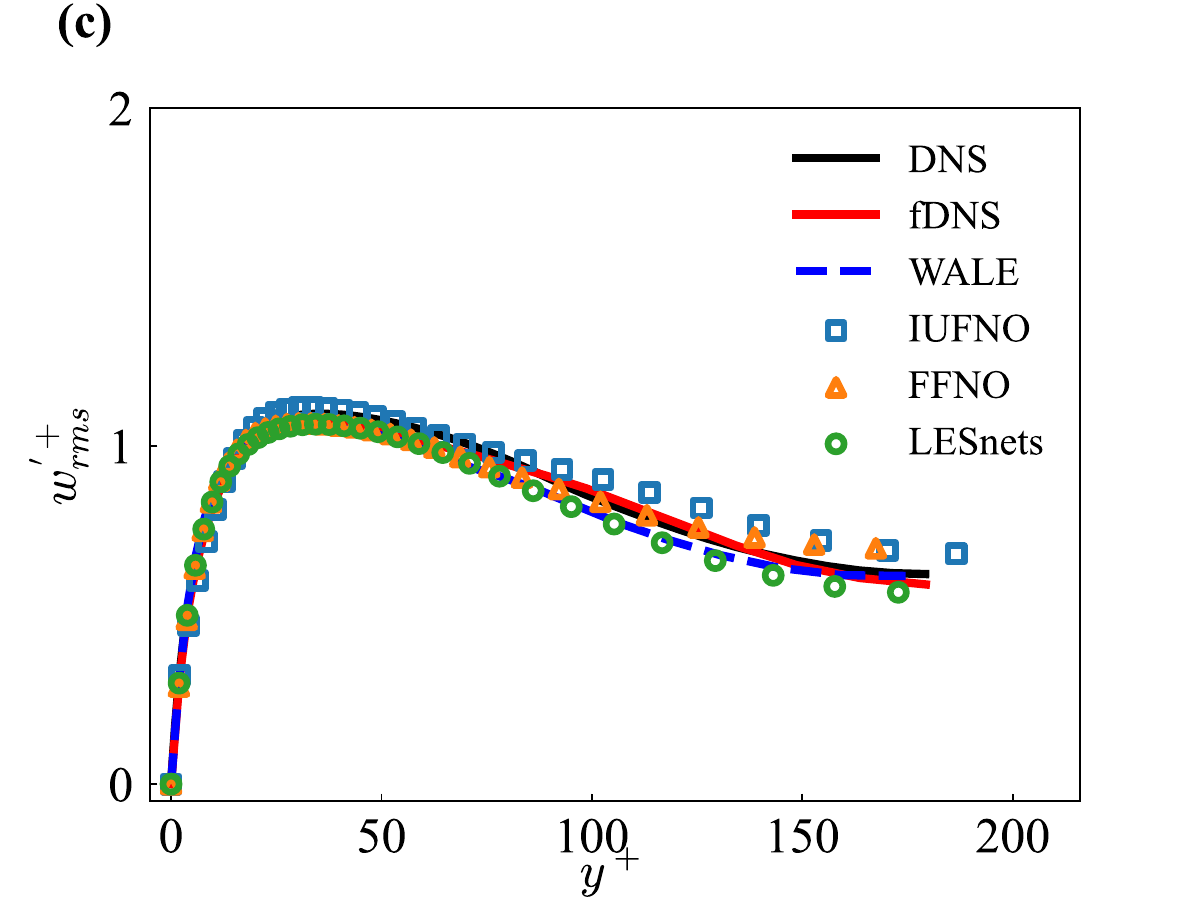}}
\end{minipage}
\caption{Comparison between DNS, fDNS, WALE, and three machine-learning models of turbulent channel flow at friction Reynolds number $Re_{\tau}\approx180$. Root-mean-square (RMS) fluctuations of \textbf{(a)} streamwise velocity; \textbf{(b)} wall-normal velocity; \textbf{(c)} spanwise velocity.}
\label{fig:re180-rms-fluctuations}
\end{figure}

\begin{figure}[htbp]
\centering
\begin{minipage}{0.49\linewidth}
\centerline{\includegraphics[width=\textwidth]{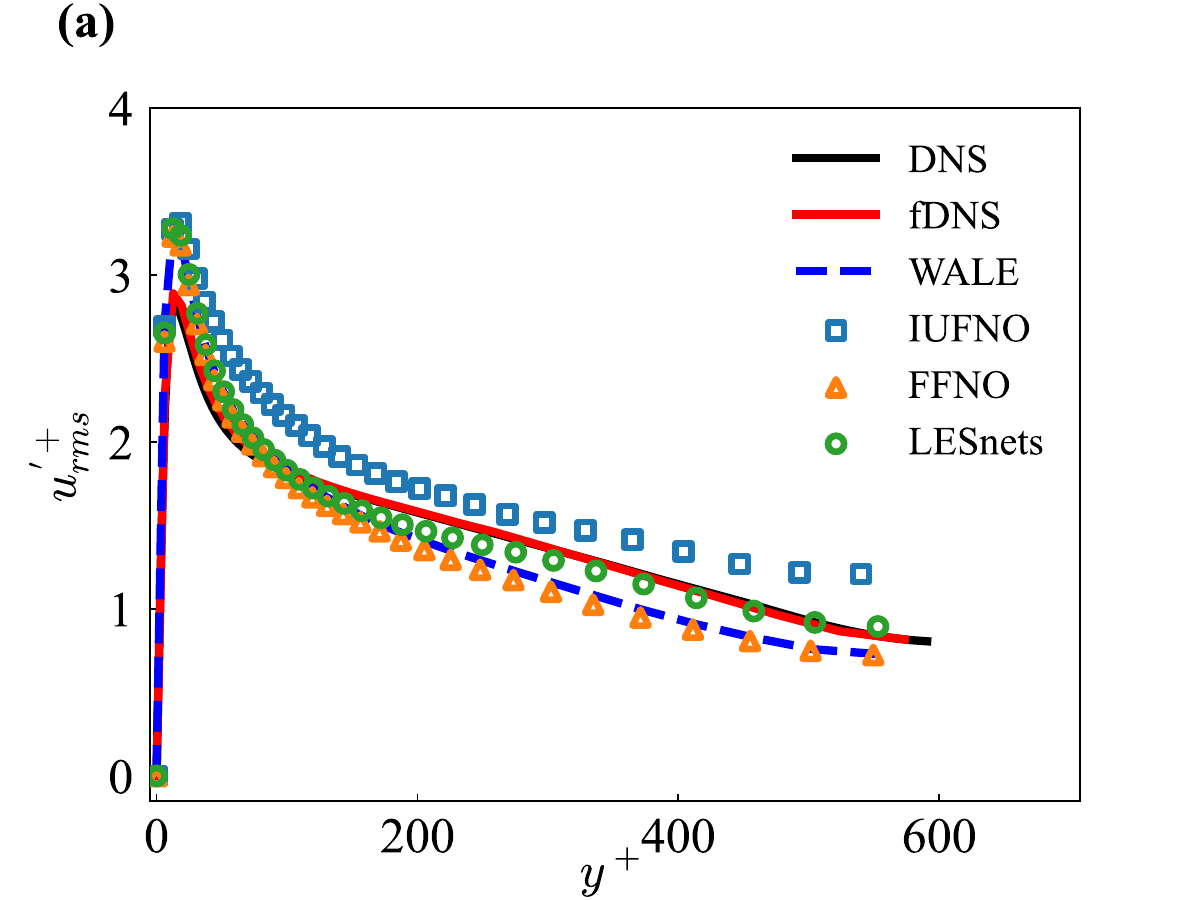}}
\end{minipage}
\hspace{-2pt}
\begin{minipage}{0.49\linewidth}
\centerline{\includegraphics[width=\textwidth]{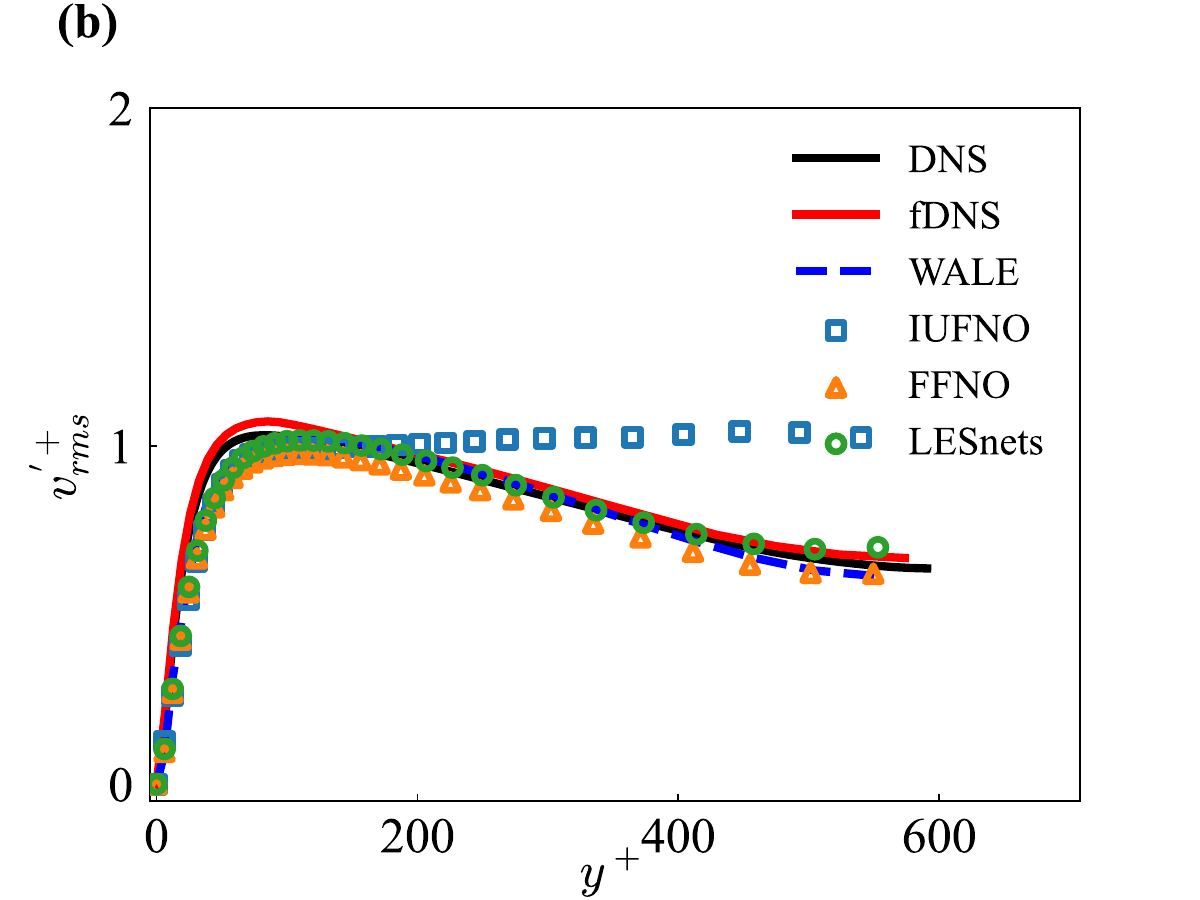}}
\end{minipage}

\begin{minipage}{0.49\linewidth}
\centerline{\includegraphics[width=\textwidth]{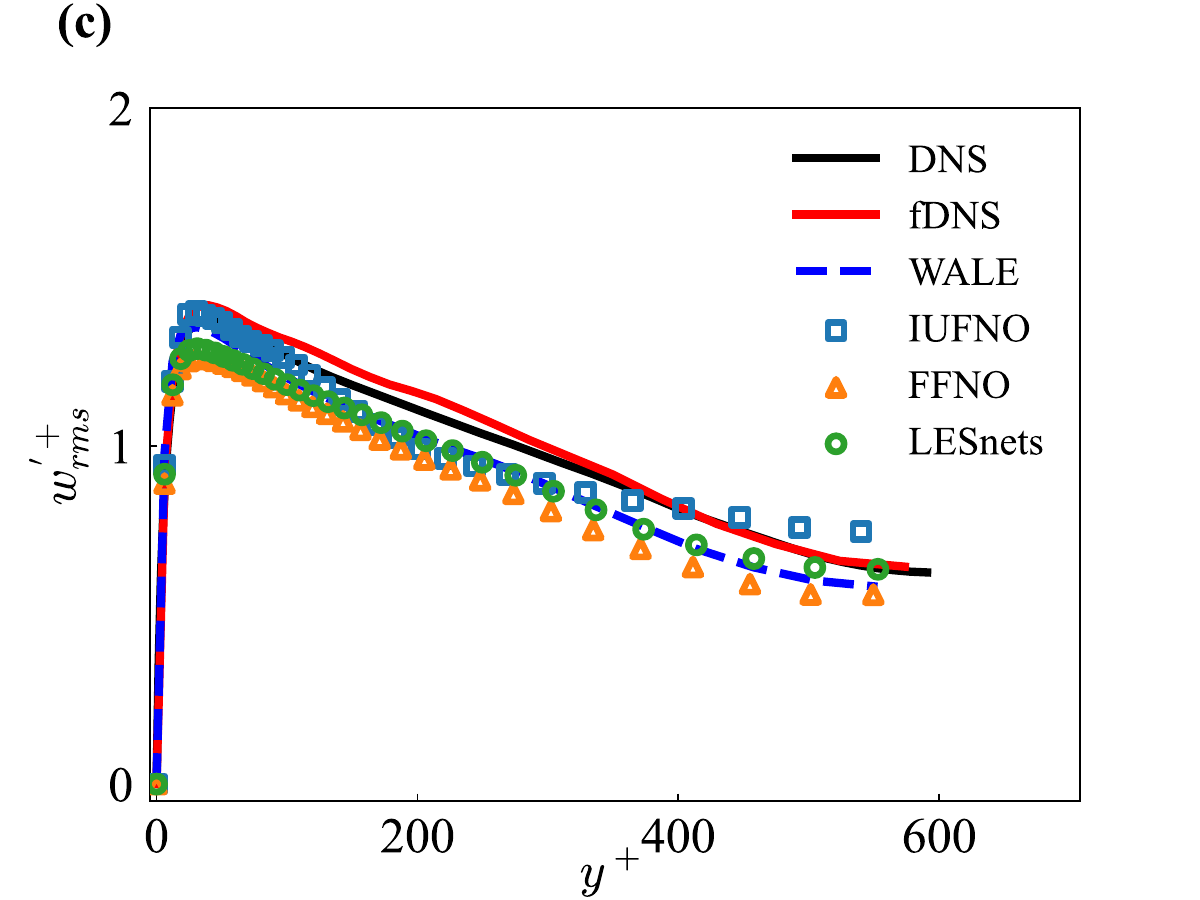}}
\end{minipage}
\caption{Comparison between DNS, fDNS, WALE, and three machine-learning models of turbulent channel flow at friction Reynolds number $Re_{\tau}\approx590$. RMS fluctuations of \textbf{(a)} streamwise velocity; \textbf{(b)} wall-normal velocity; \textbf{(c)} spanwise velocity.}
\label{fig:re590-rms-fluctuations}
\end{figure}

\section{Results and discussion}
\label{sec:results-and-discussion}

The methodology presented in Section \ref{sec:methodology} is tested here on two test cases: 3D incompressible turbulent channel flows in Section \ref{subsec:case-1}, and the 3D incompressible turbulent flow over periodic hills in Section \ref{subsec:case-2}. To provide a comparative analysis, we introduce two types of data-driven neural operators, including IUFNO and F-FNO, as baseline models. For all three machine-learning models, we impose the boundary conditions as hard constraints, as described in Section \ref{subsubsec:hard-boundary-constraints}.

Due to variations in dataset size and available computational resources, we adjust the batch size and the number of implicit iterations for the IUFNO model. All other hyper-parameters are identical to those in the previous open-source implementation, and the corresponding code is available at \href{https://github.com/Turbulence-AI/IUFNO-CHL}{https://github.com/Turbulence-AI/IUFNO-CHL}. 

To ensure fairness in the numerical experiments, an identical neural operator framework and parameter settings are used for F-FNO and our proposed LESnets. Specifically, the parameters for LESnets and F-FNO are set as follows: layers = 4, width = 80, and batch size = 4. The activation function is $tanh(x)=({e^x-e^{-x}})/({e^x+e^{-x}})$. The models are trained by minimizing the loss over $2\times10^{3}$ iterations using the SOAP optimizer \cite{SOAP}, with parameters $\beta_1 = \beta_2 = 0.99$. The initial learning rate is $1\times10^{-3}$, followed by an exponential decay with a factor of 0.8 every 200 steps. These parameters are held constant across all numerical experiments unless explicitly stated otherwise. The hyper-parameters used for training the three neural operator models are summarized in \ref{app:hyperparameters}. For the turbulent channel flows, all machine-learning models are trained on the same four datasets, namely $\mathcal{C}_{train}^{180}$, $\mathcal{C}_{train}^{590}$, $\mathcal{C}_{train}^{1000}$, and $\mathcal{C}_{train}^{2000}$, each of which contains $T_{train}=200$ time steps. For the turbulent flow over periodic hills, all machine-learning models are trained on a single dataset, $\mathcal{H}_{train}^{5600}$, which contains $T_{train}=800$ time steps. For consistency, all models are trained and evaluated on the same hardware: training is performed on a single NVIDIA A100 40 GB PCIe GPU, while data and model parameters are loaded on a Huawei Kunpeng-920 CPU at 3.00 GHz. It is worth emphasizing that sharing the same training datasets among the three models is intended for convenience and a fair comparison. The proposed LESnets model, however, does not rely on training datasets with a prescribed temporal ordering, as will be detailed in Section \ref{subsubsec:flexible-output}.

In the $a$ $posteriori$ test, the results obtained from the three traditional methods, including DNS, fDNS, and WALE, are used as reference values to assess the performance of the machine-learning models. For the F-FNO and LESnets, an identical initial field is provided; for the IUFNO \cite{wang2024prediction}, five consecutive flow fields are given to match the model's input and output, ensuring that the data at the last time step align with the initial fields of the other two models. All models perform iterative inference by recursively using their predicted velocity fields as the updated initial fields. The total inference horizon consists of $T_{inf}=800$ inference time steps. Unless otherwise specified, all statistical results are time-averaged over the entire inference period
covering $T_{inf}=800$ inference time steps.

\begin{figure}[htbp]
\centering
\begin{minipage}{0.49\linewidth} 
\centerline{\includegraphics[width=\textwidth]{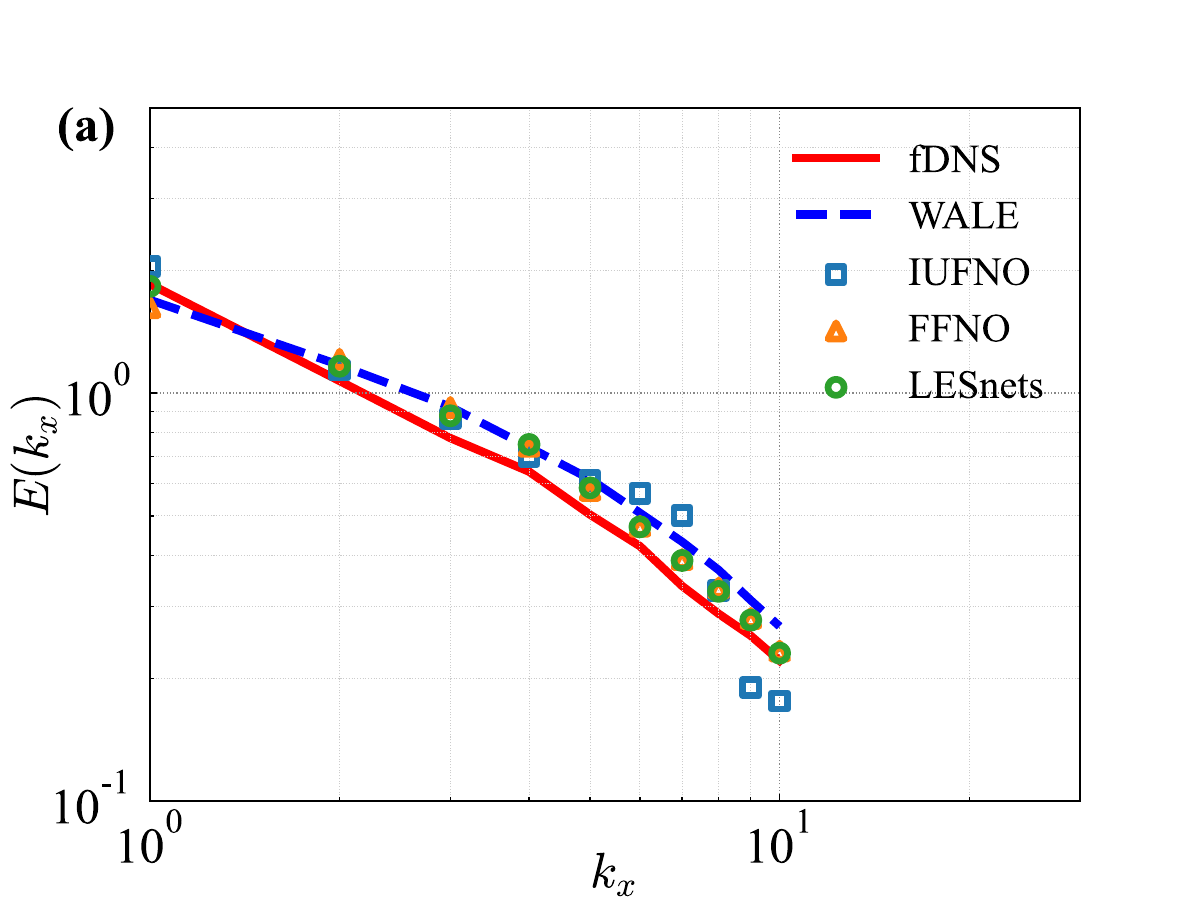}}
\end{minipage}
\hspace{-2pt} 
\begin{minipage}{0.49\linewidth} 
\centerline{\includegraphics[width=\textwidth]{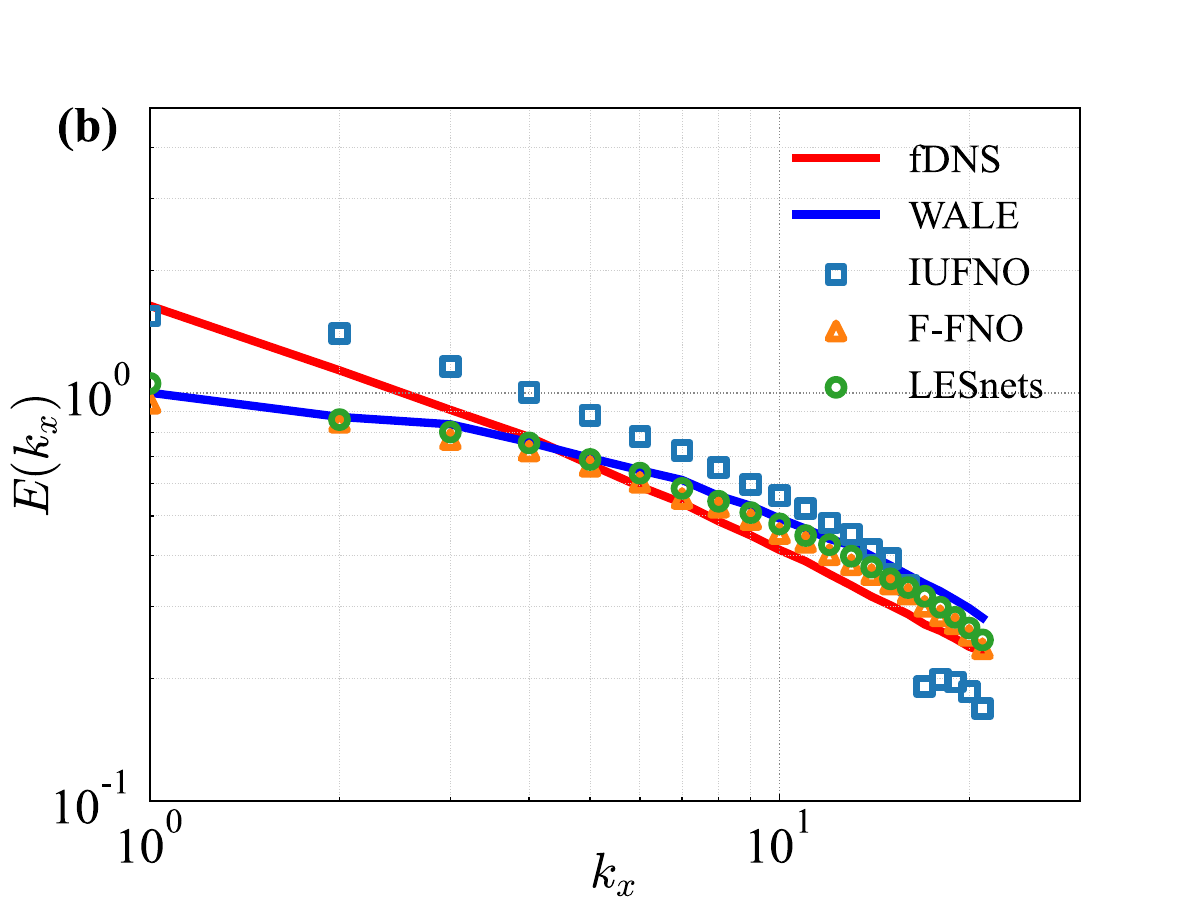}}
\end{minipage}
\begin{minipage}{0.49\linewidth} 
\centerline{\includegraphics[width=\textwidth]{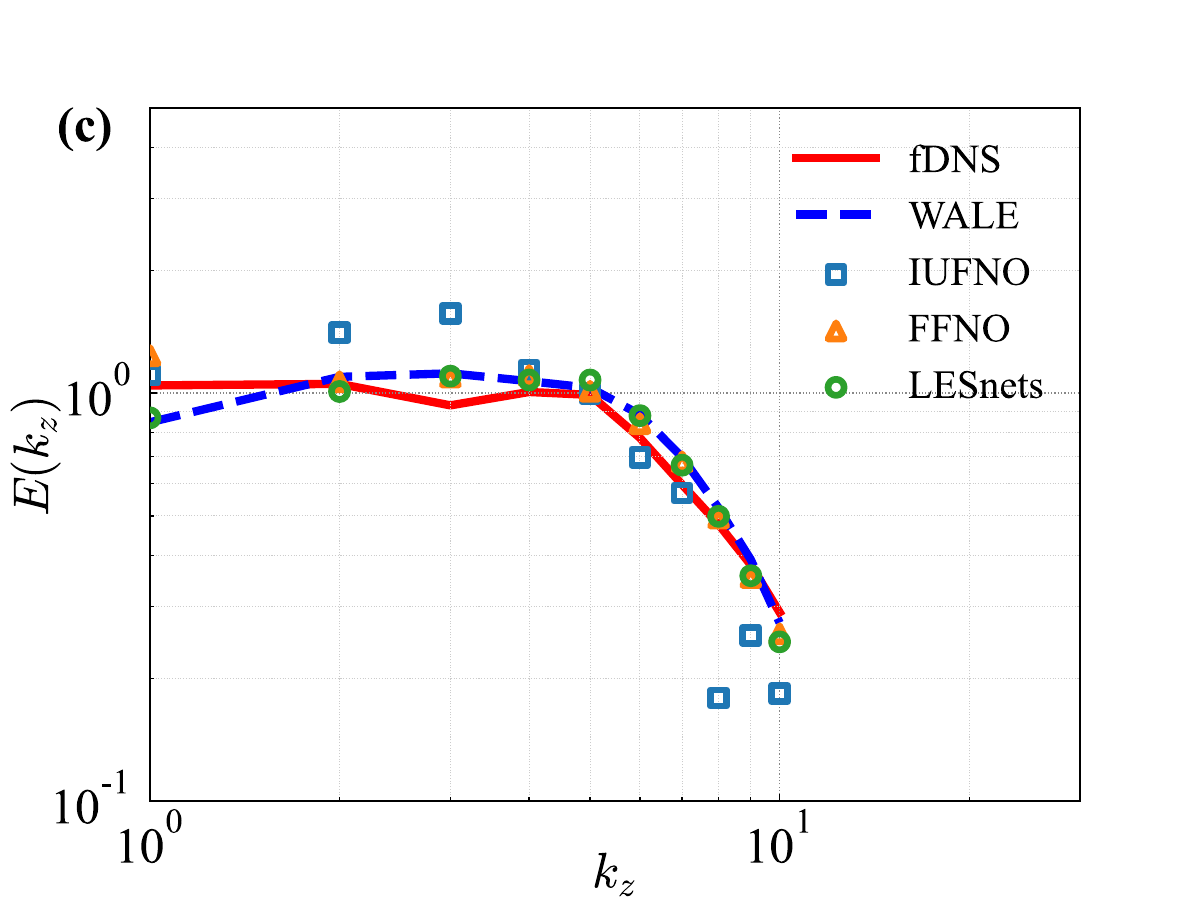}}
\end{minipage}
\hspace{-2pt} 
\begin{minipage}{0.49\linewidth} 
\centerline{\includegraphics[width=\textwidth]{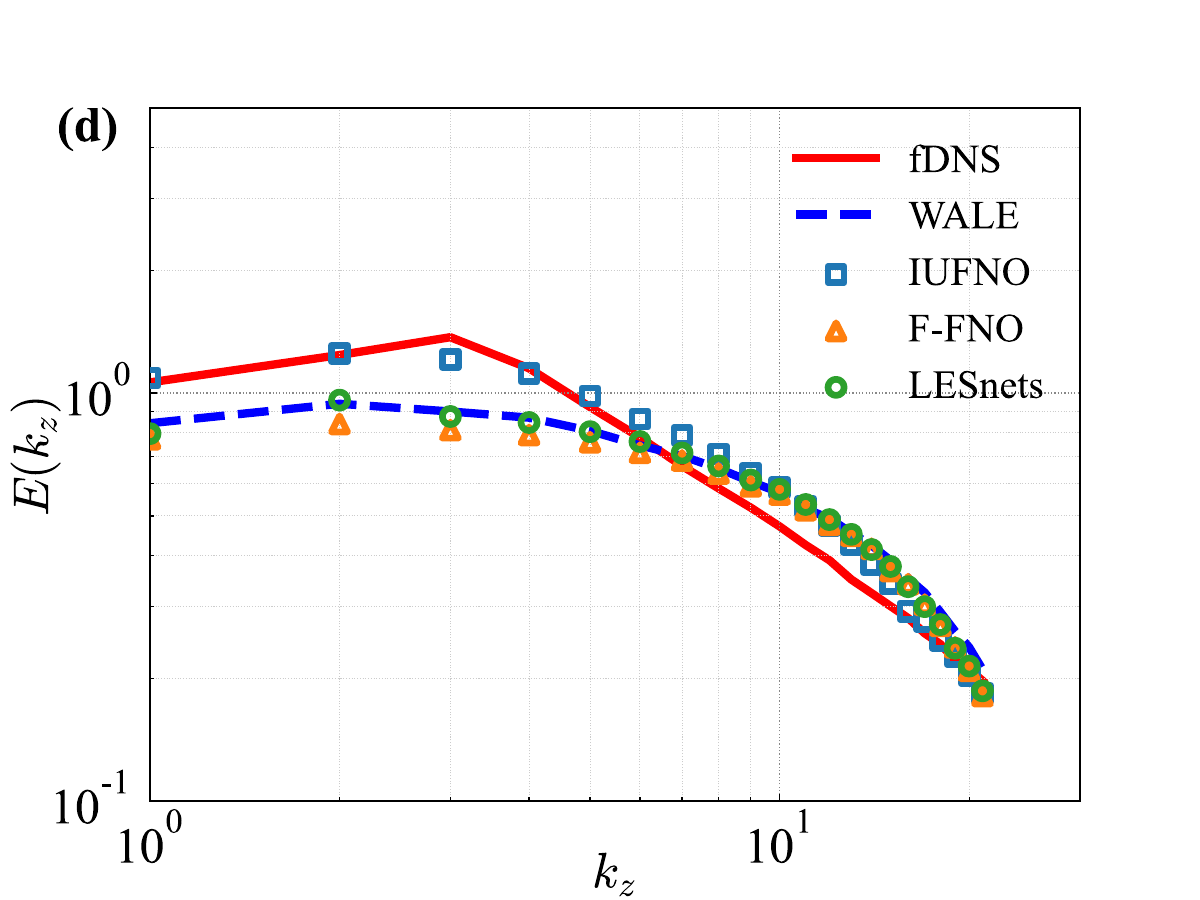}}
\end{minipage}
\caption{The kinetic energy spectra along \textbf{(a)} $Re_{\tau}\approx180$, streamwise direction ($k_x$); \textbf{(b)} $Re_{\tau}\approx590$, streamwise direction ($k_x$); \textbf{(c)} $Re_{\tau}\approx180$, spanwise direction ($k_z$); \textbf{(d)} $Re_{\tau}\approx590$, spanwise direction ($k_z$) comparison between fDNS, WALE, and three machine-learning models of turbulent channel flow at two friction Reynolds numbers $Re_{\tau}\approx180$ and $590$.}
\label{fig:energy-spectra-re180-re590}
\end{figure}

In Section \ref{subsubsec:model-training-evaluation}, we evaluate the proposed LESnets model at friction Reynolds numbers of $Re_{\tau}\approx180$ and $590$. In Section \ref{subsubsec:lesnets-wall-model}, we integrate a wall model into LESnets (referred to as LESnets-WM) to enable accurate and efficient simulations of wall-bounded turbulence at high Reynolds number on coarse grids, and assess the models at $Re_{\tau}\approx1000$ and $2000$. In Section \ref{subsec:case-2}, we examine the turbulent flow over periodic hills, whose wall boundary is not aligned with the computational grid and is treated with the immersed boundary method of Section \ref{subsec:IBM}. In Section \ref{subsec:framework-properties-and-analysis}, we further investigate the properties of the framework: its flexibility with respect to the choice of output time interval (Section \ref{subsubsec:flexible-output}), a learning strategy that simultaneously optimizes the underlying operator and the SGS model coefficient by leveraging available fDNS data (Section \ref{subsubsec:learning-les-coefficient}), and the computational efficiency of the three machine-learning models relative to the WALE (Section \ref{subsubsec:computational-efficiency}).


\subsection{Test case 1: 3D incompressible turbulent channel flows}
\label{subsec:case-1}

\subsubsection{Model training and evaluations at \texorpdfstring{$Re_{\tau} \approx 180$}{} and \texorpdfstring{$590$}{}}
\label{subsubsec:model-training-evaluation}
The two training datasets $\mathcal{C}_{train}^{180}$ and $\mathcal{C}_{train}^{590}$ are used to train three machine-learning models (i.e., IUFNO, F-FNO, and our LESnets) at $Re_{\tau} \approx 180$ and $590$, respectively. The training datasets $\mathcal{C}_{train}^{180}$ and $\mathcal{C}_{train}^{590}$ comprise 20 groups of well-resolved LES fields with the WALE model, and are organized in tensors of dimensions $N\times T_{train} \times N_x\times N_y \times N_z \times N_d=20\times200\times32\times65\times32\times4$ and $20\times200\times64\times65\times64\times4$, respectively. Here, $N$ is the number of samples in the training dataset, $T_{train}$ is the number of training time steps, $N_x,N_y,N_z$ are the grid resolutions, and $N_d=4$ is the number of velocity components and pressure. In this subsection, we use the WALE coefficient $C_w=0.1$ \cite{nicoud1999subgrid} to compute the PDE loss. Meanwhile, the number of output time steps of LESnets and F-FNO is $T_{output}=1$, and the output time interval is $\Delta T$. Our goal is to assess whether LESnets can accurately reproduce turbulent channel flow statistics over an inference horizon of $T_{inf}$ time steps with 800 iterations.

\begin{figure}[htbp]
\centering
\includegraphics [width=1.0\textwidth]{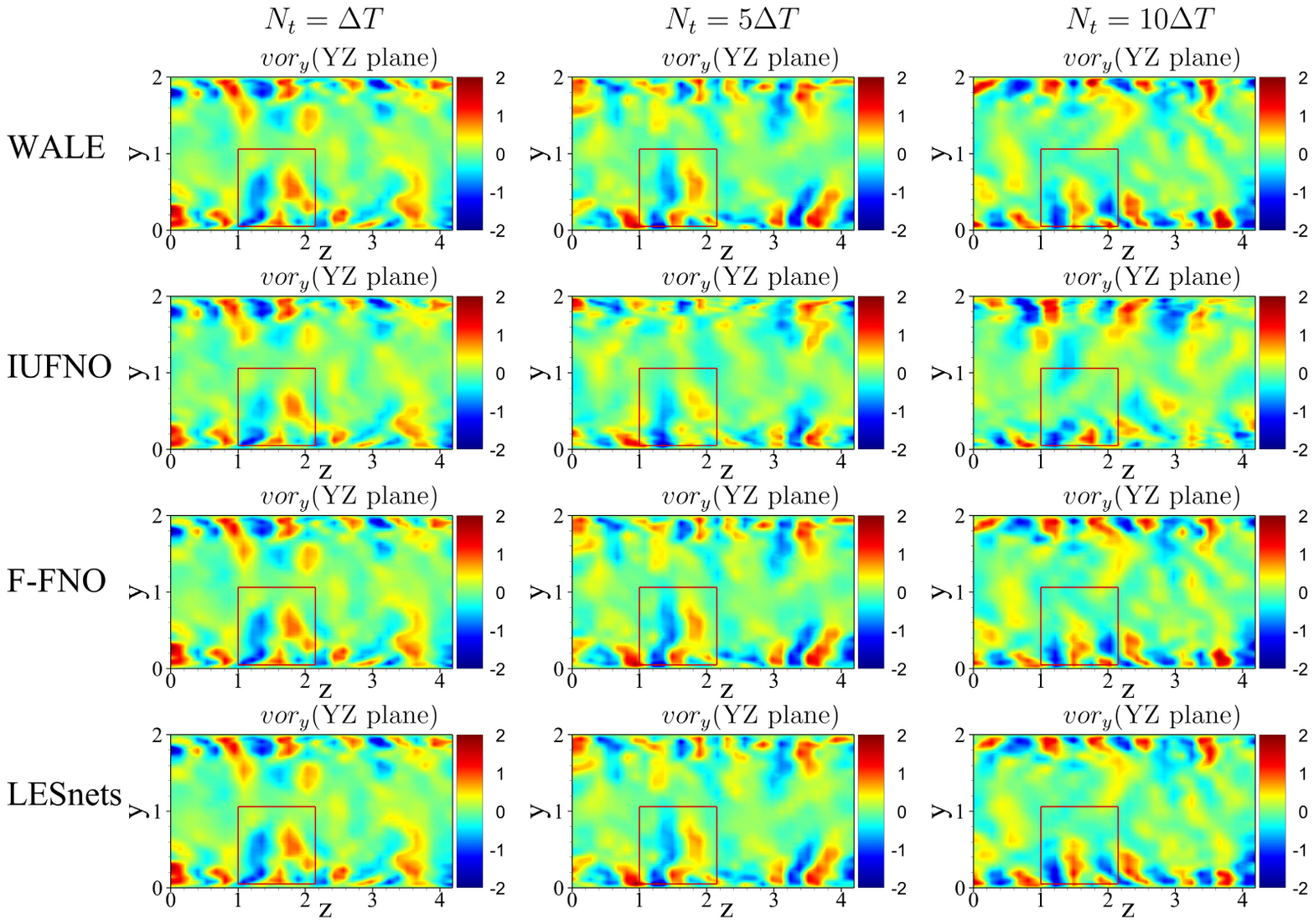}
\caption{Wall-normal vorticity field slices of turbulent channel flow at friction Reynolds number $Re_{\tau}\approx180$ in time series $N_t=\Delta T$, $N_t=5\Delta T$, and $N_t=10\Delta T$ (at the center YZ plane).}
\label{fig:re180-vorticity-evolution}
\end{figure}
\begin{figure}[htbp]
\centering
\includegraphics [width=1.0\textwidth]{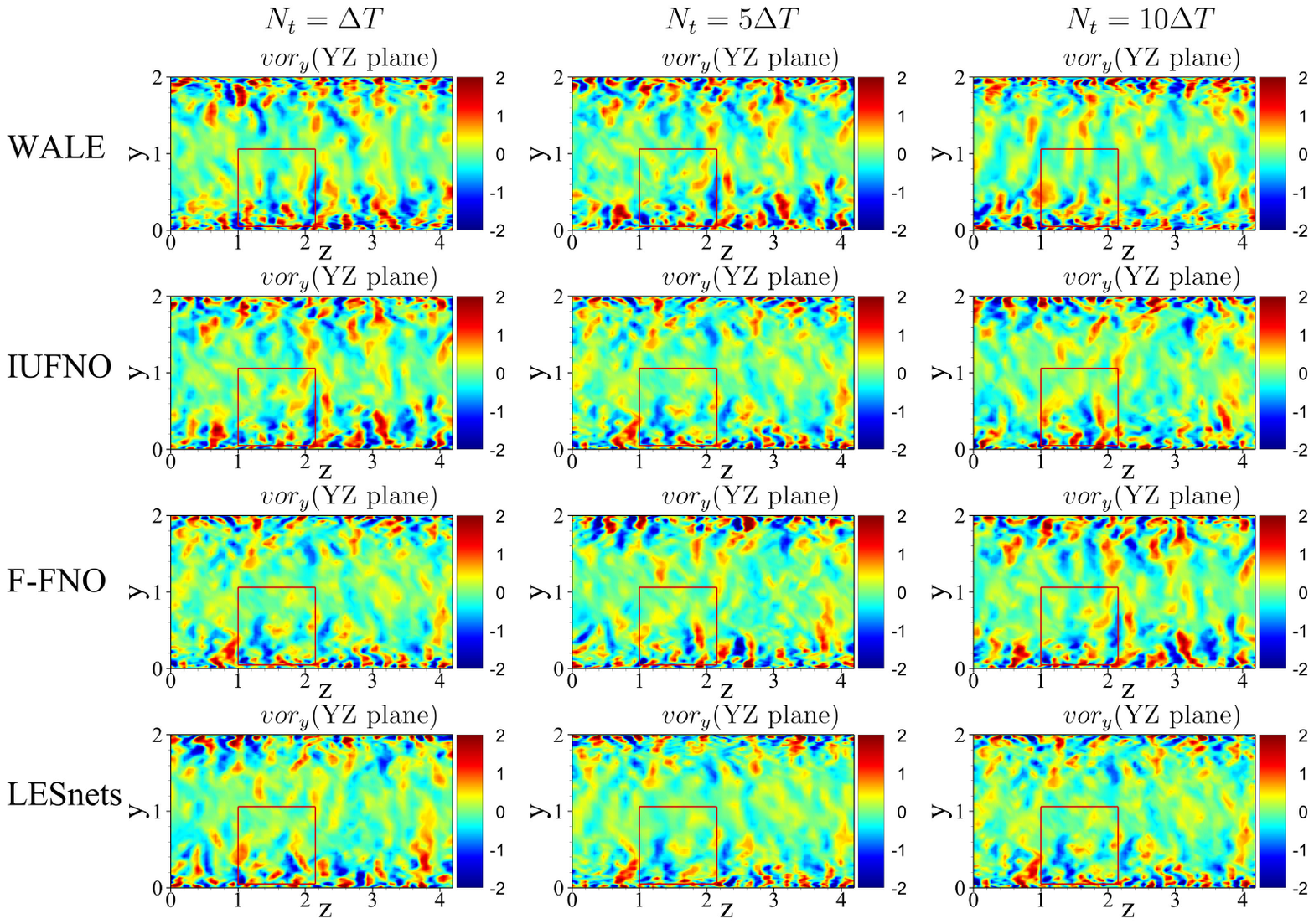}
\caption{Wall-normal vorticity field slices of turbulent channel flow at friction Reynolds number $Re_{\tau}\approx590$ in time series $N_t=\Delta T$, $N_t=5\Delta T$, and $N_t=10\Delta T$ (at the center YZ plane).}
\label{fig:re590-vorticity-evolution}
\end{figure}

We compare the flow statistics predicted by three machine-learning models at two friction Reynolds numbers $Re_{\tau}\approx180$ and $590$ against results of DNS, fDNS, and WALE. Figs. \ref{fig:mean-velocity-reynolds-stress}(a) and (b) show the normalized mean streamwise velocity profiles, normalized by the square of the friction velocity, i.e., $U^+ = \langle \overline{u}\rangle/u_{\tau}$. All three machine-learning approaches are consistent with the results from three traditional methods in the viscous sub-layer and buffer layer. However, F-FNO overestimates and IUFNO underestimates the streamwise velocity in the log-law region $y^+\in[40,180]$ at friction Reynolds number $Re_{\tau}\approx180$. Moreover, IUFNO overestimates the streamwise velocity in the log-law region $y^+\in[30,400]$ at friction Reynolds number $Re_{\tau}\approx590$. Figs. \ref{fig:mean-velocity-reynolds-stress}(c) and (d) present the Reynolds shear stress $-\langle u'v'\rangle^+$ normalized by the square of the friction velocity $u_{\tau}^2$. F-FNO performs slightly better than LESnets at $Re_{\tau} \approx 180$. In contrast, F-FNO performs slightly worse than LESnets at $Re_{\tau} \approx 590$. IUFNO shows deviations at both Reynolds numbers. The results obtained from the three traditional methods are nearly identical in terms of the mean streamwise velocity profiles and Reynolds shear stress.



As shown in Figs. \ref{fig:re180-rms-fluctuations} and \ref{fig:re590-rms-fluctuations}, the root-mean-square (RMS) values of fluctuating velocities $u',v',w'$ predicted by the DNS, fDNS, WALE, IUFNO, F-FNO, and LESnets are displayed at two friction Reynolds numbers $Re_{\tau}\approx180$ and $590$. It can be seen that the LESnets model gives consistent results with WALE, the F-FNO shows a slight deviation, and IUFNO overestimates the RMS fluctuating velocities at two friction Reynolds numbers. 

Fig. \ref{fig:energy-spectra-re180-re590} shows the kinetic energy spectrum in streamwise and spanwise directions at friction Reynolds numbers $Re_{\tau}\approx180$ and $590$. IUFNO exhibits non-physical behavior, which is also observed in \cite{wang2024prediction}. In contrast, the results from F-FNO and LESnets are almost identical to those of WALE, except for a slight deviation of F-FNO at the first wavenumber at friction Reynolds number $Re_{\tau}\approx180$ and at the low wavenumbers $k\in[1,6]$ at friction Reynolds number $Re_{\tau}\approx590$. This observation demonstrates the ability of the LESnets model to accurately capture the energy distribution at different scales.

Furthermore, we compare the wall-normal vorticity component slices from $N_t = \Delta T$ to $N_t = 10\Delta T$ at two friction Reynolds numbers $Re_{\tau}\approx180$ and $590$ at the center YZ plane in Figs. \ref{fig:re180-vorticity-evolution} and \ref{fig:re590-vorticity-evolution}. All three machine-learning models have a good performance at the first inference step $N_t = \Delta T$. At $N_t = 10\Delta T$, IUFNO starts to deviate, and the predictions by LESnets remains the closest to WALE. The flow structures at friction Reynolds number $Re_{\tau}\approx590$ are more diverse and complex than those at a lower Reynolds number $Re_{\tau}\approx180$, which also places a higher demand on the predictive capability of the models. 




\subsubsection{Model training and evaluations at \texorpdfstring{$Re_{\tau} \approx 1000$}{} and \texorpdfstring{$2000$}{}}
\label{subsubsec:lesnets-wall-model}

The WMLES methods have been widely used as effective ways to reduce the computational cost of wall-bounded turbulence by avoiding the explicit resolution of small-scale turbulence structures near the wall. This idea of incorporating the law of the wall into turbulence modeling has also been adopted in previous machine-learning-based approaches \cite{Bae2022, Lozano_Duran_Bae_2023, zhou23_POF}.

Inspired by the previous works, we incorporate the wall model into LESnets (LESnets-WM) to simulate turbulent channel flows at higher friction Reynolds numbers $Re_{\tau}\approx1000$ and $2000$ using coarse grids. We approximate the law of the wall using the three-layer matching model proposed by Inagaki et al. \cite{Inagaki02} (given by Eq. \eqref{eq:wall-model}). The fourth grid point away from the wall in the wall-normal direction is employed as the sampling location for the wall model. Using the wall-normal distance and the local velocity at this point, we calculate the friction velocity $u_{\tau}$ from which the modeled wall shear stress $\tau_w$ is subsequently obtained.

\begin{figure}[htbp]
\centering
\begin{minipage}{0.49\linewidth}
\centerline{\includegraphics[width=\textwidth]{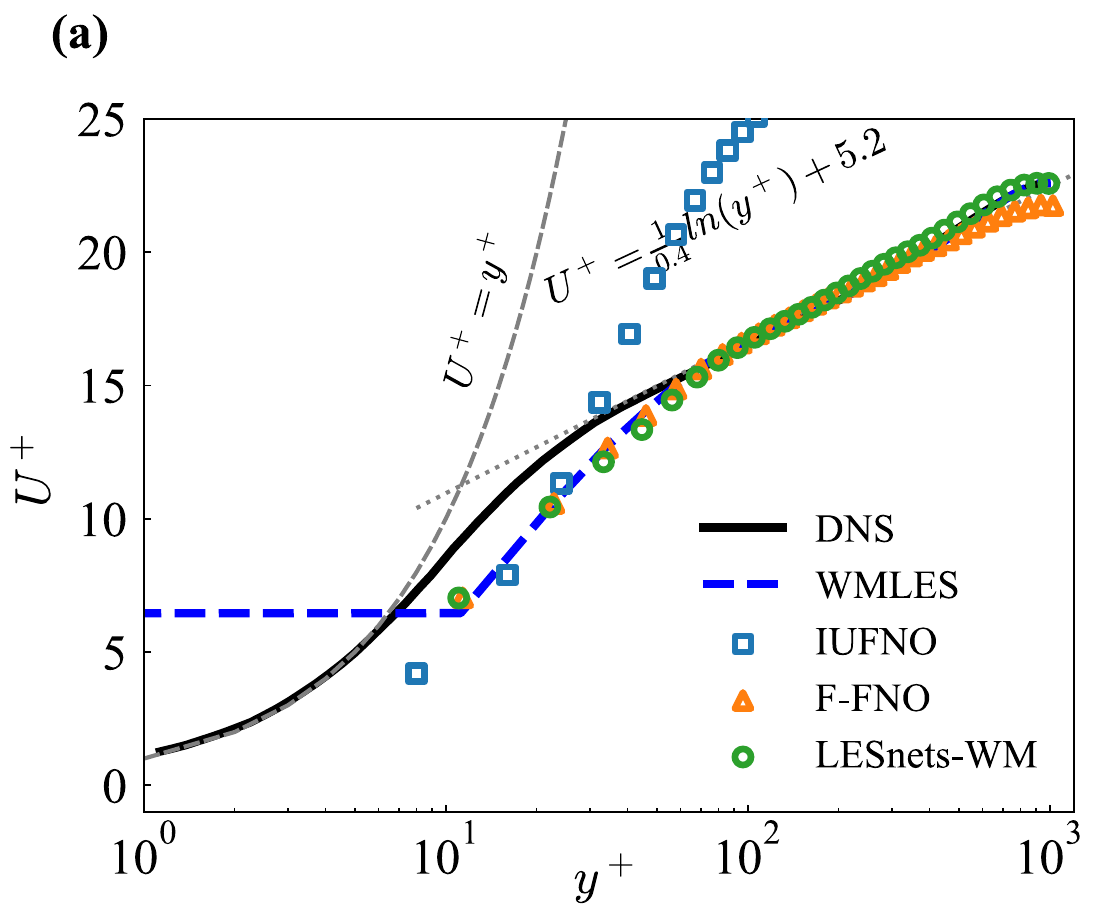}}
\end{minipage}
\hspace{-2pt}
\begin{minipage}{0.49\linewidth}
\centerline{\includegraphics[width=\textwidth]{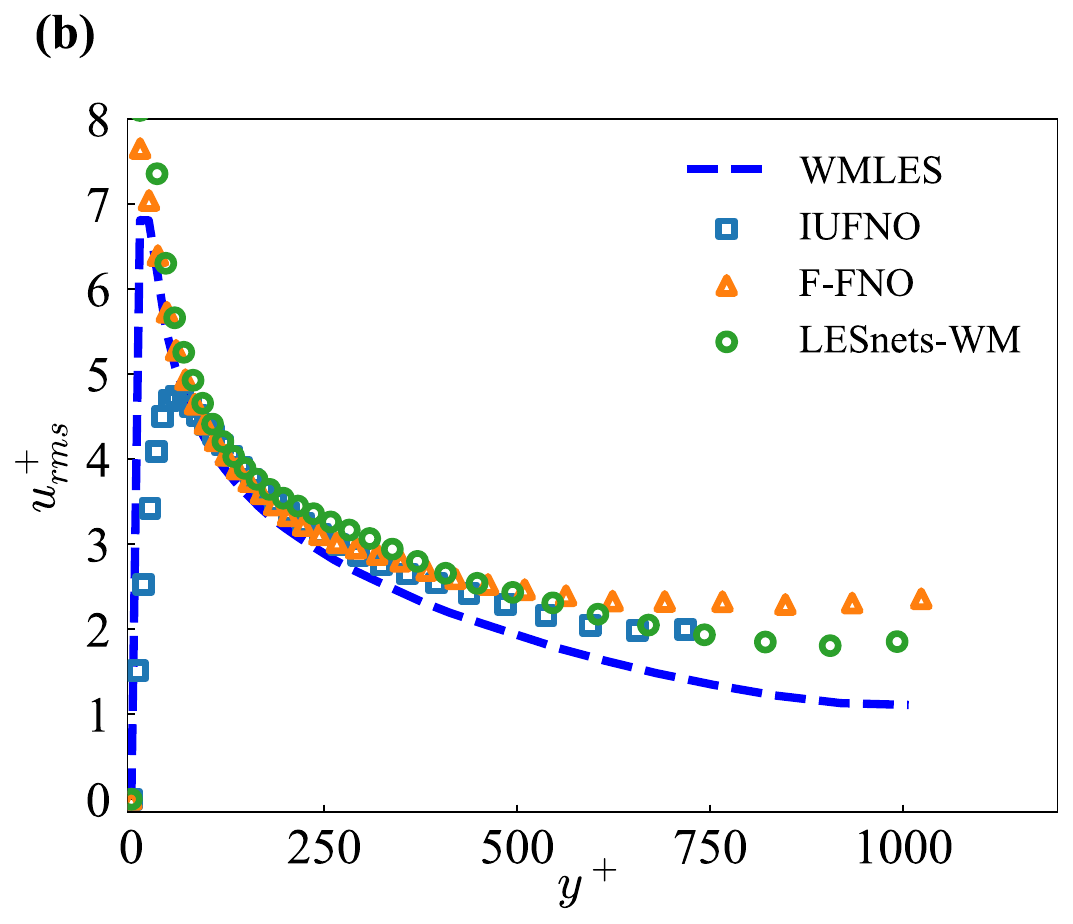}}
\end{minipage}

\begin{minipage}{0.49\linewidth}
\centerline{\includegraphics[width=\textwidth]{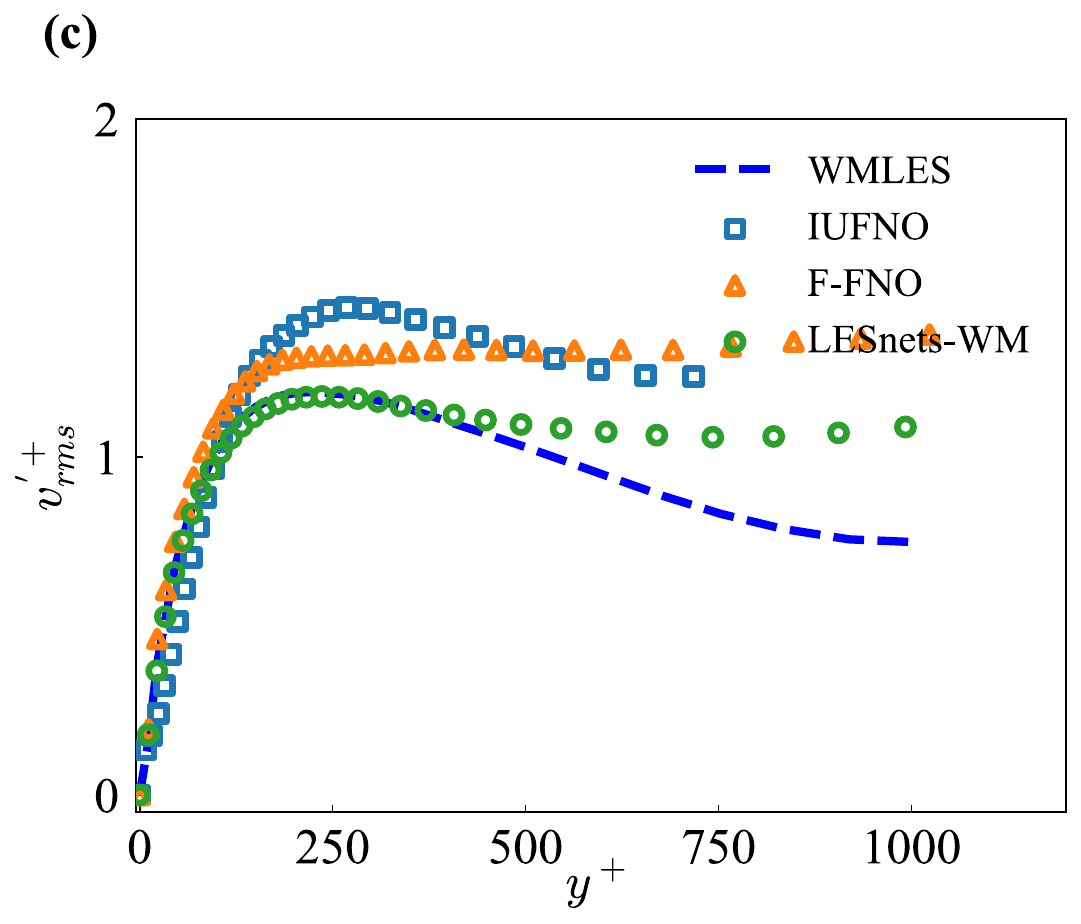}}
\end{minipage}
\hspace{-2pt}
\begin{minipage}{0.49\linewidth}
\centerline{\includegraphics[width=\textwidth]{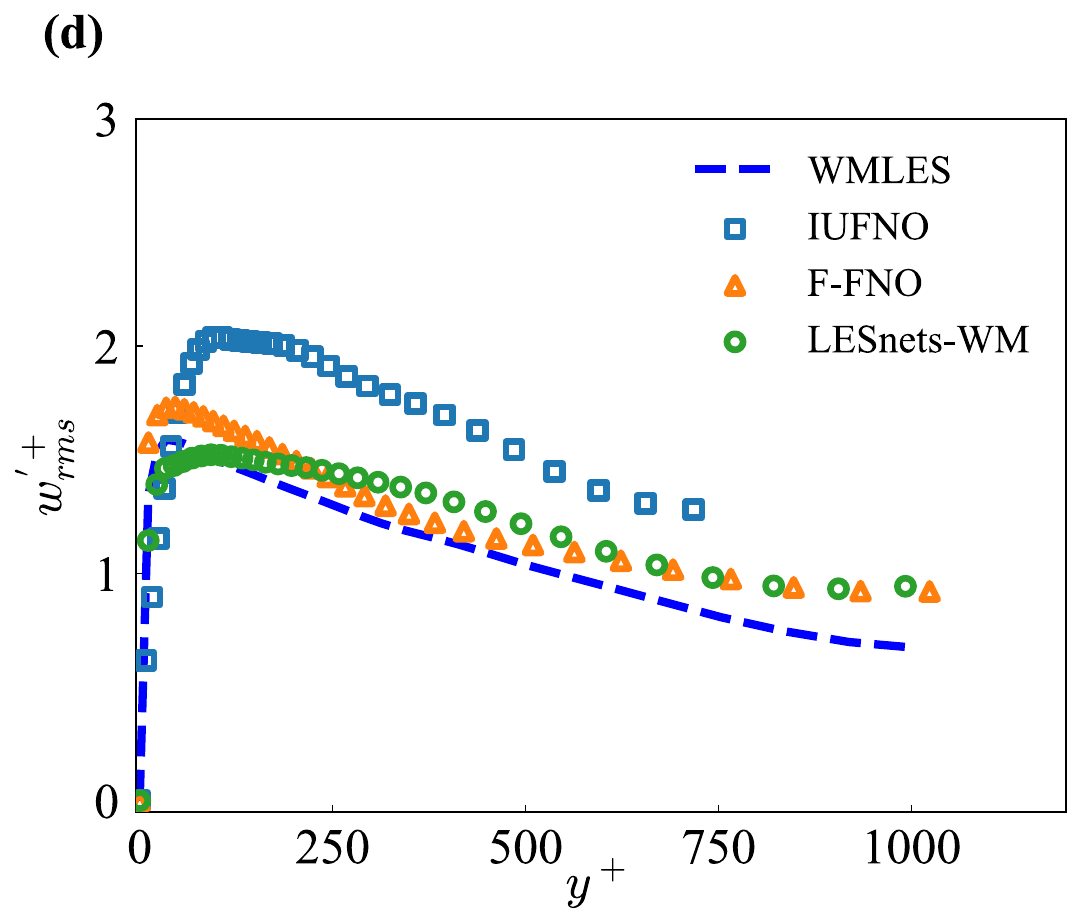}}
\end{minipage}
\caption{Comparison between DNS, WMLES, and three machine-learning models of turbulent channel flow at friction Reynolds number $Re_{\tau}\approx1000$. \textbf{(a)} Mean streamwise velocity $U^+ = \langle \overline{u}\rangle/u_{\tau}$ profile, normalized in wall units; \textbf{(b)} RMS fluctuations of streamwise velocity; \textbf{(c)} RMS fluctuations of wall-normal velocity; \textbf{(d)} RMS fluctuations of spanwise velocity.}
\label{fig:re1000-wall-model-statistics}
\end{figure}

Here, we use the training datasets $\mathcal{C}_{train}^{1000}$ and $\mathcal{C}_{train}^{2000}$ for the three machine-learning methods. IUFNO is not applied to turbulent channel flows at $Re_{\tau}\approx1000$ or $Re_{\tau}\approx2000$ in \cite{wang2024prediction}. However, the grid resolutions for these two cases correspond to those used at $Re_{\tau}\approx180$ and $Re_{\tau}\approx590$, respectively. Therefore, IUFNO is trained using the same parameters as those used for the corresponding $Re_{\tau}\approx180$ and $Re_{\tau}\approx590$ cases, respectively. For a fair comparison, we use the same parameters as those used in the simulation of $Re_{\tau} = 180$ and $590$ to train F-FNO and LESnets-WM, while halving the time interval to $\Delta t_{\mathrm{train}}=\Delta t_{inf} = 0.5\Delta T = 100\Delta t = 0.5$ at $Re_{\tau}\approx2000$, setting the WALE coefficient $C_w=0.5$, setting the width to 120, and setting an exponential decay with a factor of $0.9$. The hyper-parameters used for training the three operators are summarized in \ref{app:hyperparameters}. All the remaining settings are identical to those described at the beginning of this section. The DNS \cite{Lee_Moser_2015} and WMLES statistics are used as the benchmark results to validate the three machine-learning models. The SGS model is the WALE model with $C_w=0.5$.

\begin{figure}[htbp]
\centering
\begin{minipage}{0.49\linewidth}
\centerline{\includegraphics[width=\textwidth]{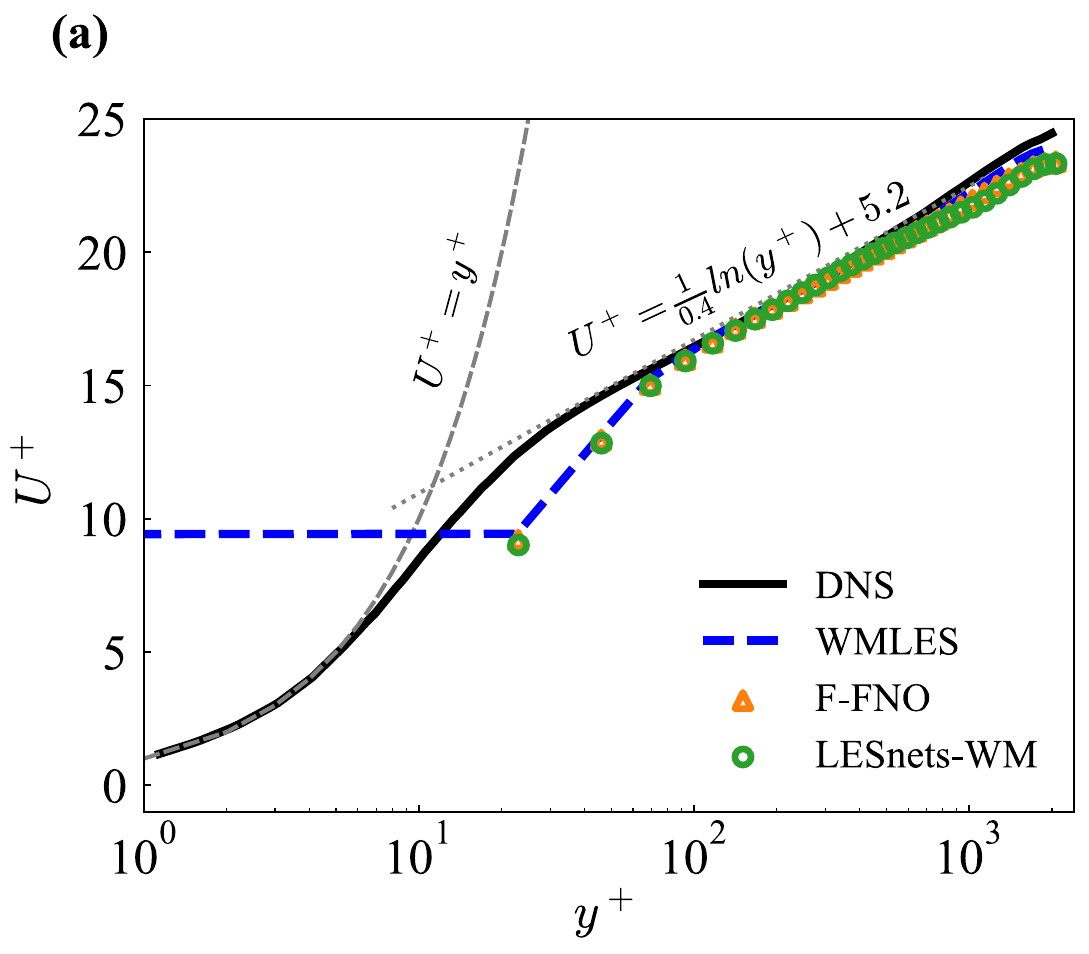}}
\end{minipage}
\hspace{-2pt}
\begin{minipage}{0.49\linewidth}
\centerline{\includegraphics[width=\textwidth]{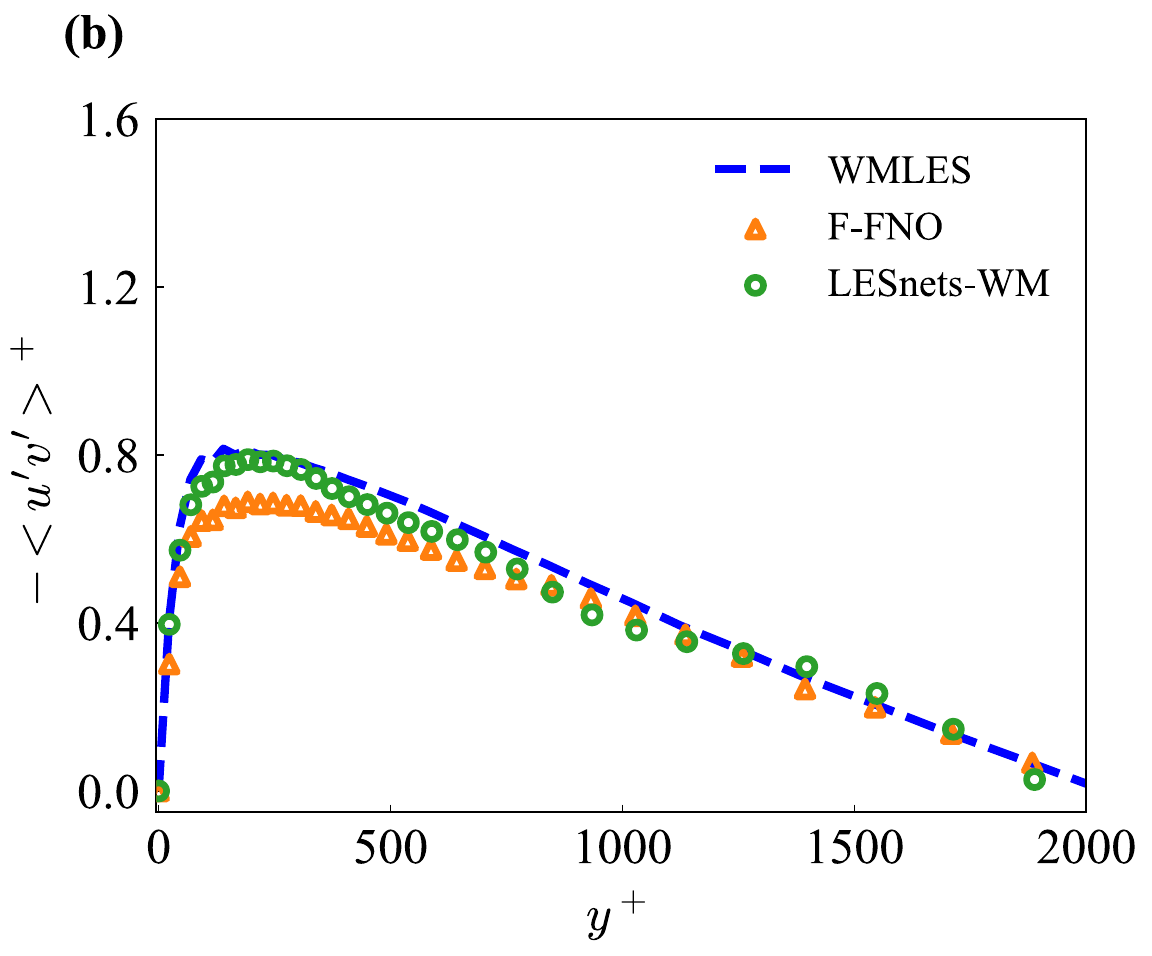}}
\end{minipage}

\begin{minipage}{0.49\linewidth}
\centerline{\includegraphics[width=\textwidth]{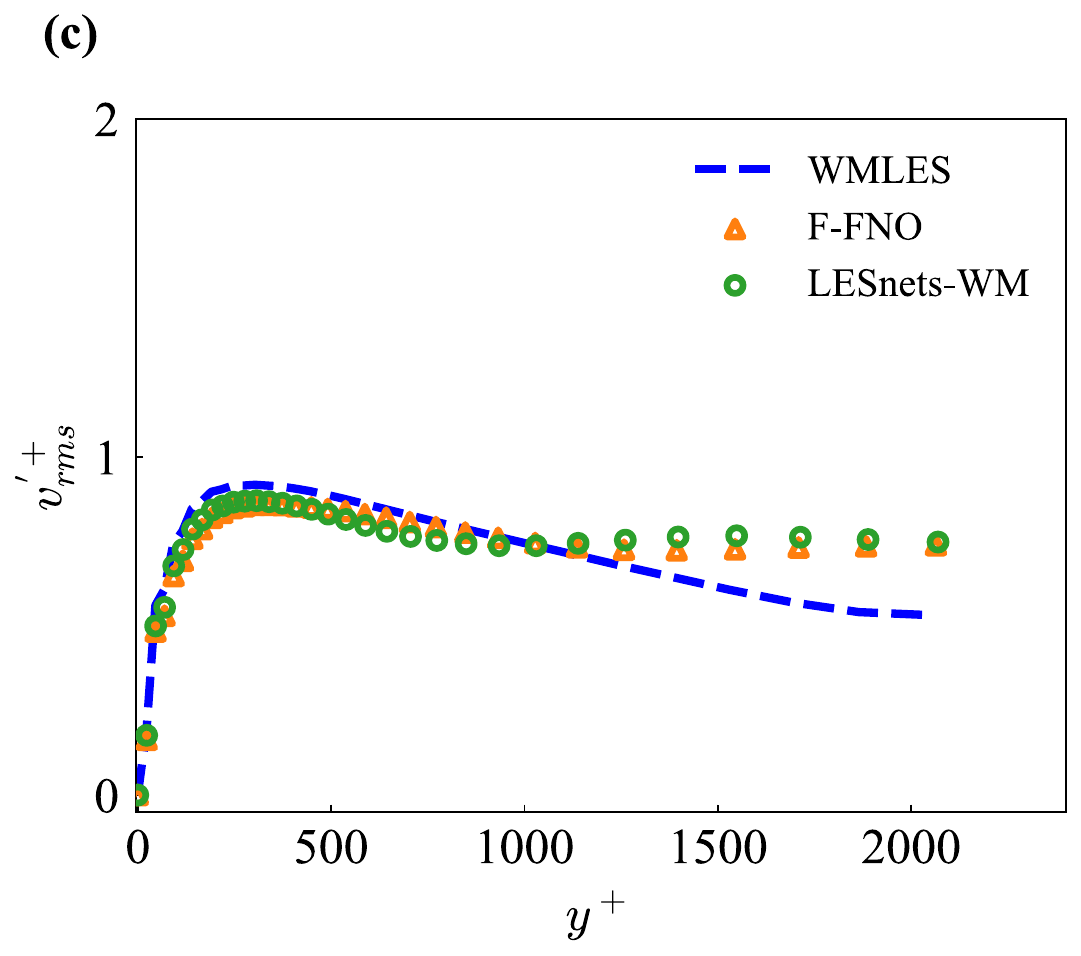}}
\end{minipage}
\hspace{-2pt}
\begin{minipage}{0.49\linewidth}
\centerline{\includegraphics[width=\textwidth]{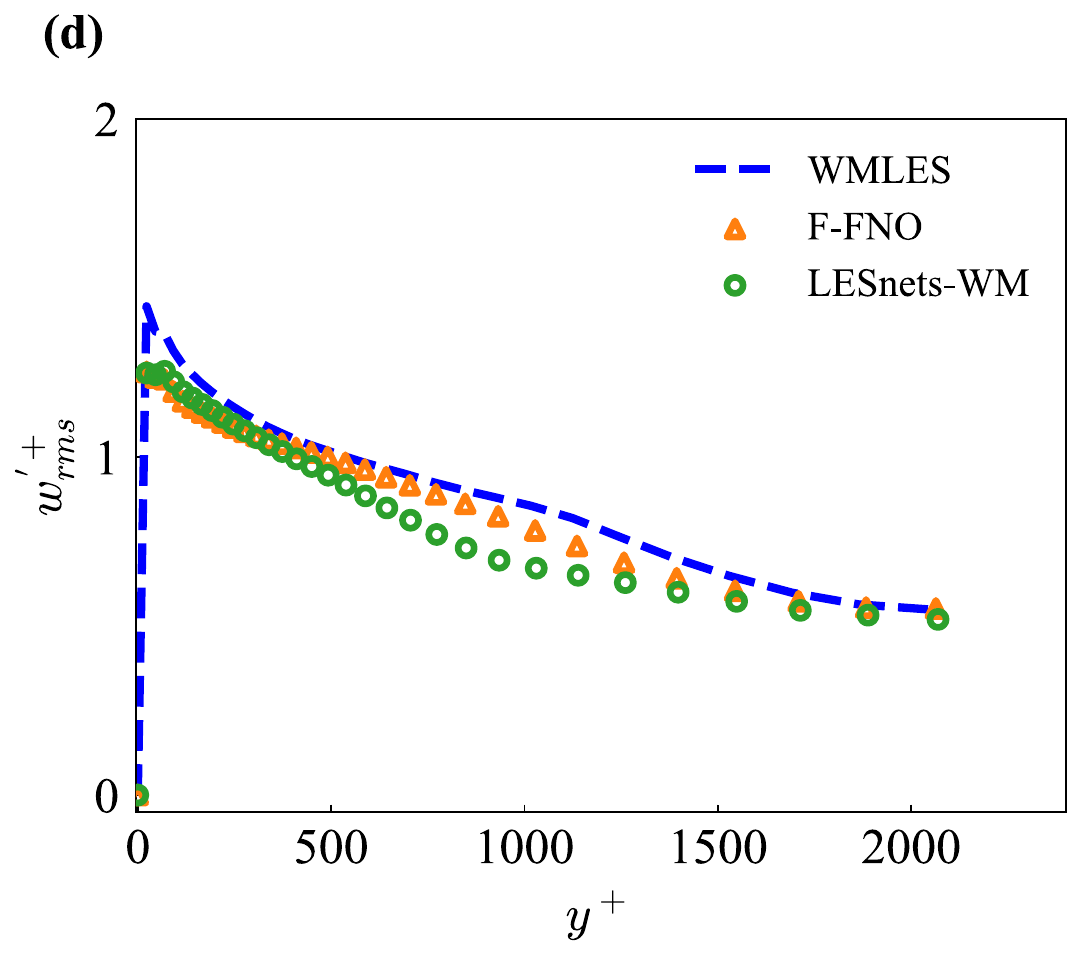}}
\end{minipage}
\caption{Comparison between DNS, WMLES, and two machine-learning models of turbulent channel flow at friction Reynolds number $Re_{\tau}\approx2000$. \textbf{(a)} Mean streamwise velocity $U^+ = \langle \overline{u}\rangle/u_{\tau}$ profile, normalized in wall units; \textbf{(b)} Reynolds shear stress $-\langle u'v'\rangle^+$, normalized by the square of respective friction velocity $u_{\tau}^2$; \textbf{(c)} RMS fluctuations of wall-normal velocity; \textbf{(d)} RMS fluctuations of spanwise velocity.}
\label{fig:re2000-wall-model-statistics}
\end{figure}

Fig. \ref{fig:re1000-wall-model-statistics} compares the turbulent statistics predicted by WMLES and the three machine-learning models with the DNS data at $Re_{\tau}\approx1000$. Fig. \ref{fig:re1000-wall-model-statistics}(a) shows the normalized mean streamwise velocity, $U^+ = \langle \overline{u}\rangle/u_{\tau}$. Within the first four off-wall grid points, WMLES underpredicts the DNS velocity because of its inherent limitations in the near-wall region. We therefore focus primarily on model performance in the log-law region. In this region, IUFNO overpredicts the streamwise velocity, whereas F-FNO underpredicts it. By contrast, LESnets-WM agrees well with both the DNS and WMLES results. Figs. \ref{fig:re1000-wall-model-statistics}(b) to (d) show the RMS fluctuations of the streamwise, wall-normal, and spanwise velocities, respectively. Although LESnets-WM deviates slightly from WMLES in the outer region, it generally reproduces the DNS and WMLES statistics more accurately than F-FNO and IUFNO. These results demonstrate that incorporating the wall model improves the long-term predictions of LESnets relative to the two data-driven baseline models at $Re_{\tau}\approx1000$. Moreover, the $Re_{\tau}\approx1000$ and $Re_{\tau}\approx180$ simulations use the same grid configuration. Consequently, they have comparable inference-stage computational efficiency, indicating that introducing the wall model incurs no additional computational cost during inference.

Fig. \ref{fig:re2000-wall-model-statistics} compares the turbulent statistics predicted by WMLES and two machine-learning models with the DNS data at $Re_{\tau}\approx2000$. IUFNO is excluded from this comparison because it fails to produce valid predictions at $Re_{\tau}\approx2000$. Fig. \ref{fig:re2000-wall-model-statistics}(a) shows the normalized mean streamwise velocity, $U^+ = \langle \overline{u}\rangle/u_{\tau}$. Both LESnets-WM and F-FNO agree well with the DNS and WMLES results in the log-law region. Fig. \ref{fig:re2000-wall-model-statistics}(b) presents the Reynolds shear stress, $-\langle u'v'\rangle^+$, normalized by the square of the friction velocity $u_{\tau}^2$. For this statistic, LESnets-WM slightly outperforms F-FNO. Figs. \ref{fig:re2000-wall-model-statistics}(c) and (d) show the RMS fluctuations of the wall-normal and spanwise velocities, respectively, for which LESnets-WM and F-FNO yield nearly identical results. Overall, LESnets-WM achieves performance comparable to that of F-FNO for turbulent channel flow at $Re_{\tau}\approx2000$.

\subsection{Test case 2: 3D incompressible turbulent flow over periodic hills}
\label{subsec:case-2}

\begin{figure}[htbp]
\centering
\begin{minipage}{0.49\linewidth} 
\centerline{\includegraphics[width=\textwidth]{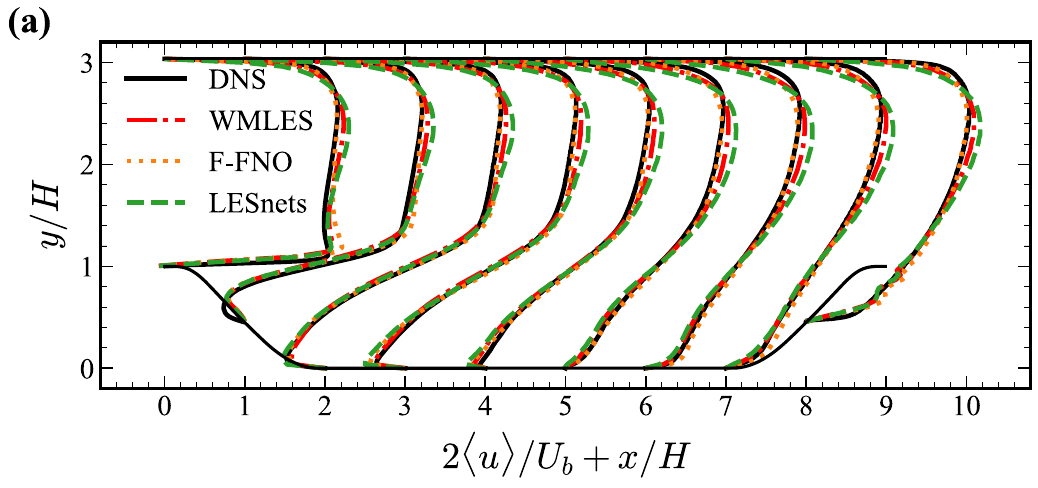}}
\end{minipage}
\hspace{-2pt} 
\begin{minipage}{0.49\linewidth} 
\centerline{\includegraphics[width=\textwidth]{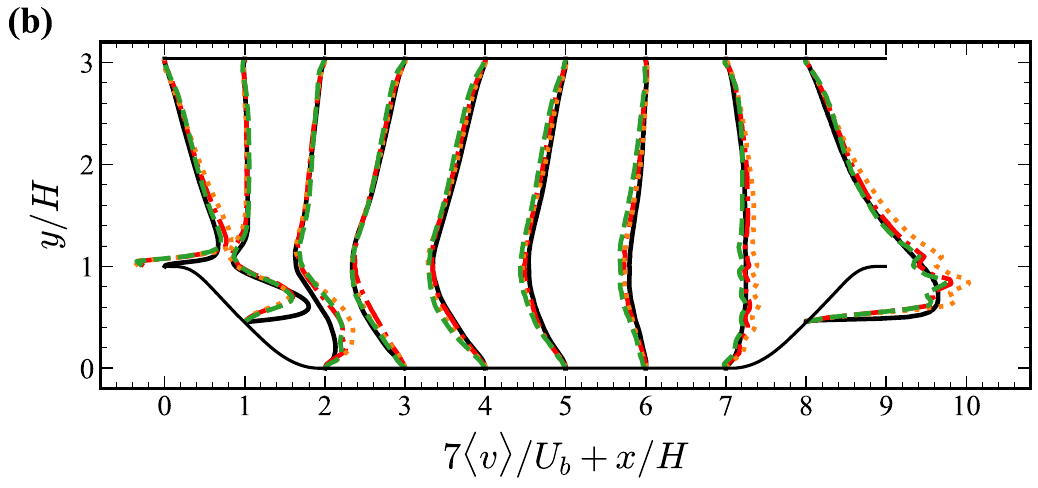}}
\end{minipage}

\begin{minipage}{0.49\linewidth} 
\centerline{\includegraphics[width=\textwidth]{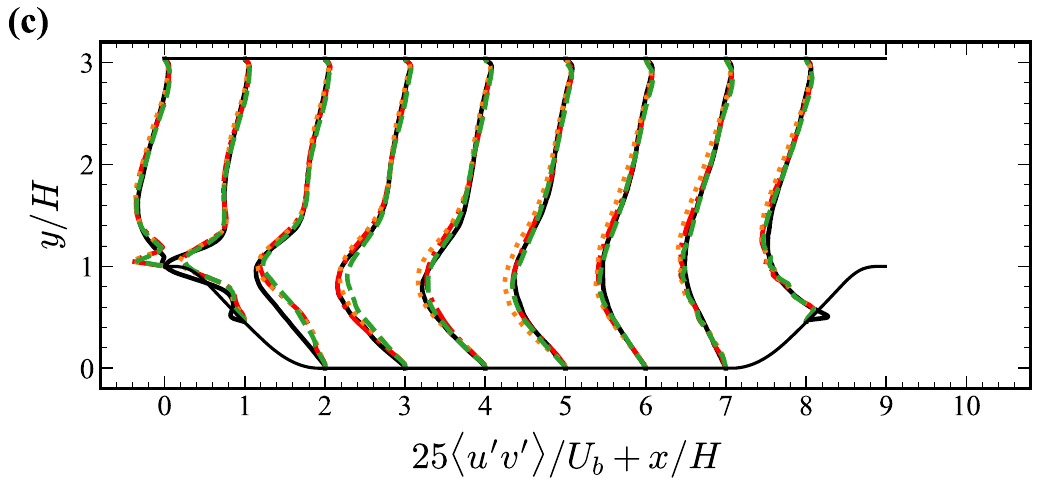}}
\end{minipage}
\hspace{-2pt} 
\begin{minipage}{0.49\linewidth} 
\centerline{\includegraphics[width=\textwidth]{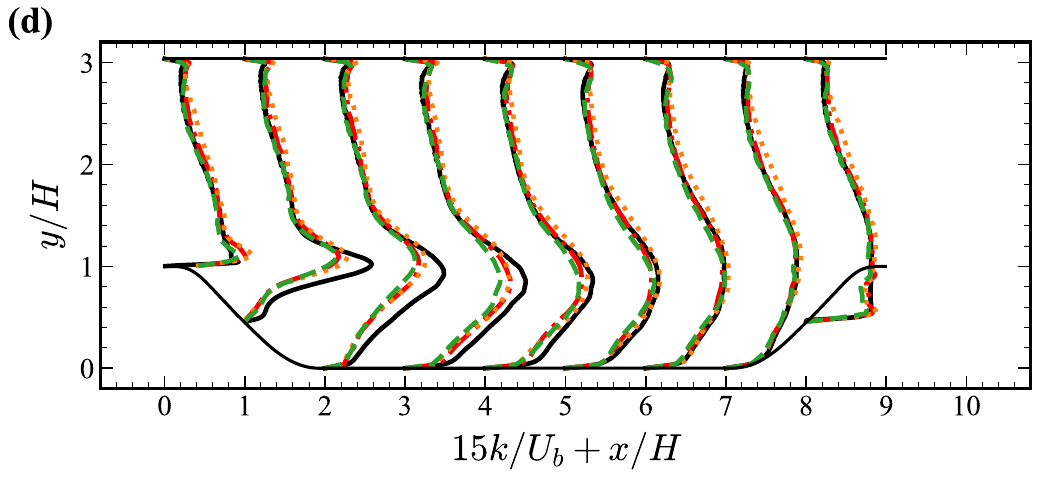}}
\end{minipage}
\caption{Comparison between DNS, WMLES, F-FNO and LESnets of turbulent flow over periodic hills at Reynolds number $Re_{h}\approx5600$. Wall-normal profiles of the time-averaged \textbf{(a)} streamwise velocity $\langle u\rangle$; \textbf{(b)} wall-normal velocity $\langle v\rangle$; \textbf{(c)} Reynolds shear stress $\langle u'v'\rangle$; \textbf{(d)} turbulence kinetic energy (TKE) $k=\frac{1}{2}\langle u'u'+v'v'+w'w'\rangle$.}
\label{fig:Hill_Re5600_u_v_uv_k}
\end{figure}

\begin{figure}[htbp]
\centering
\begin{minipage}{0.49\linewidth} 
\centerline{\includegraphics[width=\textwidth]{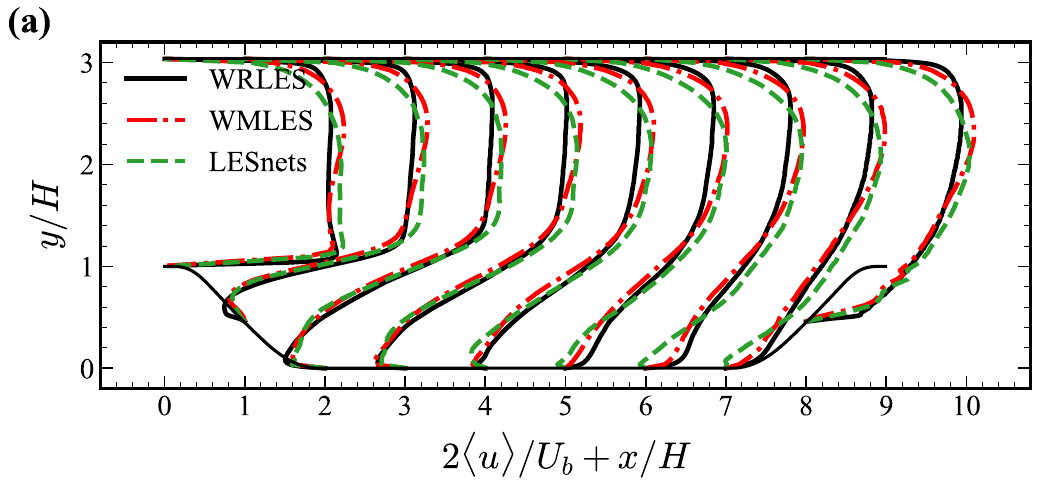}}
\end{minipage}
\hspace{-2pt} 
\begin{minipage}{0.49\linewidth} 
\centerline{\includegraphics[width=\textwidth]{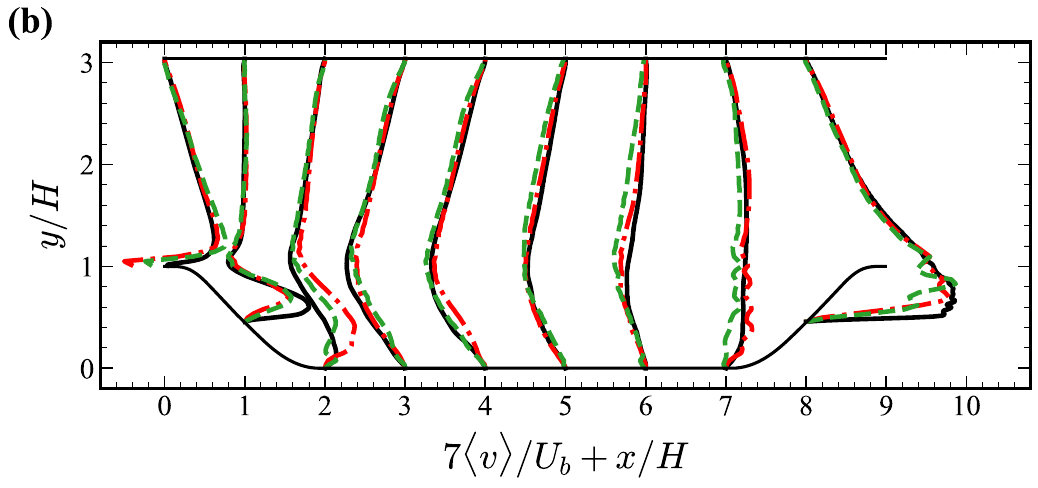}}
\end{minipage}

\begin{minipage}{0.49\linewidth} 
\centerline{\includegraphics[width=\textwidth]{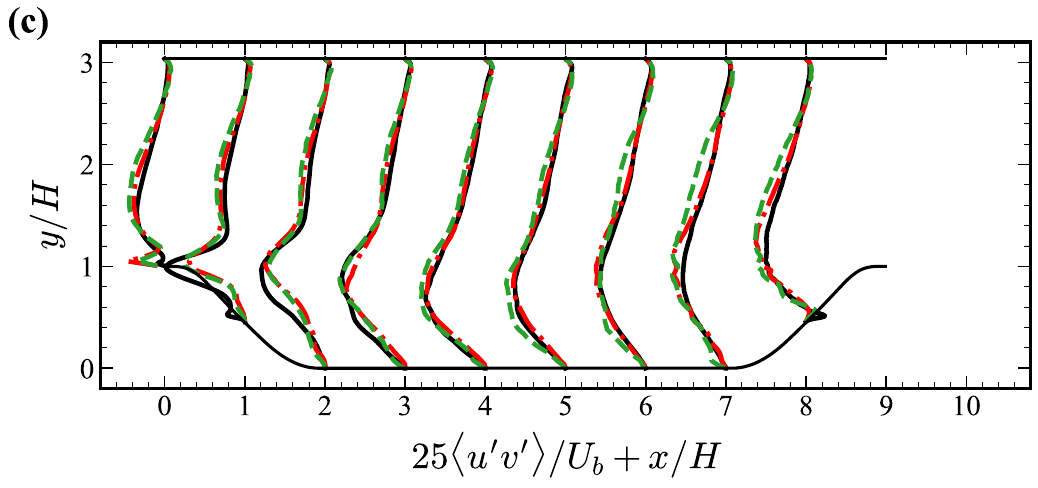}}
\end{minipage}
\hspace{-2pt} 
\begin{minipage}{0.49\linewidth} 
\centerline{\includegraphics[width=\textwidth]{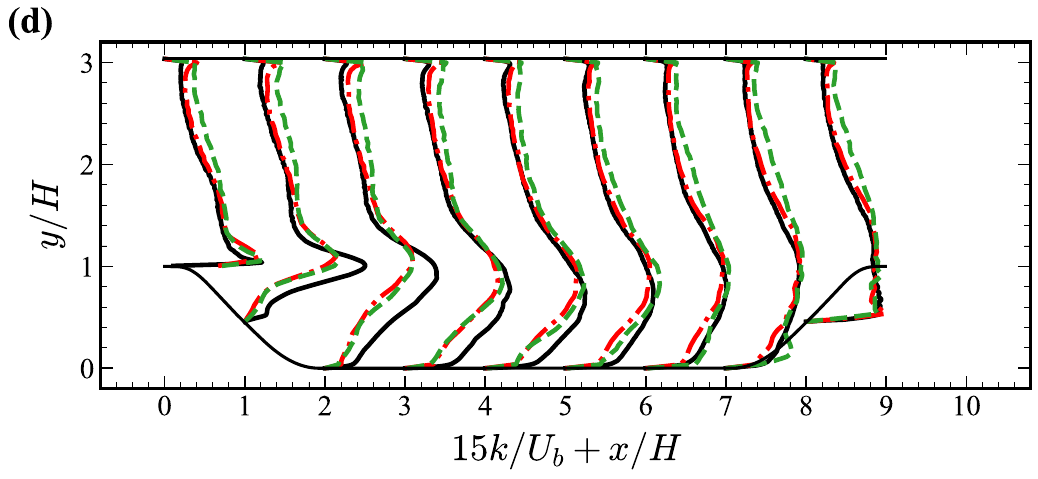}}
\end{minipage}
\caption{Comparison between WRLES, WMLES, and LESnets of turbulent flow over periodic hills at Reynolds number $Re_{h}\approx10595$. Wall-normal profiles of the time-averaged \textbf{(a)} streamwise velocity $\langle u\rangle$; \textbf{(b)} wall-normal velocity $\langle v\rangle$; \textbf{(c)} Reynolds shear stress $\langle u'v'\rangle$;  \textbf{(d)} turbulence kinetic energy (TKE) $k=\frac{1}{2}\langle u'u'+v'v'+w'w'\rangle$.}
\label{fig:Hill_Re10595_u_v_uv_k}
\end{figure}

\begin{figure}[htbp]
\centering
\begin{minipage}{0.49\linewidth} 
\centerline{\includegraphics[width=\textwidth]{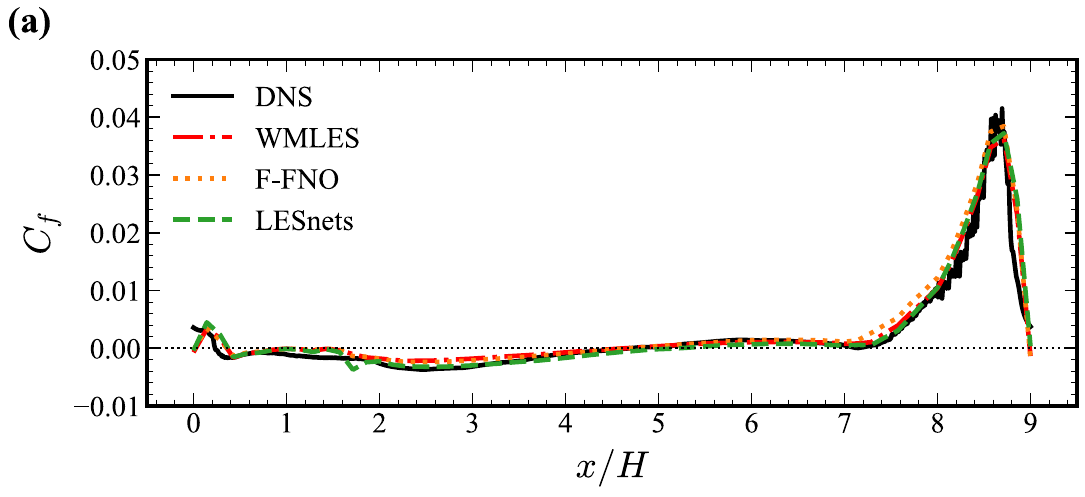}}
\end{minipage}
\hspace{-2pt} 
\begin{minipage}{0.49\linewidth} 
\centerline{\includegraphics[width=\textwidth]{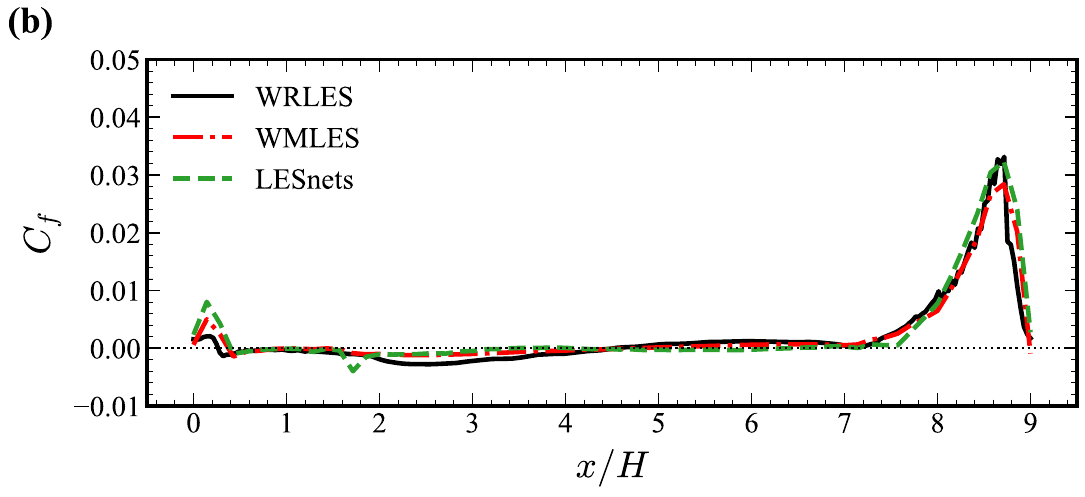}}
\end{minipage}
\caption{Comparison of the time-averaged skin friction coefficient at \textbf{(a)} $Re_{h}\approx5600$, and \textbf{(b)} $Re_{h}\approx10595$.}
\label{fig:Hill_Cf}
\end{figure}

\begin{figure}[htbp]
\centering
\begin{minipage}{0.49\linewidth} 
\centerline{\includegraphics[width=\textwidth]{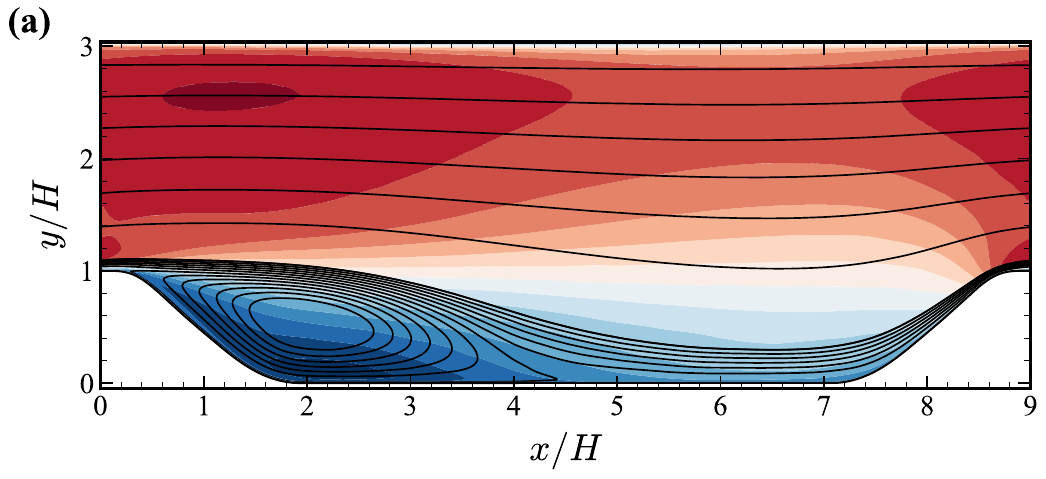}}
\end{minipage}
\hspace{-2pt} 
\begin{minipage}{0.49\linewidth} 
\centerline{\includegraphics[width=\textwidth]{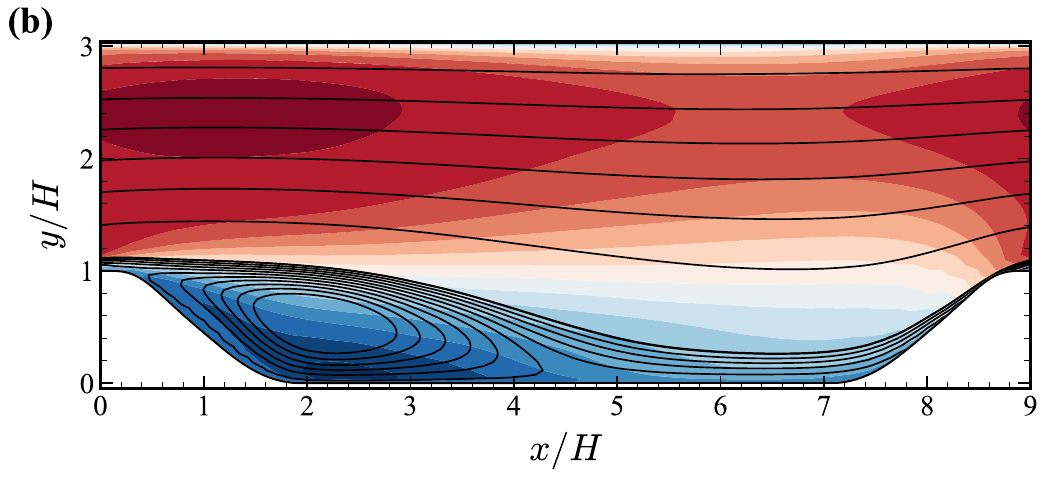}}
\end{minipage}

\begin{minipage}{0.49\linewidth} 
\centerline{\includegraphics[width=\textwidth]{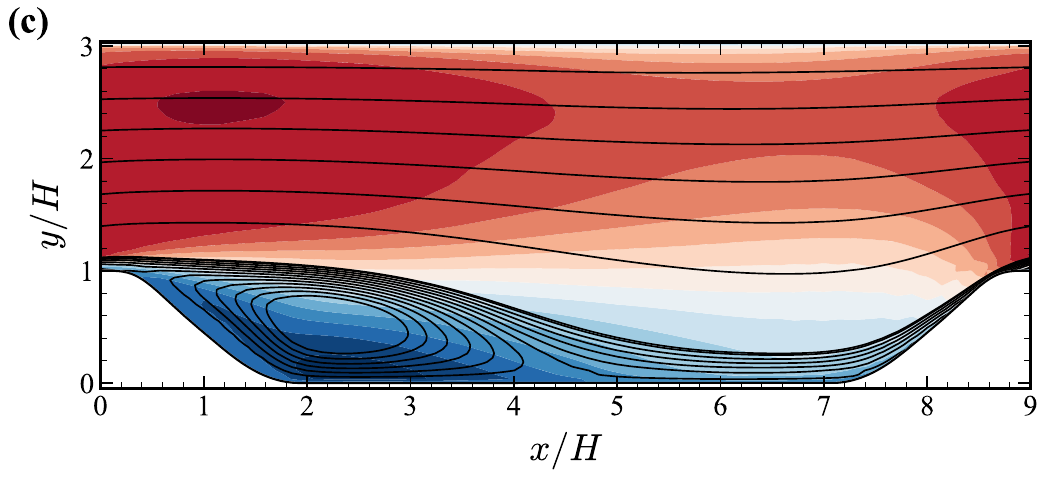}}
\end{minipage}
\hspace{-2pt} 
\begin{minipage}{0.49\linewidth} 
\centerline{\includegraphics[width=\textwidth]{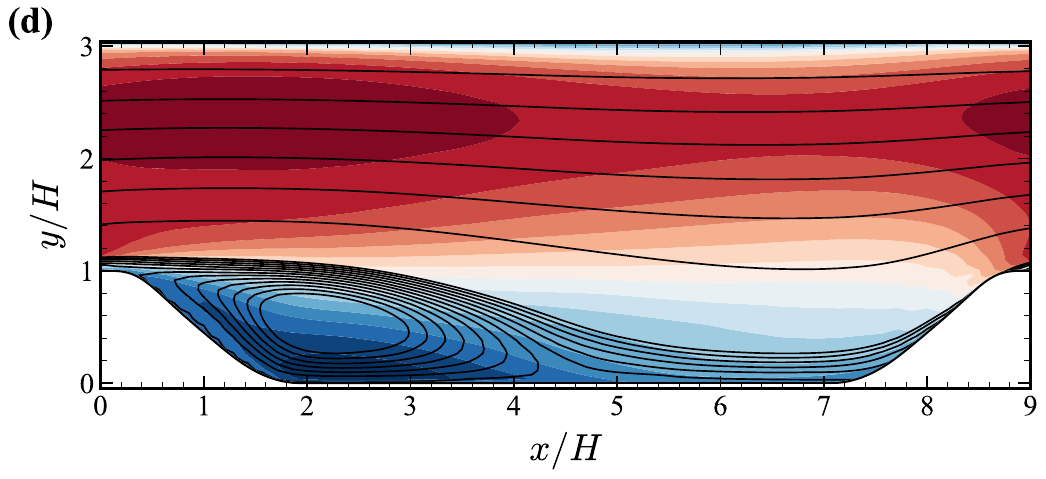}}
\end{minipage}
\caption{Contours of time-averaged streamwise velocity with streamlines at Reynolds number $Re_{h}\approx5600$ obtained from \textbf{(a)} DNS; \textbf{(b)} WMLES; \textbf{(c)} F-FNO; \textbf{(d)} LESnets.}
\label{fig:Hill_Re5600_u_stream}
\end{figure}

\begin{figure}[htbp]
\centering
\begin{minipage}{0.49\linewidth} 
\centerline{\includegraphics[width=\textwidth]{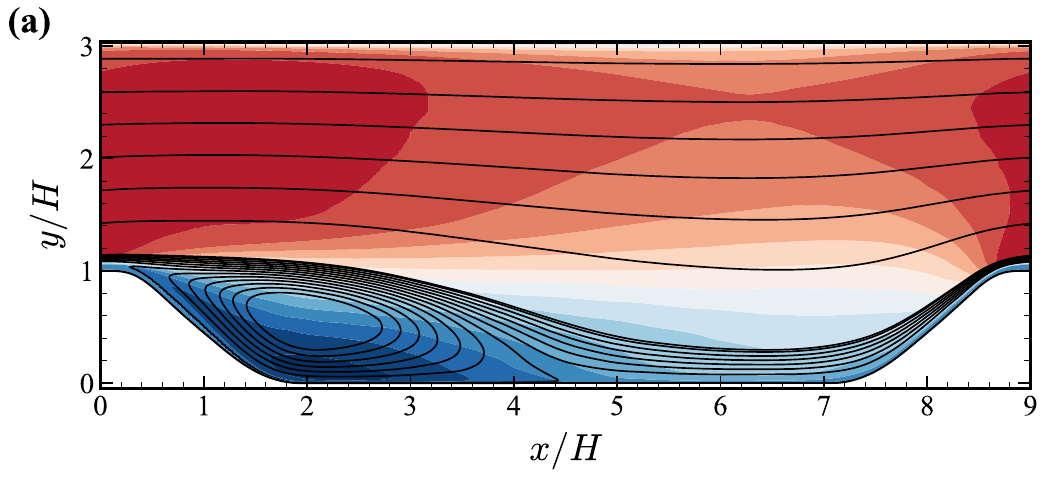}}
\end{minipage}
\hspace{-2pt} 
\begin{minipage}{0.49\linewidth} 
\centerline{\includegraphics[width=\textwidth]{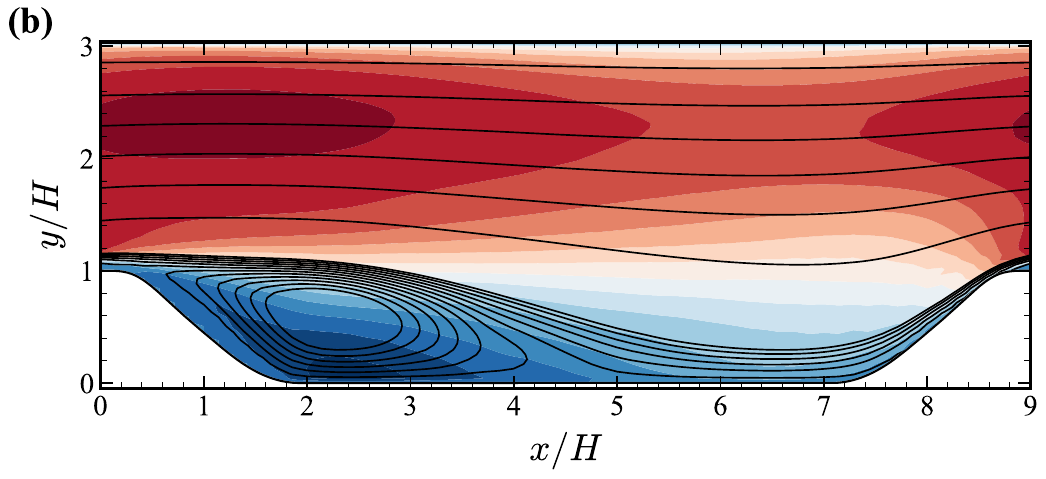}}
\end{minipage}

\begin{minipage}{0.49\linewidth} 
\centerline{\includegraphics[width=\textwidth]{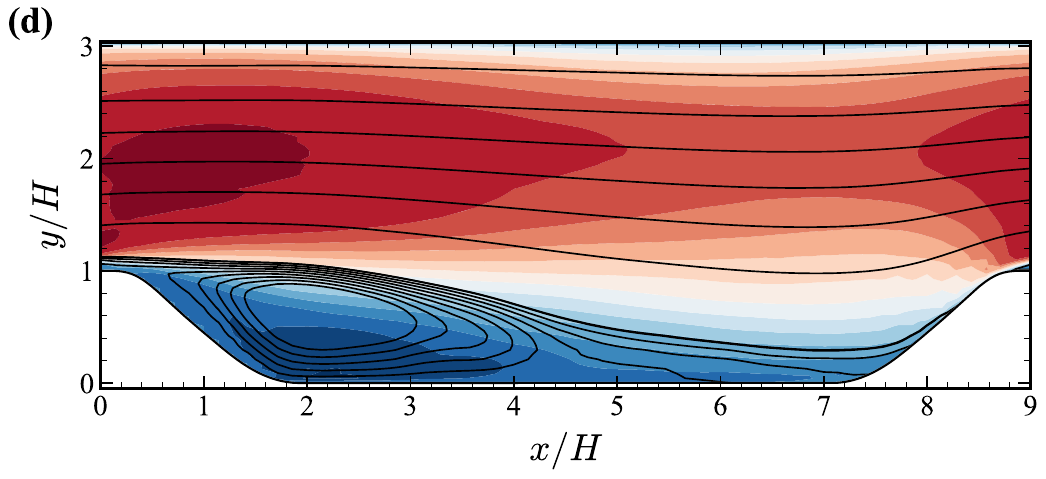}}
\end{minipage}

\caption{Contours of time-averaged streamwise velocity with streamlines at Reynolds number $Re_{h}\approx10595$ obtained from \textbf{(a)} DNS; \textbf{(b)} WMLES; \textbf{(c)} LESnets.}
\label{fig:Hill_Re10595_u_stream}
\end{figure}

The second test case is the 3D incompressible turbulent flow over periodic hills, a canonical configuration for studying separation and reattachment over a curved, non-grid-aligned boundary. Unlike channel flow, the curved hill surface cannot be align with the Cartesian grid, so this case applies the immersed boundary method (IBM) introduced in Section \ref{subsec:IBM} along with the hard boundary constraints described in Section \ref{subsubsec:hard-boundary-constraints}. The flow separates on the leeward side of the hill, forms a large recirculation region, and reattaches on the flat bottom wall. Reproducing this separation--reattachment sequence is therefore a stringent test of both the predictive accuracy of the model and the treatment of the non-grid-aligned boundary. The hill geometry, the computational domain, and the numerical setup are specified in Section \ref{sec:governing-equations-data}.

As described in Subsection \ref{subsubsec:periodic-hill}, two LESnets models are trained independently on the same WMLES dataset $\mathcal{H}_{\mathrm{train}}^{5600}$, which supplies only the initial fields. Since no data term enters the loss, the same dataset can be reused at a different Reynolds number without generating any additional training data. In the training stage, the physics-informed loss is evaluated with the kinematic viscosity $\nu$ of the target Reynolds number, so that the two models enforce the governing equations at $Re_h\approx5600$ and $Re_h\approx10595$, respectively, while sharing the same model architecture, hard boundary constraints, immersed boundary method, and training hyper-parameters. In the prediction stage, both models are run recursively, with the velocity field predicted at each step fed back as the updated initial field. Starting from a WMLES field at the corresponding Reynolds number, each model advances the flow for $T_{inf}=800$ time steps with an inference time interval of $\Delta t_{\mathrm{inf}}=\Delta T/5=100\Delta t=0.2$, and the statistics are averaged over the entire inference period. The predictions at $Re_h\approx5600$ are compared against DNS, WMLES, and F-FNO, whereas those at $Re_h\approx10595$ are compared against WRLES and WMLES. IUFNO is not considered in this case because it yields poor results on the turbulent flow over periodic hills, and F-FNO is not considered at $Re_h\approx10595$ because a data-driven model requires data at the target Reynolds number, which are not available here.
Fig. \ref{fig:Hill_Re5600_u_v_uv_k} compares the turbulent statistics at $Re_h\approx5600$ among DNS, WMLES, F-FNO, and LESnets, namely the wall-normal profiles of the time-averaged streamwise velocity $\langle u\rangle$, the wall-normal velocity $\langle v\rangle$, the Reynolds shear stress $\langle u'v'\rangle$, and the turbulent kinetic energy (TKE) $k=\frac{1}{2}\left(\langle u'u'\rangle+\langle v'v'\rangle+\langle w'w'\rangle\right)$. The accuracy of LESnets is comparable to that of F-FNO, and both are in good agreement with the WMLES results.

Fig. \ref{fig:Hill_Re10595_u_v_uv_k} compares the wall-normal profiles of the flow statistics at $Re_h\approx10595$ among WRLES, WMLES, and LESnets. Notably, LESnets attains this without preparing any additional training data and without changing the network architecture or the training hyper-parameters.

Fig. \ref{fig:Hill_Cf} presents the mean skin friction coefficient $C_f=2\tau_w/(\rho {U_b}^2)$ along the lower wall predicted by DNS, WRLES, WMLES, and LESnets at $Re_h\approx5600$ and $10595$. The downstream zero crossing of the DNS curve marks the mean reattachment point, and the LESnets prediction is in good agreement with the WMLES results at most streamwise locations.

Fig. \ref{fig:Hill_Re5600_u_stream} displays the contours of time-averaged streamwise velocity with streamlines at $Re_h\approx5600$. The DNS solution serves as the reference for WMLES, whereas the two machine-learning models are benchmarked against WMLES. Both F-FNO and LESnets show only slight deviations from WMLES and reproduce the overall distribution faithfully.

Fig. \ref{fig:Hill_Re10595_u_stream} displays the contours of time-averaged streamwise velocity with streamlines at $Re_h\approx10595$. The WRLES solution serves as the reference for WMLES, whereas LESnets is benchmarked against WMLES. LESnets shows only slight deviations from WMLES and reproduces the overall distribution faithfully.

Overall, LESnets is able to predict the flow characteristics of the turbulent flow over periodic hills well over a long time horizon. The agreement with WMLES at $Re_h\approx5600$ and $Re_h\approx10595$ indicates that the accuracy achieved in the channel-flow case is retained when the wall surface becomes curved and non-grid-aligned, and that the immersed boundary method does not compromise the long-term stability of the prediction. The results therefore suggest that the framework can be extended to flows bounded by more complex geometries, which is a natural direction for future work.

\subsection{Framework properties and analysis}
\label{subsec:framework-properties-and-analysis}

\subsubsection{Flexible output}
\label{subsubsec:flexible-output}

\begin{figure}[htbp]
\centering
\includegraphics [width=1.0\textwidth]{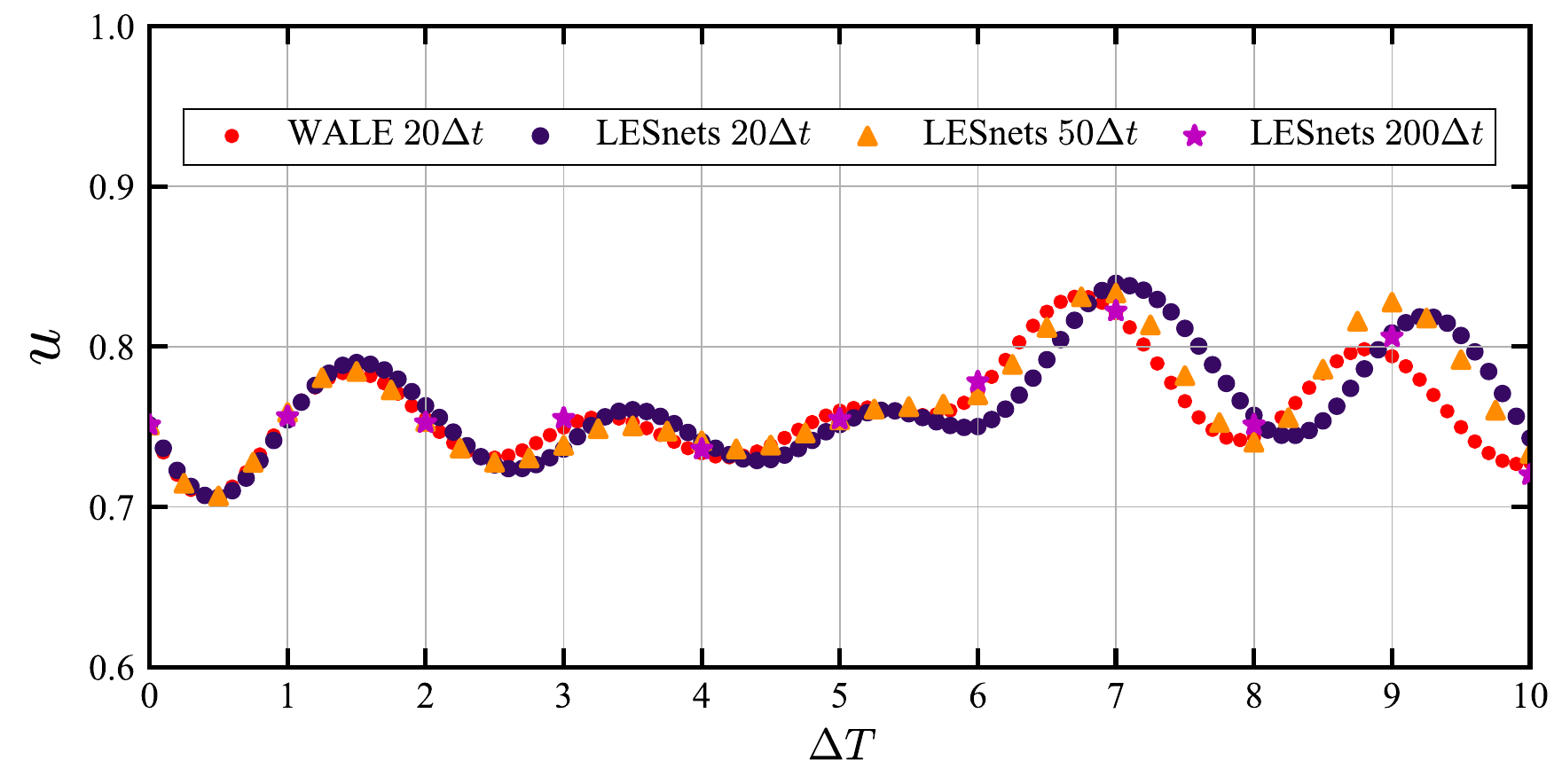}
\caption{Temporal evolution of WALE and LESnets outputs at three time intervals, $20\Delta t$, $50\Delta t$, and $200\Delta t$, for turbulent channel flow at friction Reynolds number $Re_{\tau} \approx 180$ at the spatial location $[2\pi, 0, 2\pi/3]$ in the temporal domain $[0,10\Delta T]$.}
\label{fig:re180-flexible-output-velocity}
\end{figure}

\begin{figure}[htbp]
\centering
\begin{minipage}{0.49\linewidth} 
\centerline{\includegraphics[width=\textwidth]{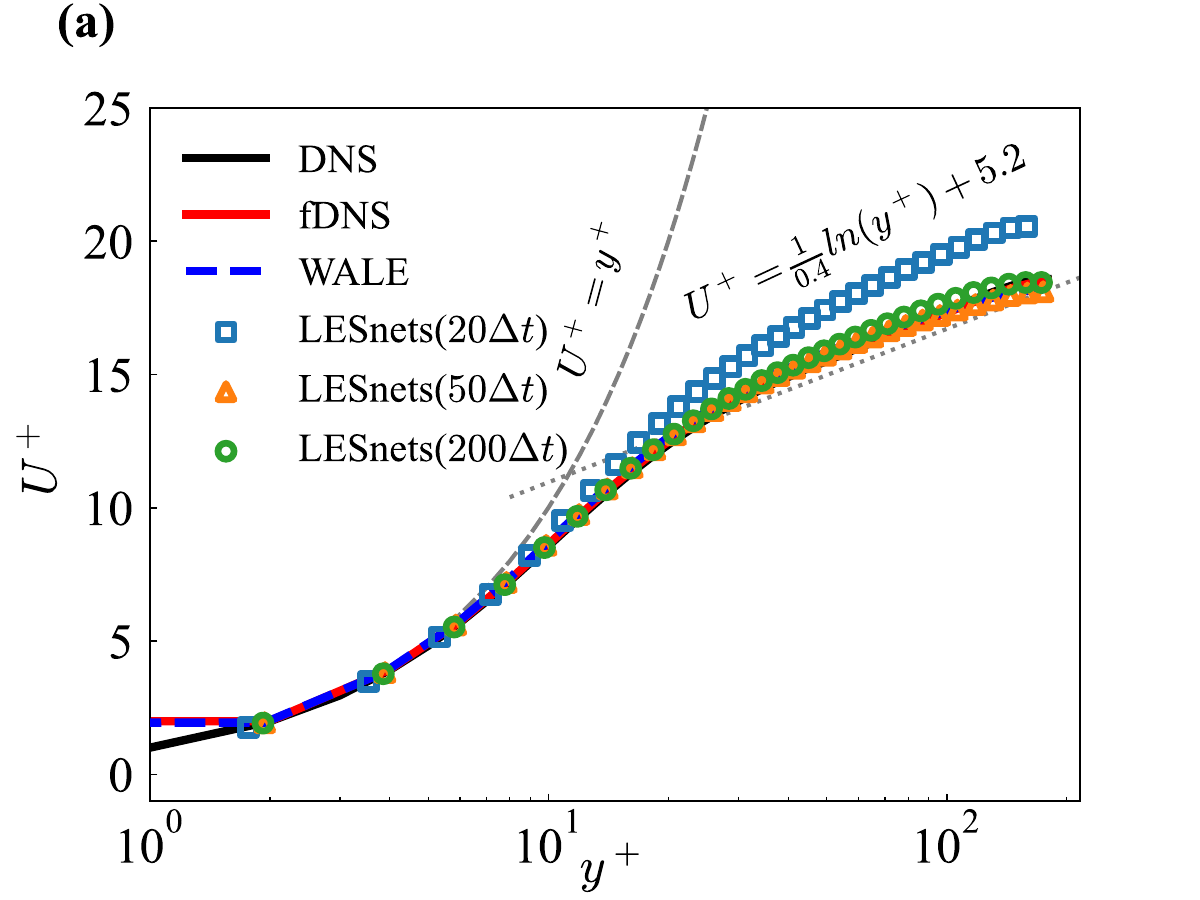}}
\end{minipage}
\hspace{-2pt} 
\begin{minipage}{0.49\linewidth} 
\centerline{\includegraphics[width=\textwidth]{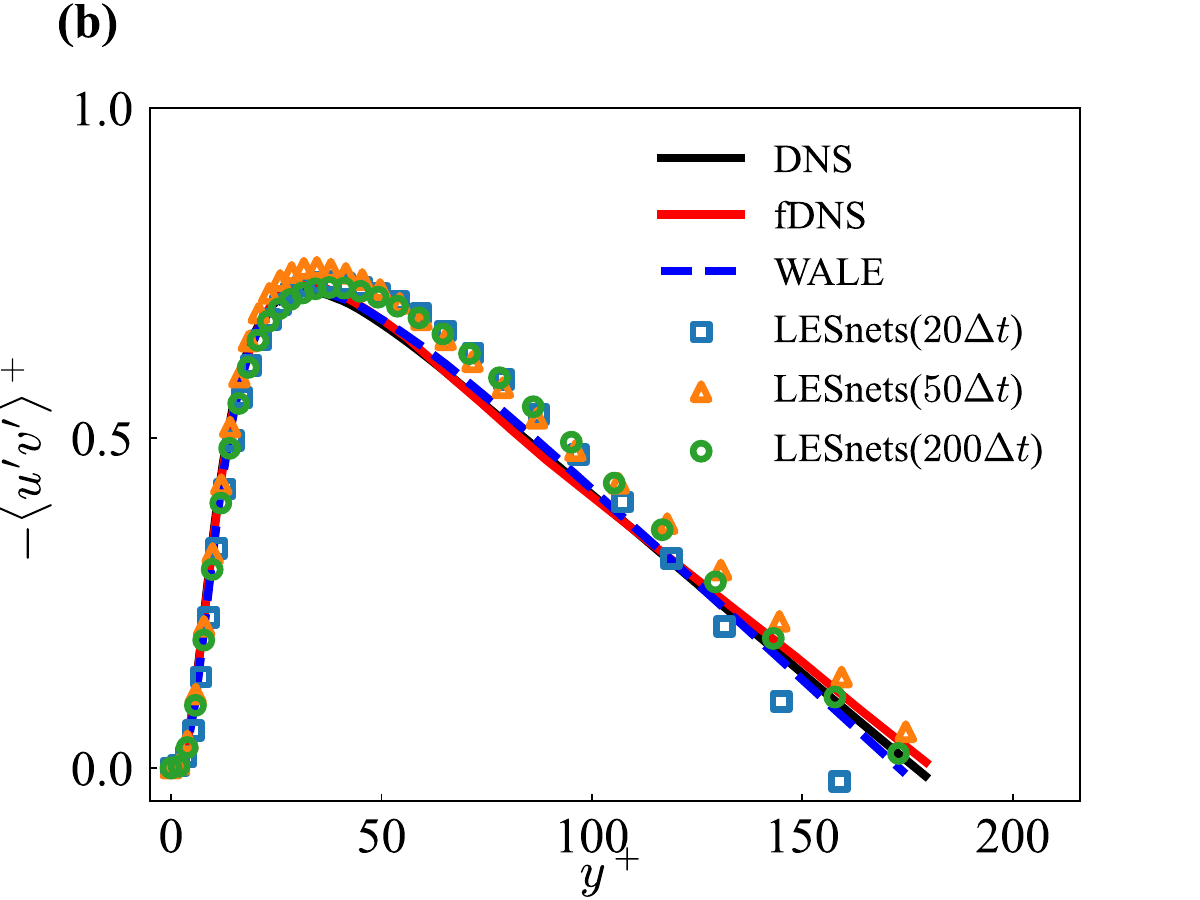}}
\end{minipage}
\caption{Comparison between WALE and LESnets outputs at three time intervals, $20\Delta t$, $50\Delta t$, and $200\Delta t$, for turbulent channel flow at friction Reynolds number $Re_{\tau} \approx 180$. \textbf{(a)} Mean streamwise velocity $U^+ = \langle \overline{u}\rangle/u_{\tau}$ profile, normalized in wall units, and \textbf{(b)} Reynolds shear stress $-\langle u'v'\rangle^+$, normalized by the square of respective friction velocity $u_{\tau}^2$.}
\label{fig:re180-flexible-output-statistics}
\end{figure}

\begin{figure}[htbp]
\centering
\begin{minipage}{0.49\linewidth} 
\centerline{\includegraphics[width=\textwidth]{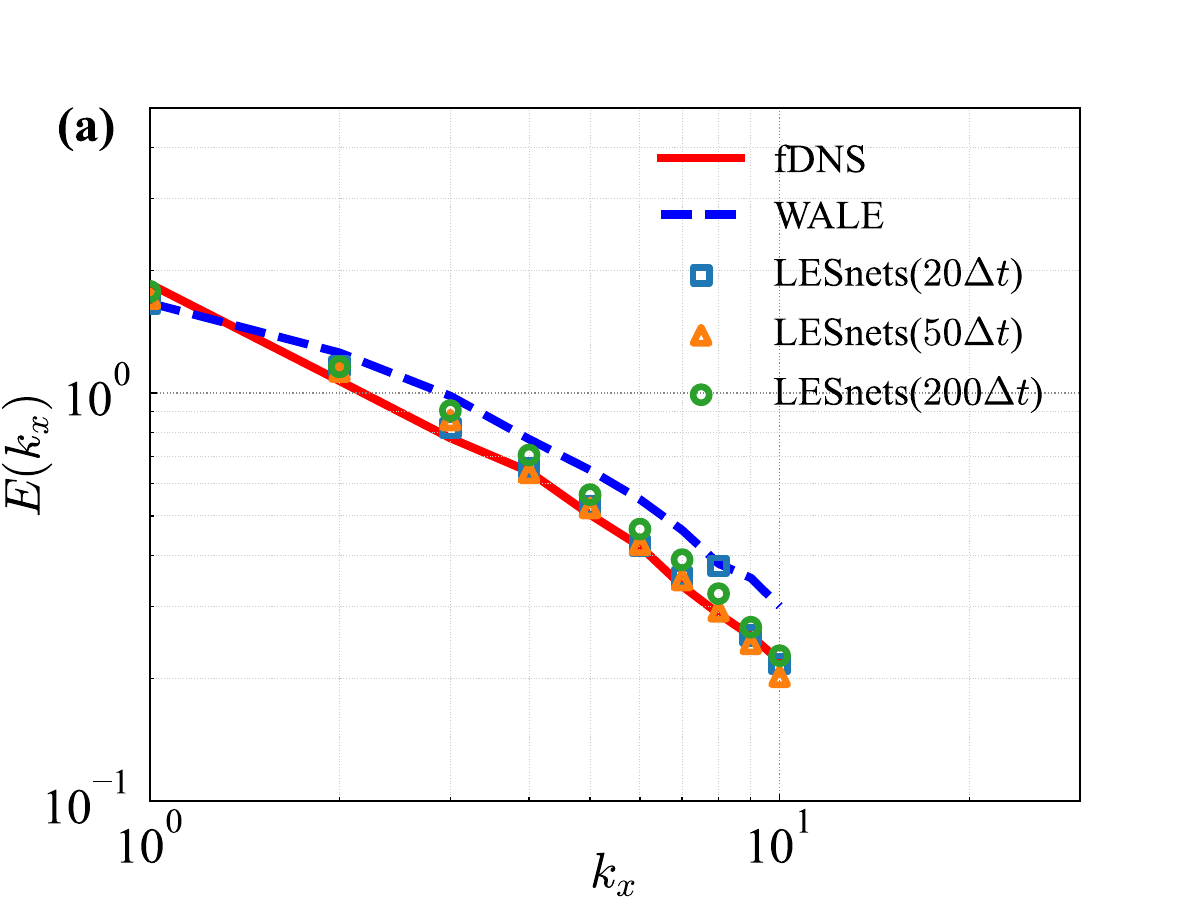}}
\end{minipage}
\hspace{-2pt} 
\begin{minipage}{0.49\linewidth} 
\centerline{\includegraphics[width=\textwidth]{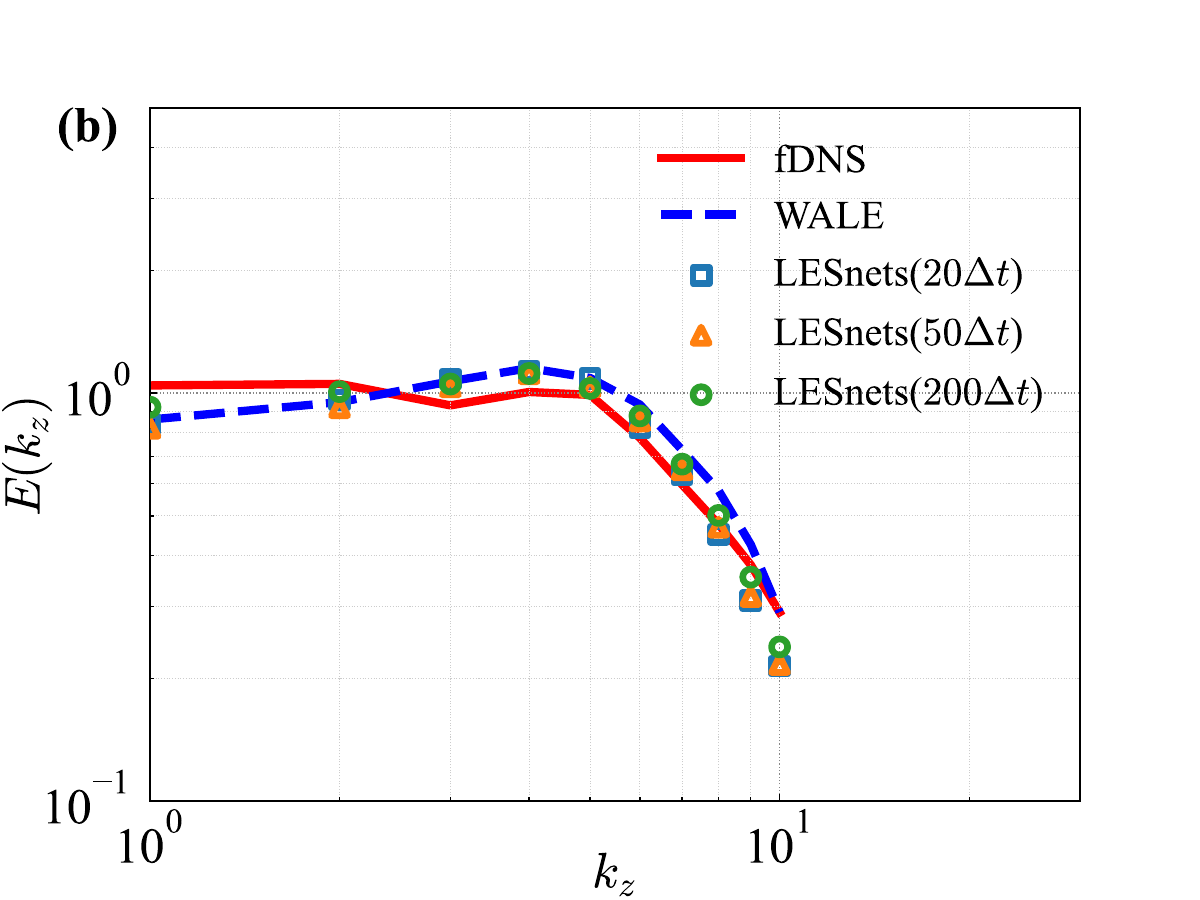}}
\end{minipage}
\caption{Streamwise velocity energy spectrum along \textbf{(a)} streamwise direction ($k_x$), and \textbf{(b)} spanwise direction ($k_z$), comparison between WALE and LESnets outputs at three time intervals, $20\Delta t$, $50\Delta t$, and $200\Delta t$, for turbulent channel flow at friction Reynolds number $Re_{\tau} \approx 180$.}
\label{fig:re180-flexible-output-spectra}
\end{figure}

In this subsubsection, we discuss the flexible capability of the LESnets model to generate time series outputs at a given time interval. In Section \ref{subsubsec:model-training-evaluation}, we set the number of output time steps of LESnets to $T_{output}=1$ and the default time interval of $\Delta t_{\mathrm{train}}=\Delta t_{inf}=\Delta T=200\Delta t = 1.0$. In this subsection, we employ the same training dataset $\mathcal{C}_{train}^{180}$ and the same WALE coefficient $C_w=0.1$. Then we train and verify the LESnets at two new configurations of the numbers of output time steps and time intervals: $T_{output} = 10$, $\Delta t_{\mathrm{train}}=\Delta t_{inf} = 0.1\Delta T = 20\Delta t = 0.1$, and $T_{output} = 4$, $\Delta t_{\mathrm{train}}=\Delta t_{inf} = 0.25\Delta T = 50\Delta t = 0.25$. The training parameters are the same as described at the beginning of this section. In three configurations, the total time span represented by each output is fixed at one $\Delta T$, which enables a direct and consistent comparison of the resulting predictions. Here, it should be emphasized that, to ensure a fair comparison and maintain consistency, we generate data with the same time interval used in the IUFNO model \cite{IUFNO}, i.e., $\Delta t_{\mathrm{train}} = \Delta T = 1.0$. In contrast, the LESnets model can be trained using only the initial field, as detailed in \cite{LESnets}. For data-driven models, generating outputs at different numbers of output time steps or with different time intervals requires preparing training data again. For the case with the number of output time steps of $T_{output} = 10$ and the time interval of $\Delta t_{\mathrm{train}}=20\Delta t$, we set the inference time steps to $T_{inf}=8000$. For the case with the number of output time steps of $T_{output} = 4$ and the time interval of $\Delta t_{\mathrm{train}}=50\Delta t$, we set the inference time steps to $T_{inf} = 3200$. This ensures that the physical inference time is consistently $800\Delta T$ across the three time interval configurations. Thanks to the time-marching finite difference framework, the accuracy of the PDE loss computation can be maintained almost unchanged across different numbers of output time steps.

Fig. \ref{fig:re180-flexible-output-velocity} shows the temporal evolution of the streamwise velocity component at friction Reynolds number $Re_{\tau}\approx180$ at the spatial location $[2\pi,0,2\pi/3]$ in the temporal domain $[0,10\Delta T]$ in three different time intervals. It can be seen that, regardless of the chosen time interval, LESnets remains in good agreement with the WALE results over short time horizons, with only a slight lag observed when multiple time step outputs are used.

Fig. \ref{fig:re180-flexible-output-statistics}(a) shows the mean streamwise velocity profiles at friction Reynolds number $Re_{\tau}\approx180$ for three time intervals. It can be seen that LESnets with $\Delta t_{\mathrm{train}}=50\Delta t$ and $200\Delta t$ exhibits almost the same results compared with three traditional methods, and LESnets with $\Delta t_{\mathrm{train}}=20\Delta t$ slightly overestimates the streamwise velocity in the log-law region. Fig. \ref{fig:re180-flexible-output-statistics}(b) presents the Reynolds shear stress $-\langle u'v'\rangle^+$ at friction Reynolds number $Re_{\tau}\approx180$ for three time intervals. LESnets with $\Delta t_{\mathrm{train}}=20\Delta t$, $50\Delta t$, and $200\Delta t$ show good agreement with the WALE data. This indicates that outputting time series at excessively high temporal resolution poses a challenge to the model's long-term predictive capability.

Fig. \ref{fig:re180-flexible-output-spectra} shows the kinetic energy spectrum in streamwise and spanwise directions at friction Reynolds number $Re_{\tau}\approx180$ in three different time intervals. The results obtained from the three time intervals are nearly identical in terms of the kinetic energy spectrum in streamwise and spanwise directions.

\subsubsection{Automatically learning the LES coefficient}
\label{subsubsec:learning-les-coefficient}

The empirical coefficients in traditional SGS models play a critical role in model accuracy. In the preceding discussion, the WALE coefficient $C_w$ has been treated as $a$ $priori$. However, this assumption substantially limits the applicability of the LESnets framework. To mitigate the dependence on model coefficients, data assimilation has been employed as an efficient complementary tool to improve SGS closures \cite{wang2023ensemble, Zhou25_FEL}. In addition, PINNs have been used in inverse problems to learn unknown PDE parameters by constructing suitable PDE loss functions using neural networks \cite{chen2021physics}. Inspired by previous data assimilation and PINN methods, Zhao et al. \cite{LESnets} proposed an approach to automatically learn the SGS model coefficient during the training process of PINO. This approach treats the SGS model coefficient as a model parameter and incorporates a subset of the LES dataset during training to enable its automatic learning. It should be emphasized that, since the flow-field mapping and the SGS coefficient are learned simultaneously, the learned coefficient forms part of a joint optimum and is not required to match the $a$ $priori$ value; the appropriate criterion is instead whether the turbulence statistics predicted in the prediction stage are accurate.

In this subsection, we incorporate one group of the fDNS dataset $\mathcal{C}_{fDNS}^{180}$ (obtained by applying Eq. \eqref{eq:spatial-filter} to the DNS data), organized in tensors of dimensions $N_s\times T_{fDNS} \times N_x\times N_y \times N_z \times N_d=1\times20\times32\times65\times32\times4$, where $N_s$ is the number of samples of the dataset $\mathcal{C}_{fDNS}^{180}$, $T_{fDNS}$ is the number of time steps of the dataset $\mathcal{C}_{fDNS}^{180}$. The time interval of $\mathcal{C}_{fDNS}^{180}$ is $\Delta t_{\mathrm{fDNS}} = \Delta t = 5.00\times 10^{-3}$. The fDNS dataset $\mathcal{C}_{fDNS}^{180}$ is used as additional input data in the training process of LESnets. Then, the physics-informed loss function is given by Eq. \eqref{eq:learnable-sgs-loss}. We set the hyper-parameter $\lambda=5$, similar to our previous work \cite{LESnets}. The other training parameters are the same as described at the beginning of this section.

\begin{figure}[htbp]
\centering
\begin{minipage}{0.49\linewidth}
\centerline{\includegraphics[width=\textwidth]{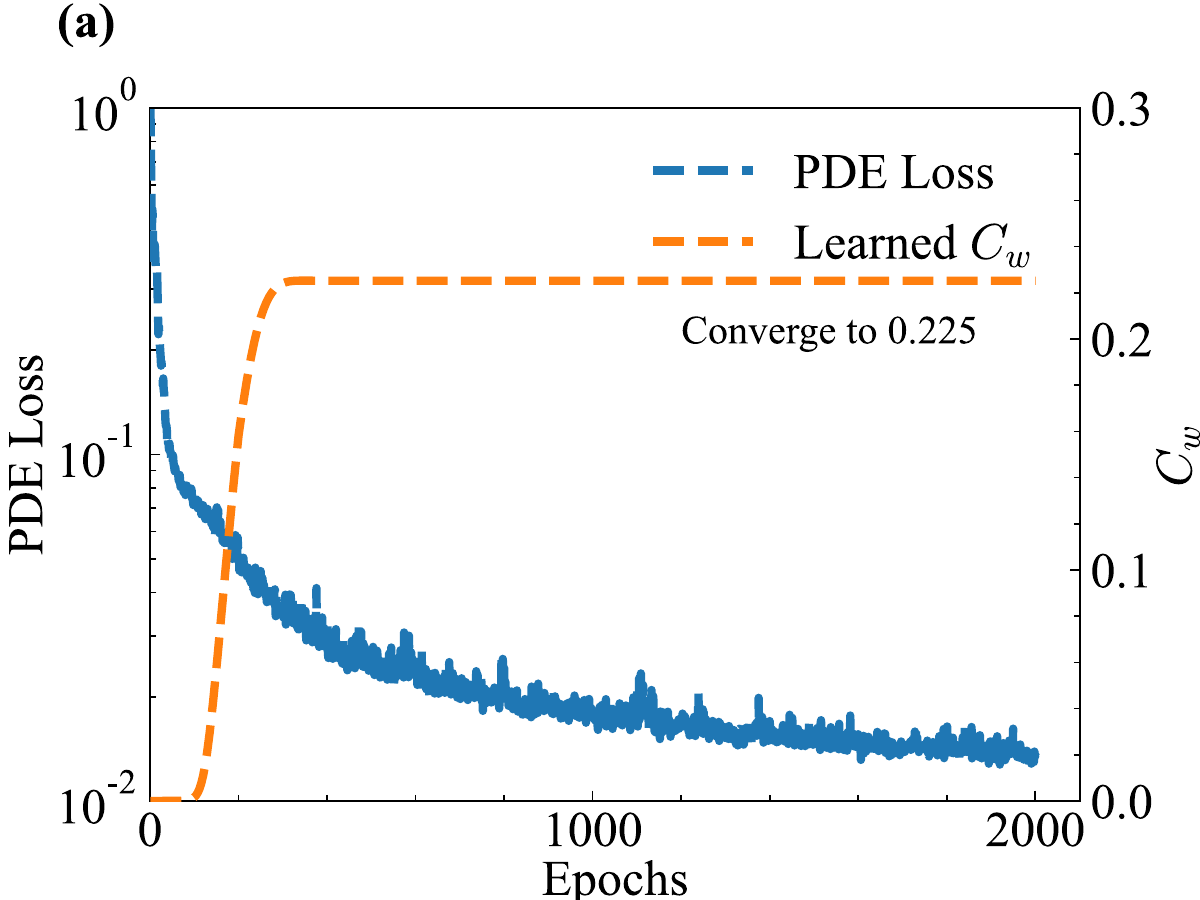}}
\end{minipage}
\hspace{-2pt}
\begin{minipage}{0.49\linewidth}
\centerline{\includegraphics[width=\textwidth]{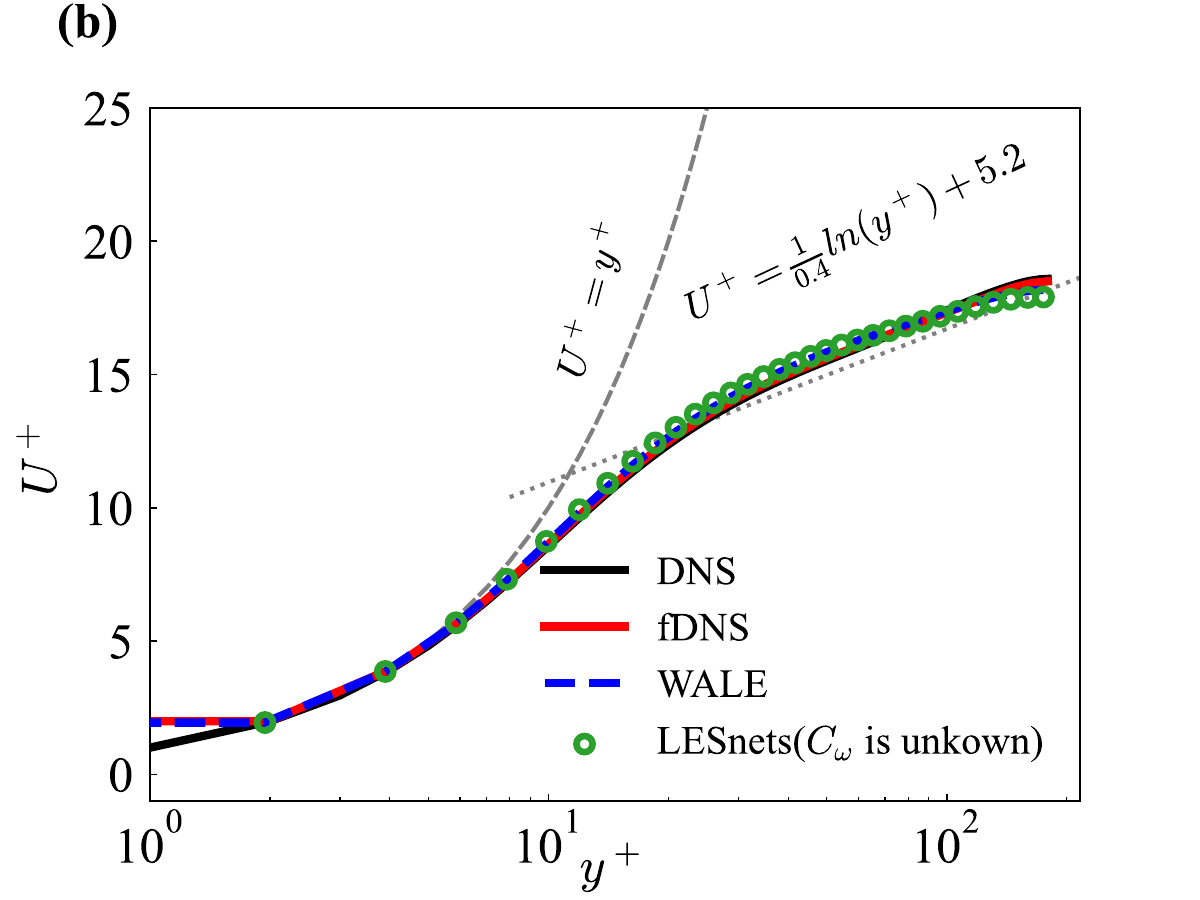}}
\end{minipage}

\begin{minipage}{0.49\linewidth}
\centerline{\includegraphics[width=\textwidth]{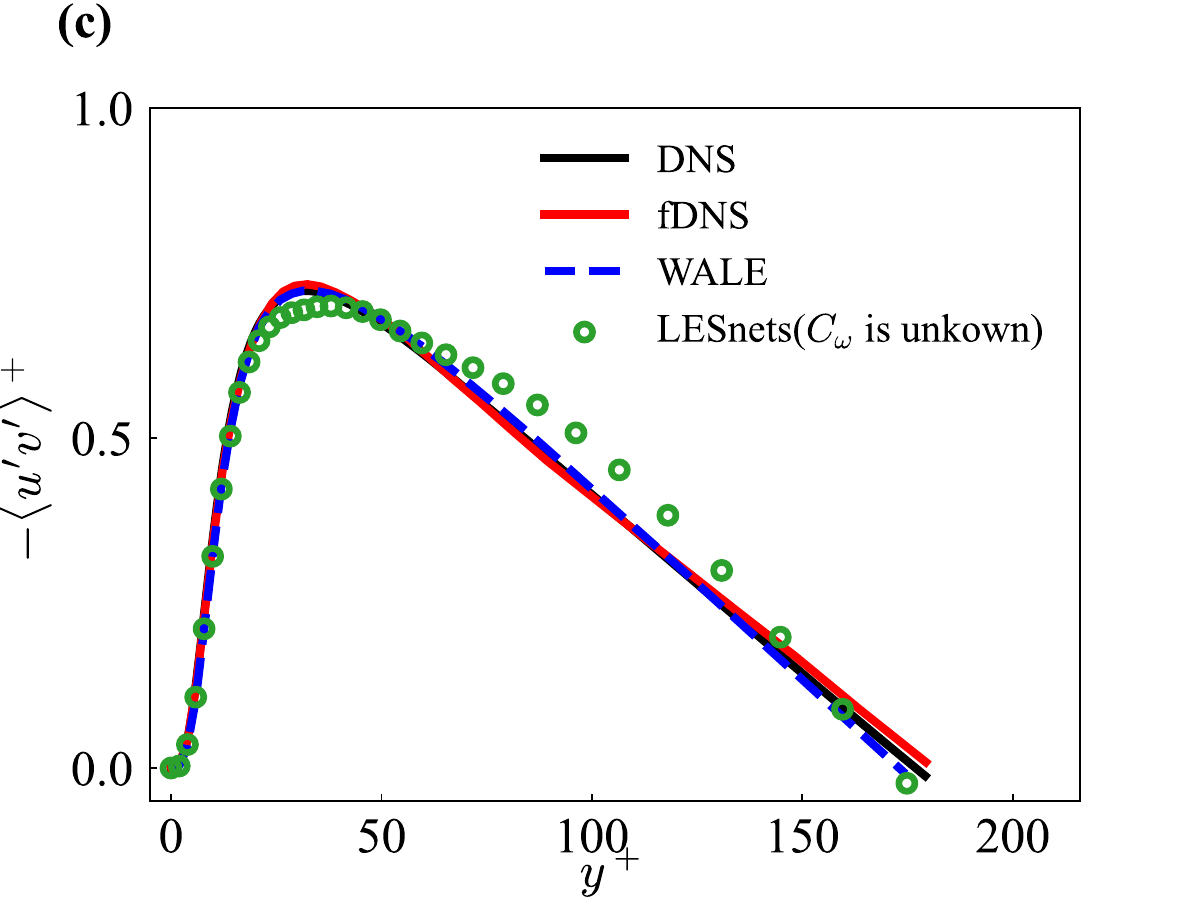}}
\end{minipage}
\hspace{-2pt}
\begin{minipage}{0.49\linewidth}
\centerline{\includegraphics[width=\textwidth]{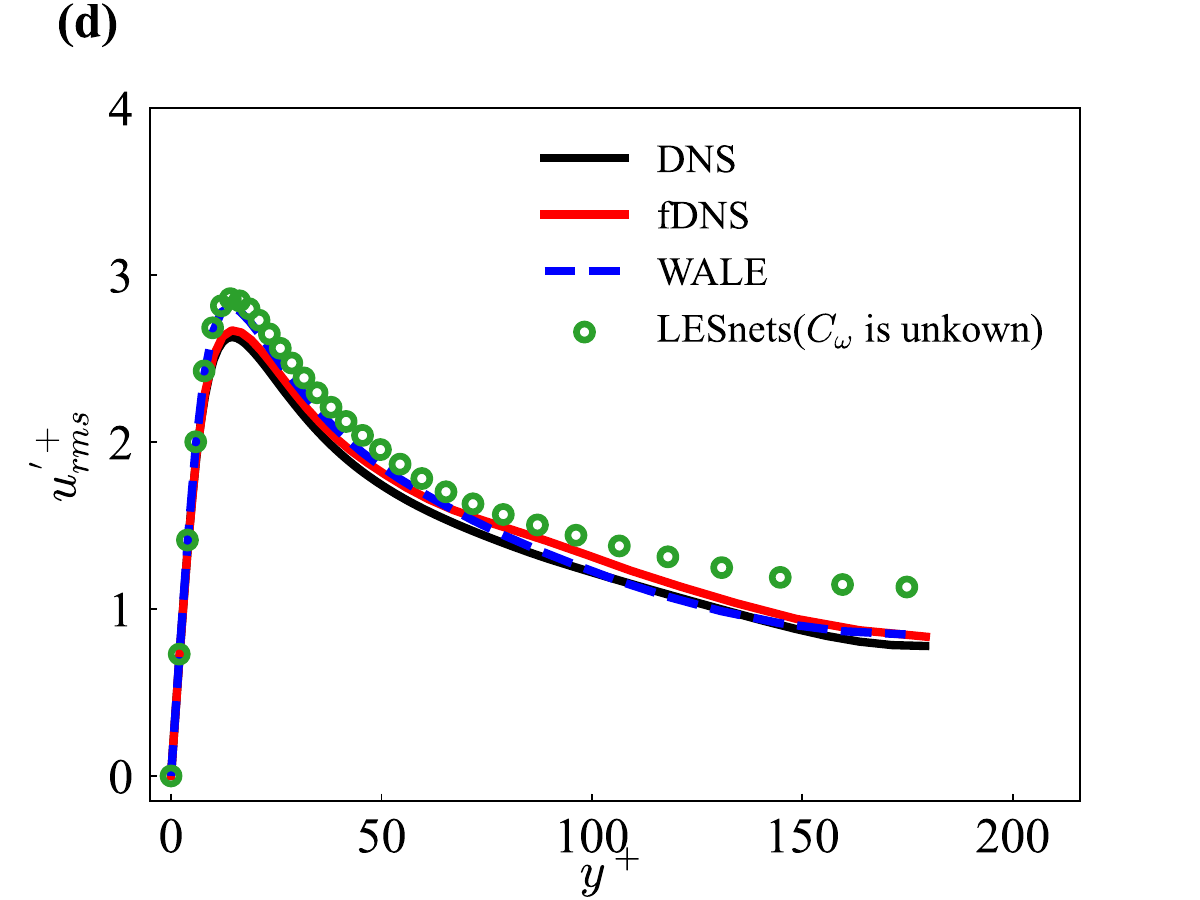}}
\end{minipage}
\caption{Comparison between DNS, fDNS, WALE, and LESnets with unknown $C_w$ coefficient of turbulent channel flow at friction Reynolds number $Re_{\tau}\approx180$. \textbf{(a)} PDE Loss and learned $C_w$ value curves; \textbf{(b)} Mean streamwise velocity $U^+ = \langle \overline{u}\rangle/u_{\tau}$ profile, normalized in wall units; \textbf{(c)} Reynolds shear stress $-\langle u'v'\rangle^+$, normalized by the square of respective friction velocity $u_{\tau}^2$; \textbf{(d)} RMS fluctuations of streamwise velocity.}
\label{fig:learned-wale-coefficient}
\end{figure}

Fig. \ref{fig:learned-wale-coefficient} shows the PDE loss and learned $C_w$ value curves in the training process. It can be seen that the additional loss term does not affect the decrease of the PDE loss. The final learned value of $C_w$ is 0.225, which differs from the value of 0.1 used earlier. The results in Figs. \ref{fig:learned-wale-coefficient}(b) to (d) show the mean streamwise velocity, the Reynolds shear stress, and the RMS fluctuations of streamwise velocity. The results are essentially consistent with three traditional methods.

\subsubsection{Computational efficiency}
\label{subsubsec:computational-efficiency}

A comparison between a neural operator and a numerical solver is meaningful only when the two methods preserve the same physical time interval, are run on the hardware on which each method is implemented, and are assessed at the accuracy that each method actually delivers. As pointed out by Huang et al. \cite{huang2026costaccuracy}, the cost of a neural operator should be counted together with the offline expense of generating the training data and of training the model. A benefit over the numerical solver is then expected only if a single inference is less expensive than the corresponding numerical solution, and only if the offline expense is amortized over a sufficiently large number of queries. Both aspects are examined below. Tables \ref{tab:computational-cost} and \ref{tab:computational-cost-hill} report the cost of advancing the solution over $10^4$ numerical time steps, that is, over $50$ and $20$ dimensionless time units for the channel and the periodic hill flows, respectively. The LES is run on $16$ or $32$ CPU cores, as indicated in the tables, whereas the machine-learning models are evaluated on a single GPU; no conversion between the two is applied.

For the channel flows, LESnets requires about one tenth of the cost of the WALE at the same time step, the speedup being $9.2$, $9.3$, $12.0$, and $5.9$ at $Re_\tau\approx180$, $590$, $1000$, and $2000$. LESnets and F-FNO share the same F-FNO backbone, so that their inference costs are identical. The benefit of the physics-informed training therefore lies not in a lower cost but in a greater freedom in the choice of the output interval. Because the physics-informed loss does not rely on strictly time-ordered data as labels, the number of output time steps and their interval can be chosen freely (Section \ref{subsubsec:flexible-output}), whereas a data-driven operator must be trained on consecutive snapshots separated by a fixed interval. IUFNO is $2.4$ to $6.9$ times more expensive than LESnets, and it diverges at $Re_\tau\approx2000$. Two mechanisms underlie this reduction. The first is hardware efficiency. The factorized Fourier layers consist of dense tensor operations that are executed efficiently on a GPU, whereas the finite difference solver is an MPI-parallel CPU code. The second is more structural and concerns the depth of the computation. LESnets advances the solution by $\Delta t_{\mathrm{inf}}=\Delta T=200\Delta t=1.0$ in a single evaluation, so that $50$ evaluations replace $10^4$ stability-limited updates. It is this replacement of many stability-limited steps by a single learned update that Huang et al. \cite{huang2026costaccuracy} identify as the main source of the advantage of a learned time-advance map. For the periodic-hill flow, the reduction is smaller, by a factor of $3.0$, because of the smaller output time interval used for the machine-learning models in this case.

Two limitations should be acknowledged. First, subject to the CFL condition, a conventional solver can adjust its time step to achieve any prescribed temporal accuracy. By contrast, a time-advancing neural operator advances the solution over an interval fixed during training. Its temporal resolution is therefore fixed once the model has been trained; predictions at intermediate times or at a finer temporal resolution require a model trained using the corresponding time interval. Second, the tables report inference costs only. Although the training data used by LESnets in this study were generated using a numerical solver, equivalent data could instead be obtained from measured flow fields sampled at appropriate time intervals. The offline cost therefore does not necessarily include an additional simulation and is, at most, comparable to a few of the costs reported in the tables, plus the cost of model training. Consequently, a net computational saving is realized only when the same configuration is queried repeatedly. In this regime, the computational cost of LESnets is lower than that of LES; for a single long-duration simulation, no savings should be expected. The intended applications of the framework, including ensembles of initial conditions, parametric sweeps, and uncertainty quantification, fall precisely within this regime.

In summary, the present results do not support the claim that a neural operator is unconditionally cheaper than a numerical solver. However, once trained, LESnets advances a wall-bounded flow at about one tenth of the cost of the reference LES at the same time step, and this reduction stems from the replacement of stability-limited sequential updates by a single learned update rather than from any approximation of the physics.

\begin{table}[htbp]
\captionsetup{font=small,labelfont=bf,width=.62\textwidth}
\setlength{\abovecaptionskip}{0pt}
\caption{Computational cost of the LES and the machine-learning models per $10^4$ numerical time steps, i.e., per $50$ dimensionless time units, for turbulent channel flows. The LES is run on $16$ CPU cores at $Re_\tau\approx180$ and $1000$ and on $32$ CPU cores at $Re_\tau\approx590$ and $2000$; the machine-learning models are evaluated on a single GPU. No core-to-GPU conversion is applied. The last column reports the corresponding speedup of LESnets with respect to the WALE at the same time step.}
\label{tab:computational-cost}
\centering
\begin{tabular}{cccccc}
\toprule
$Re_{\tau}$ & WALE & IUFNO & F-FNO & LESnets & Speedup \\
\midrule
180  & 38.5\,s  & 28.8\,s & 4.2\,s  & 4.2\,s  & $9.2$  \\
590  & 139.8\,s & 35.3\,s & 15.0\,s & 15.0\,s & $9.3$  \\
1000 & 40.8\,s  & 14.6\,s & 3.4\,s  & 3.4\,s  & $12.0$ \\
2000 & 144.4\,s & diverged & 24.3\,s & 24.3\,s & $5.9$  \\
\bottomrule
\end{tabular}
\end{table}
\begin{table}[htbp]
\captionsetup{font=small,labelfont=bf,width=.62\textwidth}
\setlength{\abovecaptionskip}{0pt}
\caption{Computational cost of the WMLES and the machine-learning models per $10^4$ numerical time steps, i.e. per $20$ dimensionless time units, for the turbulent flow over periodic hills. The WMLES is run on $32$ CPU cores; the machine-learning models are evaluated on a single GPU. No core-to-GPU conversion is applied. The last column reports the corresponding speedup of LESnets with respect to the WMLES at the same time step.}
\label{tab:computational-cost-hill}
\centering
\begin{tabular}{ccccc}
\toprule
$Re_h$ & WMLES & F-FNO & LESnets & Speedup \\
\midrule
5600, 10595 & 53.5\,s & 18.1\,s & 18.1\,s & $3.0$ \\
\bottomrule
\end{tabular}
\end{table}

\section{Conclusion}
\label{sec:conclusion}

We propose a novel framework, LESnets, based on physics-informed neural operators for the fast and accurate prediction of wall-bounded turbulent flows. We embed the F-FNO architecture to support stable and accurate long-term predictions in the turbulent channel flow cases examined in this study. LESnets does not require labeled data to train the model, and allows the model to output results at a flexible time interval. Moreover, the coefficient of the SGS model can be optimized during training by incorporating a small amount of fDNS data. For wall-bounded turbulent flows at high Reynolds numbers, we incorporate a wall model to approximately enforce the law of the wall as a physical constraint within the LESnets model. Overall, for the tested Reynolds numbers, LESnets achieves prediction accuracy comparable to or better than the two data-driven baseline models for most of the examined flow statistics.

In the $a$ $posteriori$ test, the prediction ability of LESnets is comprehensively examined and compared against DNS, WALE, and WMLES, as well as the data-driven IUFNO and F-FNO models, in turbulent channel flows at $Re_{\tau}\approx180$ and $590$ and in wall-modeled turbulent channel flow at $Re_{\tau}\approx1000$ and $2000$. For most of the examined statistical quantities, the LESnets predictions are closer to WALE or WMLES benchmarks than those of IUFNO and F-FNO, although the relative performance depends on the Reynolds number and on the quantity considered.

The framework is further assessed for the turbulent flow over periodic hills, in which the wall is no longer aligned with the grid. The curved hill is handled by the immersed boundary method, so that no body-fitted mesh is required. We train two models on the same WMLES dataset at $Re_h\approx5600$; we validate the first against DNS at $Re_h\approx5600$ and the second against the WRLES at $Re_h\approx10595$. In both cases, LESnets accurately reproduces the mean velocity, Reynolds shear stress, turbulent kinetic energy, and separation and reattachment locations, showing good agreement with the reference solutions, while achieving accuracy comparable to that of F-FNO over the entire inference period. The same training dataset is reused at $Re_h\approx10595$, with no additional data required. These results indicate that the combination of the hard boundary constraints, the immersed boundary method, and the physics-informed loss remains stable when the geometry is not aligned with the grid.

To the best of our knowledge, the current study is among the first attempts to develop a PINO framework for large-eddy simulation of 3D wall-bounded turbulent flows. Although the present results are encouraging, further work is needed to develop the PINO method for more challenging problems at higher Reynolds numbers and in more complex geometries. In addition, the present study is limited to incompressible turbulence. Extending the framework to compressible turbulent flows is an important direction for future research.
\section*{CRediT authorship contribution statement}

\textbf{Sunan Zhao:} Conceptualization, Methodology, Investigation, Coding, Validation, Writing -- original draft preparation, Writing -- reviewing and editing. \textbf{Yunpeng Wang:} Conceptualization, Methodology, Investigation, Coding, Writing -- reviewing and editing. \textbf{Boyu Fan:} Conceptualization, Methodology, Investigation, Coding, Writing -- reviewing and editing. \textbf{Huiyu Yang:} Conceptualization, Methodology, Investigation, Coding, Writing -- reviewing and editing. \textbf{Zhihong Guo:} Conceptualization, Methodology, Investigation, Coding, Writing -- reviewing and editing. \textbf{Jianchun Wang:} Conceptualization, Methodology, Investigation, Supervision, Writing -- reviewing and editing, Project administration, Funding acquisition.

\section*{Declaration of competing interest}
The authors declare that they have no known competing financial interests or personal relationships that could have appeared to influence the work reported in this paper.

\section*{{Code availability}}
{Code and dataset in this study will be released to the public after the paper is accepted for publication.} 

\section*{ACKNOWLEDGMENTS}
This work was supported by the NSFC Excellence Research Group Program for ‘Multiscale Problems in Nonlinear Mechanics’
(No. 12588201), the National Natural Science Foundation of China (NSFC) (Grant Nos. 12302283 and 12588301); the Shenzhen Science and Technology Program (Grant Nos. SYSPG20241211173725008); and the Department of Science and Technology of Guangdong Province (Grant Nos. 2019B21203001, 2020B1212030001, and 2023B1212060001). Additional support was provided by the Center for Computational Science and Engineering of Southern University of Science and Technology.

\appendix
\setcounter{table}{0}   
\setcounter{figure}{0}
\setcounter{equation}{0}

\section{{The 1D Lagrangian interpolation reconstruction approach}}
\label{app:1D_lag}
\renewcommand{\thetable}{A\arabic{table}}
\renewcommand{\thefigure}{A\arabic{figure}}
\renewcommand{\theequation}{A\arabic{equation}}

\begin{figure}[htbp]
\centering
\includegraphics [width=1.0\textwidth]{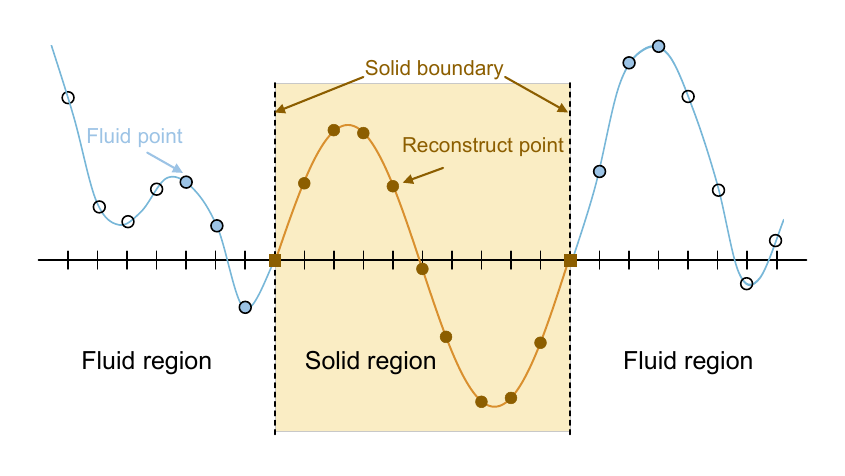}
\caption{Schematic view of the polynomial reconstruction based on a 1D Lagrangian polynomial.}
\label{fig:1d_lag}
\end{figure}

The 1D Lagrangian interpolation reconstruction approach is illustrated in Fig. \ref{fig:1d_lag} for a fully immersed situation. For each body $l$, an interpolation node set $\mathcal{N}_l=\{x_0,\dots,x_{n-1}\}$ is assembled from the two interface points $x_l^s$, $x_l^f$ together with the $n_{\mathrm{pif}}$ neighboring fluid grid points on each side, giving $n=6$ nodes (a fifth-degree polynomial) for $n_{\mathrm{pif}}=2$. The node values satisfy the no-slip condition at the interface, i.e.\ $u_j=0$ for $x_j\in\{x_l^s,x_l^f\}$, and equal the fluid velocity elsewhere. The velocity at the grid points inside the body is then reconstructed by 1D Lagrangian interpolation,
\begin{equation}
\tilde u(x_p) = \sum_{j=0}^{n-1} u_j \,\ell_j(x_p), \qquad
\ell_j(x) = \prod_{\substack{k=0 \\ k\neq j}}^{n-1}
\frac{x - x_k}{x_j - x_k},
\label{eq:lagrange}
\end{equation}

\section{{The architecture of implicit U-Net enhanced Fourier neural operator (IUFNO)}}
\label{app:iufno-architecture}
\renewcommand{\thetable}{B\arabic{table}}
\renewcommand{\thefigure}{B\arabic{figure}}
\renewcommand{\theequation}{B\arabic{equation}}
The architecture of the IUFNO model used in this study is shown in Fig. \ref{fig:iufno-architecture}. IUFNO follows the same main framework as FNO, with lifting and projection layers identical to those in FNO. The key difference lies in the Fourier layer: the IUFNO model employs a single Fourier layer with shared parameters in a looper format and incorporates a U-Net to capture small-scale flow structures. In addition, the implicit use of a shared hidden layer substantially reduces the number of trainable parameters, allowing the network to be very deep. For more information about the theories behind IUFNO, refer to \cite{IUFNO, wang2024prediction}. 

\begin{figure}[htbp]
\centering
\includegraphics [width=1.0\textwidth]{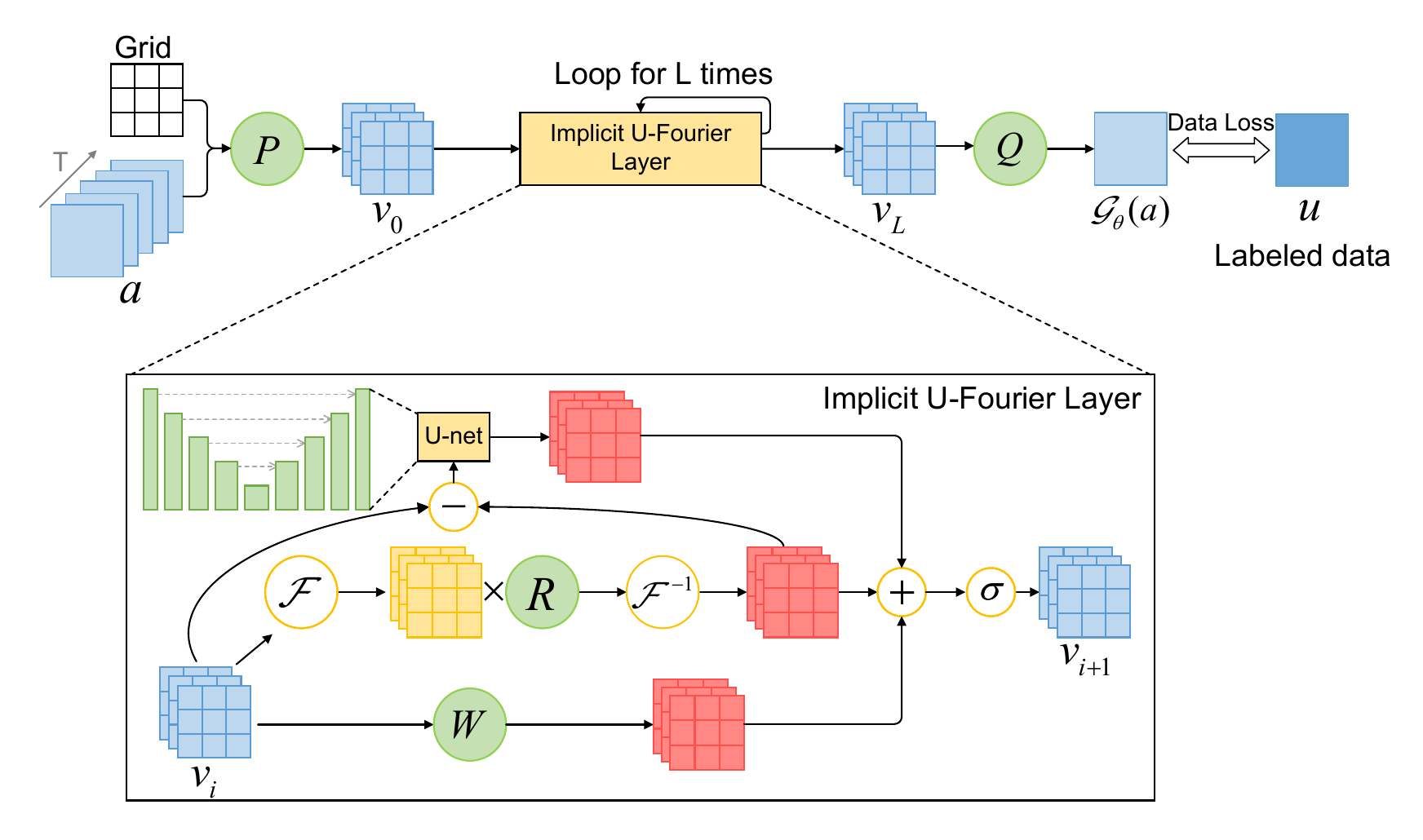}
\caption{The architecture of the implicit U-Net enhanced Fourier neural operator. The top part is a single Fourier layer with an iterative scheme. The zoomed-in operator layer is shown at the bottom.}
\label{fig:iufno-architecture}
\end{figure}

\section{{Finite difference physics-informed loss}}
\label{app:finite-difference-loss}
\renewcommand{\thetable}{C\arabic{table}}
\renewcommand{\thefigure}{C\arabic{figure}}
\renewcommand{\theequation}{C\arabic{equation}}
\begin{figure}[htbp]
\centering
\begin{minipage}{0.49\linewidth} 
\centerline{\includegraphics[width=\textwidth]{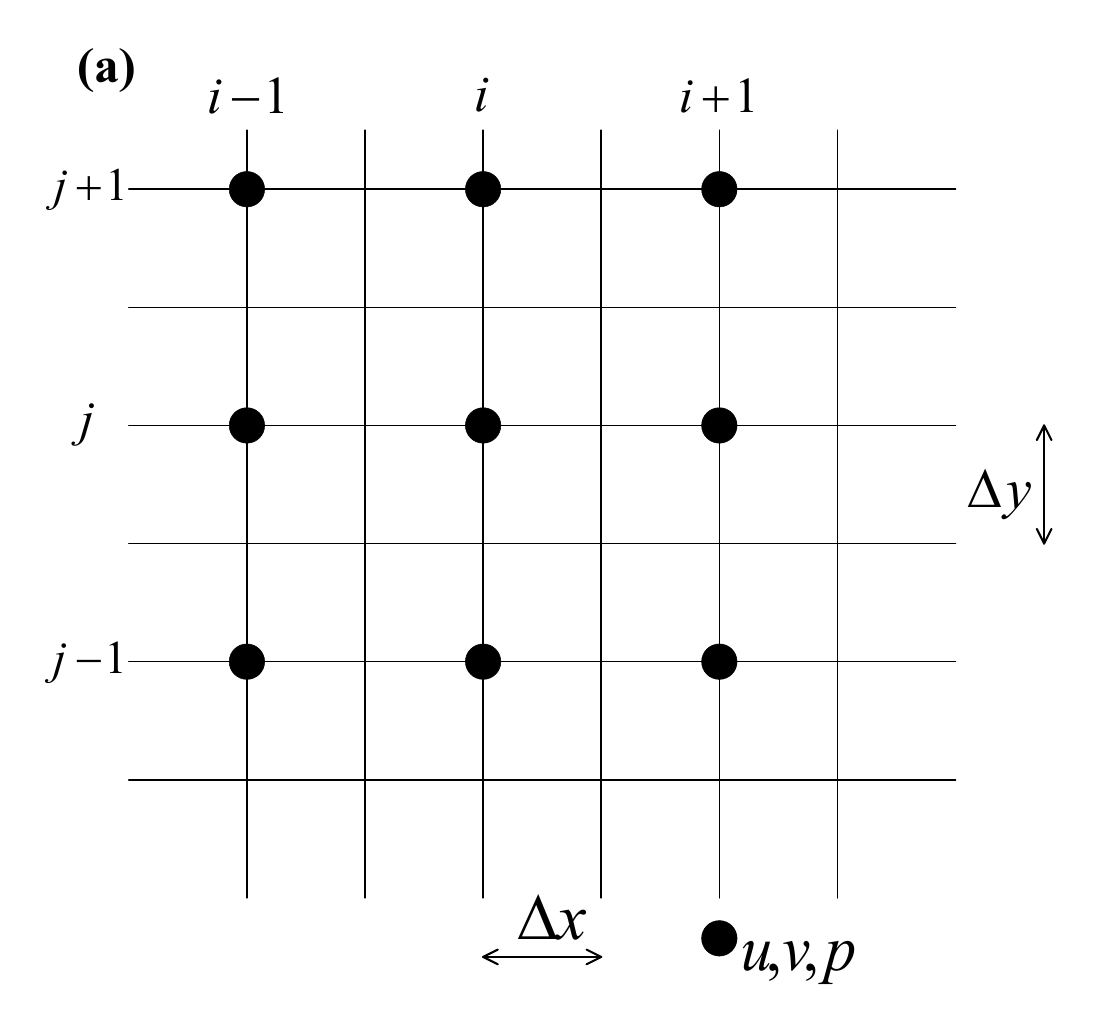}}
\end{minipage}
\hspace{-2pt} 
\begin{minipage}{0.49\linewidth} 
\centerline{\includegraphics[width=\textwidth]{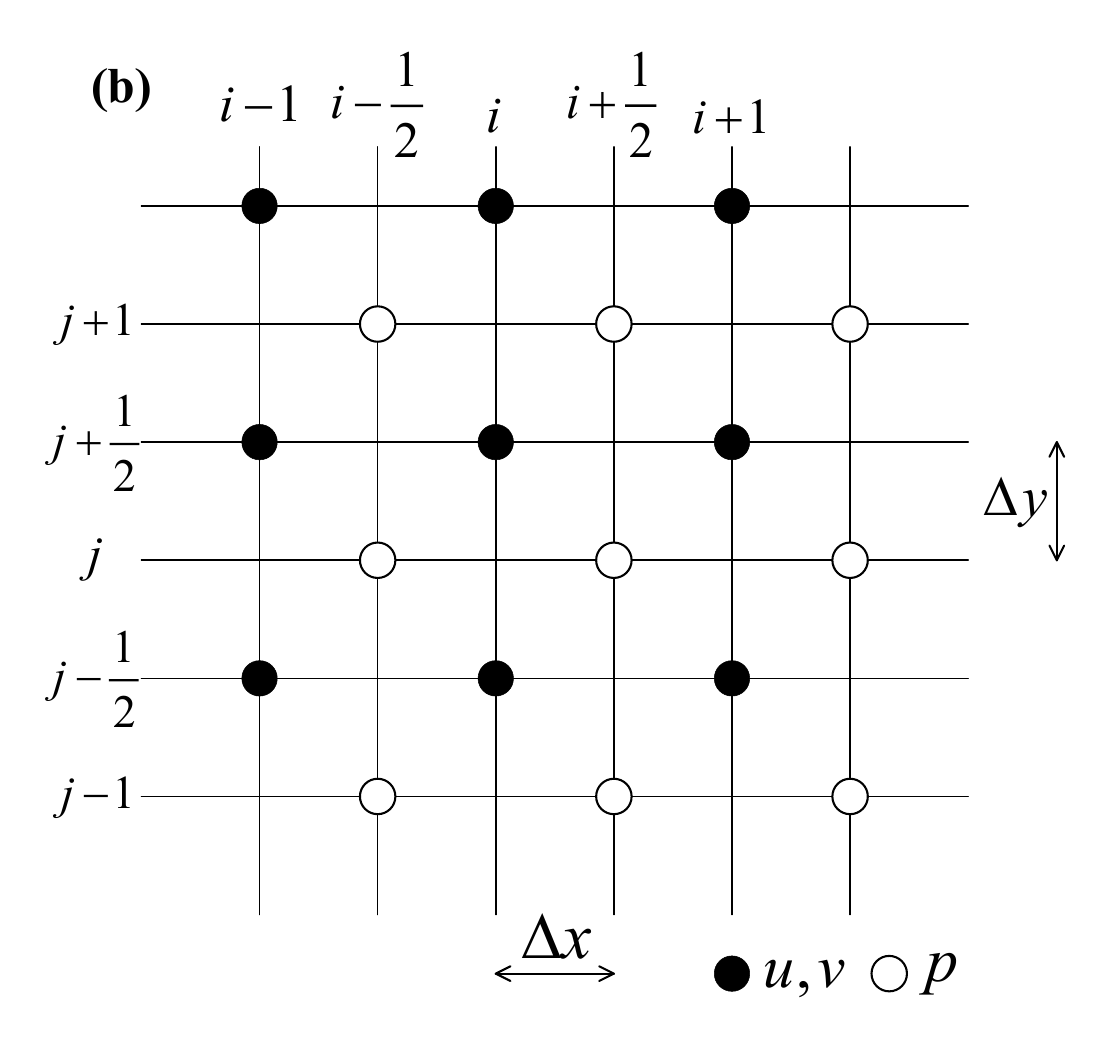}}
\end{minipage}

\caption{Arrangement of variables in two dimensions for \textbf{(a)} collocated, and \textbf{(b)} partially staggered meshes.}
\label{fig:grid-arrangements}
\end{figure}

Many traditional physics-informed machine-learning methods rely on automatic differentiation (AD) \cite{autograd} for calculating derivatives in output space. While this approach is robust, it can be computationally expensive, particularly when dealing with high-order derivatives, as it requires backpropagation through the entirety of the network. Rather than using AD, PINO models typically leverage Fourier derivatives for deriving the PDE constraints in straightforward 1D and 2D PDE scenarios, as AD is very memory-intensive for this type of architecture. However, it’s important to recognize that Fourier derivatives have strict requirements regarding boundary conditions and mesh specifications, which may limit their versatility in some applications. To improve computational efficiency, we instead employ a finite difference (FD) stencil in physical space to approximate the necessary gradients.

In the FD solver used in this work, the velocity field in the model output is always located at the same grid locations. The only staggering concerns the pressure field from the given input data, which can either be placed at the velocity grid nodes (collocated mesh, see Fig. \ref{fig:grid-arrangements}(a)) or shifted by half a grid spacing in each spatial direction (staggered mesh, see Fig. \ref{fig:grid-arrangements}(b)). This partially staggered arrangement is straightforward to implement and computationally efficient: it requires only a modification of the pressure treatment, avoids costly mid-point interpolations, and allows any collocated-mesh code to be adapted with minimal changes. 

For each sample realization $(i)$, we approximate sixth-order spatial derivatives of a certain physical quantity described in Table \ref{tab:finite-difference-schemes}.

\begin{table}[htbp]
\captionsetup{font=small,labelfont=bf, width=.80\textwidth}
\setlength{\abovecaptionskip}{0pt}

\caption{Spatial derivatives used for calculation of physics-informed losses.}
\label{tab:finite-difference-schemes}
\centering
\renewcommand{\arraystretch}{1.5}
\begin{tabular}{|l|l|l|}
\hline
\textbf{Operator} & \textbf{Difference scheme} & \textbf{Scheme parameters} \\
\hline

\multirow{3}{*}{First Derivative} & 
\multicolumn{1}{c|}{
\multirow{3}{*}{\parbox{0.4\textwidth}{
$\begin{array}{@{}r@{}l@{}}
\alpha f^{'}_{i-1}+f^{'}_{i}+\alpha f^{'}_{i+1} =a \dfrac{f_{i+1}-f_{i-1}}{2\Delta x} \\[5pt]
 +b\dfrac{f_{i+2}-f_{i-2}}{4\Delta x}
\end{array}$
}}} &
$\alpha = 1/3$, \\
& & $a=14/9$, \\
& & $b=1/9$,\\
\hline
\multirow{5}{*}{Second Derivative} & 
\multicolumn{1}{c|}{
\multirow{5}{*}{\parbox{0.4\textwidth}{
$\begin{array}{@{}r@{}l@{}}
\alpha f^{''}_{i-1}+f^{''}_{i}+\alpha f^{''}_{i+1} =a \frac{f_{i+1}-2f_{i}+f_{i-1}}{\Delta x^2} \\[5pt]
+b\frac{f_{i+2}-2f_{i}+f_{i-2}}{4\Delta x^2} \\ [5pt]
+c\frac{f_{i+3}-2f_{i}+f_{i-3}}{9\Delta x^2} \\ [5pt]
+d\frac{f_{i+4}-2f_{i}+f_{i-4}}{16\Delta x^2}
\end{array}$
}}} & $\alpha = 2/11$,\\
& & $a=12/11$,\\
& & $b=3/11$,\\
& & $c=0$,\\
& & $d=0$.\\
\hline
\end{tabular}
\end{table}

For the time derivative term, we use the explicit second-order Adams-Bashforth scheme, shown as follows:

\begin{equation}
\label{eq:adams-bashforth}
    \{\frac{\partial u}{\partial t}\}_{t_n} = \frac{1}{2}[3f(t_n,u_{t_n})-f(t_{n-1},u_{t_{n-1}})],
\end{equation}
where $f(t_n,u_{t_n})$ is the right-hand side term of Eq. \ref{eq:filtered-momentum} at time step ($t_n$), i.e., the part of the equation excluding the time derivative term.

In particular, by taking the divergence of the momentum equation Eq. \ref{eq:filtered-momentum}, we obtain the pressure Poisson equation in the following form:

\begin{equation}
\label{eq:pressure-poisson}
    \frac{\partial^2\bar{p}}{\partial x^2_i}= \frac{1}{\Delta t}\frac{\partial\bar{u}_i}{\partial x_i}.
\end{equation}

In this work, we refine the model's output velocity field to satisfy the continuity condition by solving the intermediate velocity from the pressure Poisson equation using the input pressure field. This reduces the number of loss terms and accelerates convergence of the training process.

\section{{Hyper-parameter settings for machine-learning models}}
\label{app:hyperparameters}
\renewcommand{\thetable}{D\arabic{table}}
\renewcommand{\thefigure}{D\arabic{figure}}
\renewcommand{\theequation}{D\arabic{equation}}

Tables \ref{tab:hyperparameters-re180}, \ref{tab:hyperparameters-re590}, \ref{tab:hyperparameters-re1000}, and \ref{tab:hyperparameters-re2000} show the parameters used for three machine-learning methods (i.e., IUFNO, F-FNO, and LESnets) of turbulent channel flows at four friction Reynolds numbers $Re_{\tau} \approx 180$, $590$, $1000$, and $2000$.

\begin{table}[htbp]
\captionsetup{font=small,labelfont=bf, width=.52\textwidth}
\setlength{\abovecaptionskip}{0pt}

\caption{Hyper-parameter settings of the three machine-learning models for the cases $Re_{\tau} \approx 180$.}
\label{tab:hyperparameters-re180}
\centering

\begin{tabular}{|l|l|l|l|}
\hline
\textbf{Hyper-parameter} & \textbf{IUFNO} & \textbf{F-FNO} & \textbf{LESnets} \\
\hline
Training groups & 20 & 20 & 20 \\
Fourier layers & 40 & 4 & 4 \\
Fourier width & 80 & 80 & 80 \\
Fourier modes(x-y-z) & 8-8-8 & 16-32-16 & 16-32-16 \\
Epochs & 100 & 2000 & 2000 \\
Batch size & 1 & 4 & 4 \\
Learning rate & 1e-3 & 1e-3 & 1e-3 \\
Decay factor of learning rate & 0.5 & 0.8 & 0.8 \\
Activation function & Gelu & Tanh & Tanh \\
Optimizer & Adam & SOAP & SOAP \\
\hline
\end{tabular}
\end{table}

\begin{table}[htbp]
\captionsetup{font=small,labelfont=bf, width=.52\textwidth}
\setlength{\abovecaptionskip}{0pt}

\caption{Hyper-parameter settings of the three machine-learning models for the cases $Re_{\tau} \approx 590$.}
\label{tab:hyperparameters-re590}
\centering

\begin{tabular}{|l|l|l|l|}
\hline
\textbf{Hyper-parameter} & \textbf{IUFNO} & \textbf{F-FNO} & \textbf{LESnets} \\
\hline
Training groups & 20 & 20 & 20 \\
Fourier layers & 10 & 4 & 4 \\
Fourier width & 80 & 80 & 80 \\
Fourier modes(x-y-z) & 16-16-16 & 32-32-32 & 32-32-32 \\
Epochs & 100 & 2000 & 2000 \\
Batch size & 1 & 4 & 4 \\
Learning rate & 1e-3 & 1e-3 & 1e-3 \\
Decay factor of learning rate & 0.5 & 0.8 & 0.8 \\
Activation function & Gelu & Tanh & Tanh \\
Optimizer & Adam & SOAP & SOAP \\
\hline
\end{tabular}
\end{table}

\begin{table}[htbp]
\captionsetup{font=small,labelfont=bf, width=.56\textwidth}
\setlength{\abovecaptionskip}{0pt}

\caption{Hyper-parameter settings of the three machine-learning models for the cases $Re_{\tau} \approx 1000$.}
\label{tab:hyperparameters-re1000}
\centering

\begin{tabular}{|l|l|l|l|}
\hline
\textbf{Hyper-parameter} & \textbf{IUFNO} & \textbf{F-FNO} & \textbf{LESnets-WM} \\
\hline
Training groups & 20 & 20 & 20 \\
Fourier layers & 40 & 4 & 4 \\
Fourier width & 80 & 120 & 120 \\
Fourier modes(x-y-z) & 8-8-8 & 16-32-16 & 16-32-16 \\
Epochs & 100 & 2000 & 2000 \\
Batch size & 1 & 1 & 1 \\
Learning rate & 1e-3 & 1e-3 & 1e-3 \\
Decay factor of learning rate & 0.5 & 0.9 & 0.9 \\
Activation function & Gelu & Tanh & Tanh \\
Optimizer & Adam & SOAP & SOAP \\
\hline
\end{tabular}
\end{table}

\begin{table}[htbp]
\captionsetup{font=small,labelfont=bf, width=.56\textwidth}
\setlength{\abovecaptionskip}{0pt}
\caption{Hyper-parameter settings of the three machine-learning models for the cases $Re_{\tau} \approx 2000$.}
\label{tab:hyperparameters-re2000}
\centering

\begin{tabular}{|l|l|l|l|}
\hline
\textbf{Hyper-parameter} & \textbf{IUFNO} & \textbf{F-FNO} & \textbf{LESnets-WM} \\
\hline
Training groups & 20 & 20 & 20 \\
Fourier layers & 40 & 4 & 4 \\
Fourier width & 80 & 120 & 120 \\
Fourier modes(x-y-z) & 16-16-16 & 32-32-32 & 32-32-32 \\
Epochs & 100 & 2000 & 2000 \\
Batch size & 1 & 1 & 1 \\
Learning rate & 1e-3 & 1e-3 & 1e-3 \\
Decay factor of learning rate & 0.5 & 0.9 & 0.9 \\
Activation function & Gelu & Tanh & Tanh \\
Optimizer & Adam & SOAP & SOAP \\
\hline
\end{tabular}
\end{table}

Table \ref{tab:hyperparameters-re5600-10595} shows the parameters used for two machine-learning methods (i.e., F-FNO and LESnets) for turbulent flow over periodic flows at two Reynolds numbers $Re_{\tau} \approx 5600$ and $10595$.

\begin{table}[htbp]
\captionsetup{font=small,labelfont=bf, width=.5\textwidth}
\setlength{\abovecaptionskip}{0pt}
\caption{Hyper-parameter settings of the three machine-learning models for the cases $Re_{\tau} \approx 5600$ and $10595$.}
\label{tab:hyperparameters-re5600-10595}
\centering

\begin{tabular}{|l|l|l|}
\hline
\textbf{Hyper-parameter}  & \textbf{F-FNO} & \textbf{LESnets-WM} \\
\hline
Training groups & 2 & 2 \\
Fourier layers & 4 & 4 \\
Fourier width & 120 & 120 \\
Fourier modes(x-y-z) & 32-36-16 & 32-36-16 \\
Epochs & 2000 & 2000 \\
Batch size & 1 & 1 \\
Learning rate & 1e-3 & 1e-3 \\
Decay factor of learning rate & 0.9 & 0.9 \\
Activation function & Tanh & Tanh \\
Optimizer & SOAP & SOAP \\
\hline
\end{tabular}
\end{table}

\section{{Discussion on the effect of hyper-parameters in LESnets}}
\label{app:hyperparameter-effects}
\renewcommand{\thetable}{E\arabic{table}}
\renewcommand{\thefigure}{E\arabic{figure}}
\renewcommand{\theequation}{E\arabic{equation}}
In this section, we examine how different hyper-parameters affect the performance of LESnets for turbulent channel flow at $Re_{\tau} \approx 590$ using only $T_{inf} = 10$ inference time steps, and we restrict our discussion to this specific case for brevity. In particular, we study the influence of \textbf{Fourier layers}, \textbf{Fourier width}, \textbf{Fourier modes}, \textbf{batch size}, \textbf{learning rate} and \textbf{decay factor}. Each of these hyper-parameters plays a crucial role in determining the training efficiency, predictive accuracy, and overall performance of the LESnets model. When evaluating the impact of a single hyper-parameter, all other hyper-parameters are fixed to the values listed in Table \ref{tab:hyperparameters-re590}.

\begin{figure}[htbp]
\centering
\begin{minipage}{0.45\linewidth}
\centerline{\includegraphics[width=\textwidth]{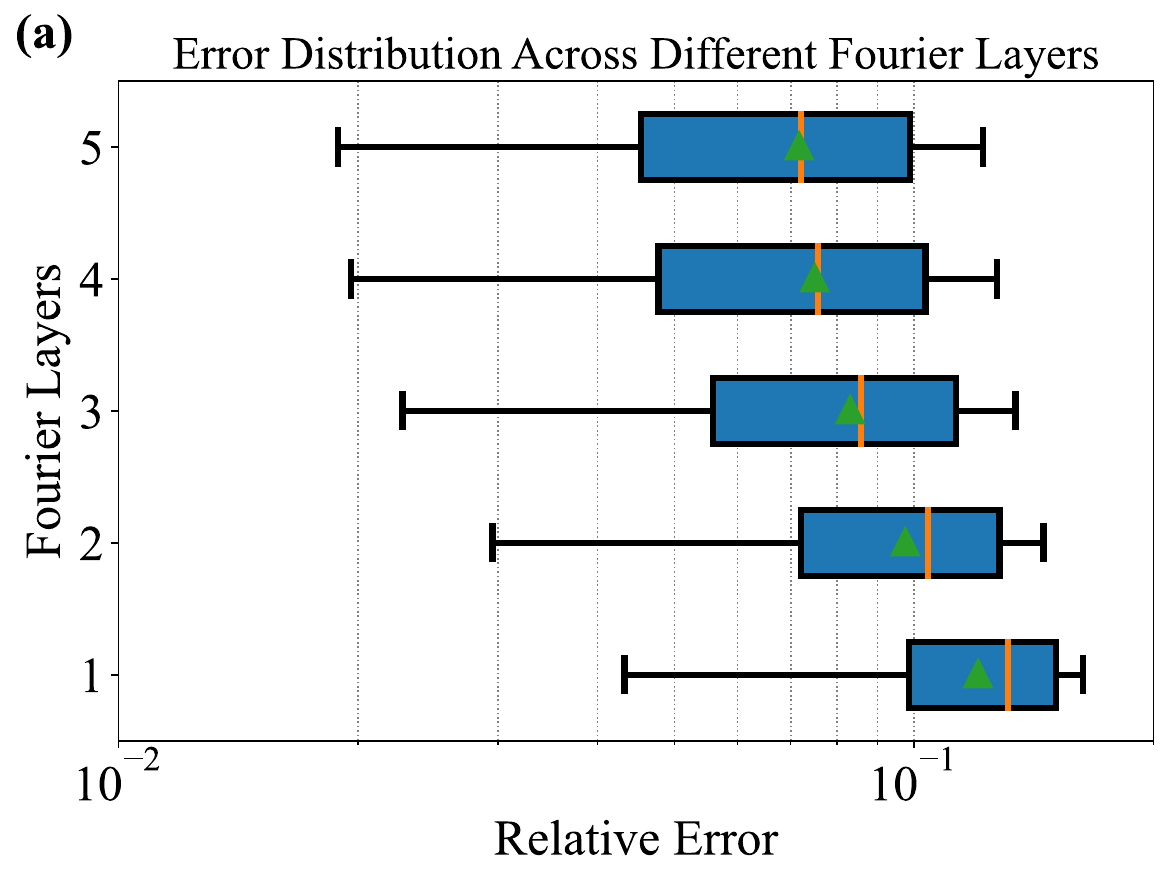}}
\end{minipage}
\hspace{1pt}
\begin{minipage}{0.45\linewidth}
\centerline{\includegraphics[width=\textwidth]{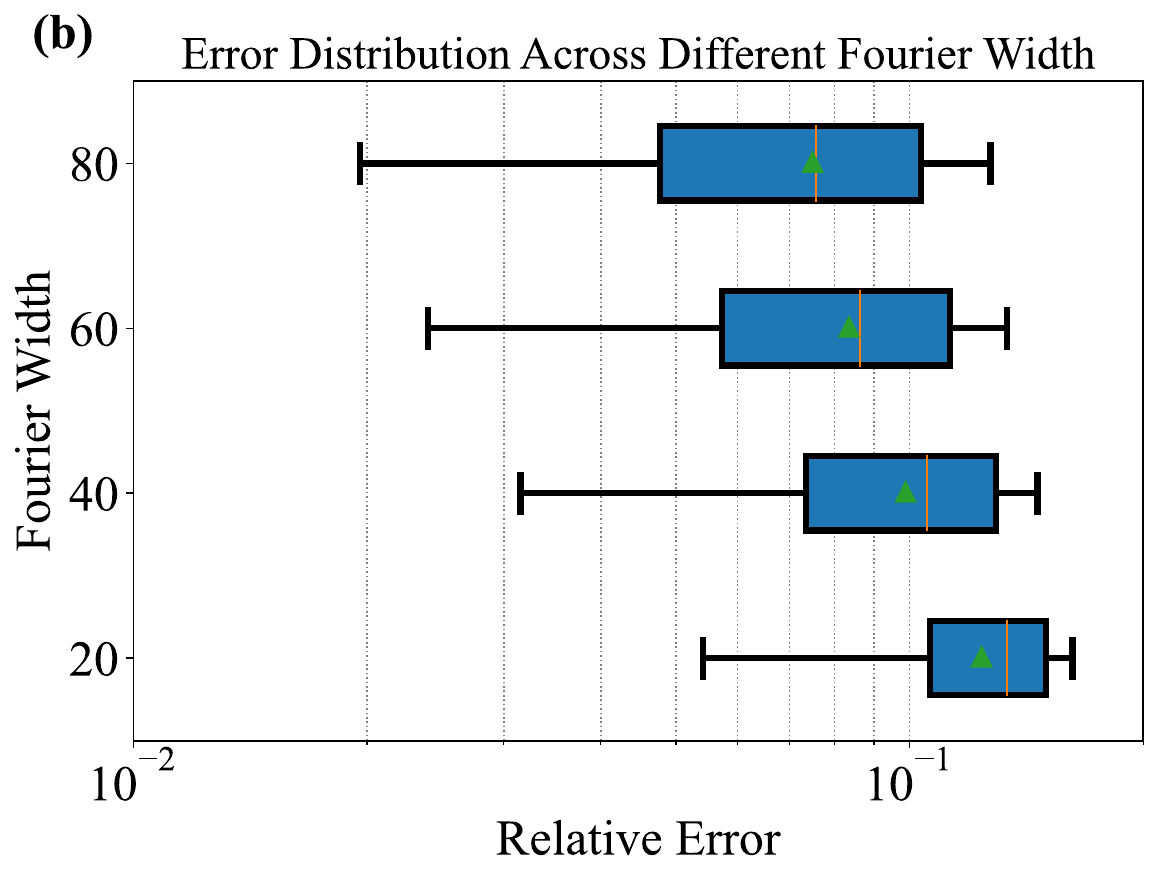}}
\end{minipage}
\vspace{-1pt}
\begin{minipage}{0.45\linewidth}
\centerline{\includegraphics[width=\textwidth]{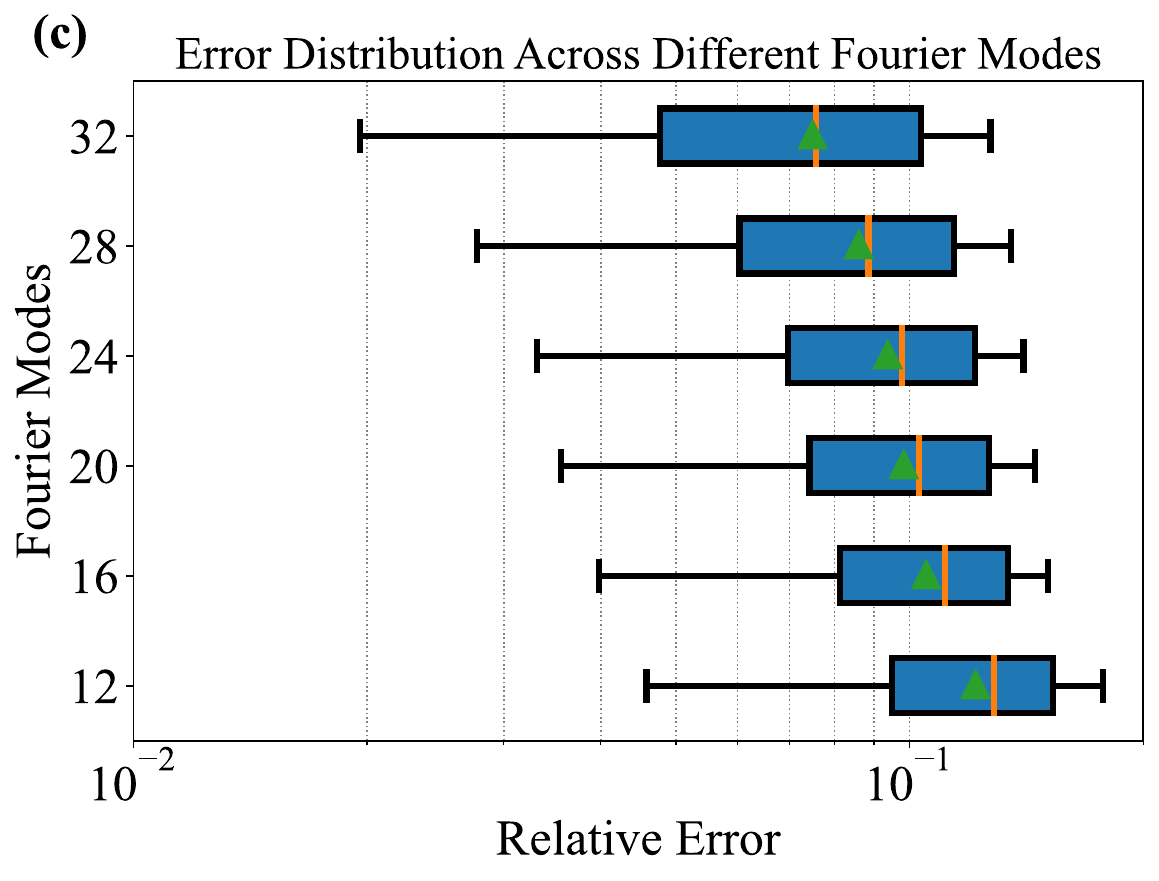}}
\end{minipage}
\hspace{1pt}
\begin{minipage}{0.45\linewidth}
\centerline{\includegraphics[width=\textwidth]{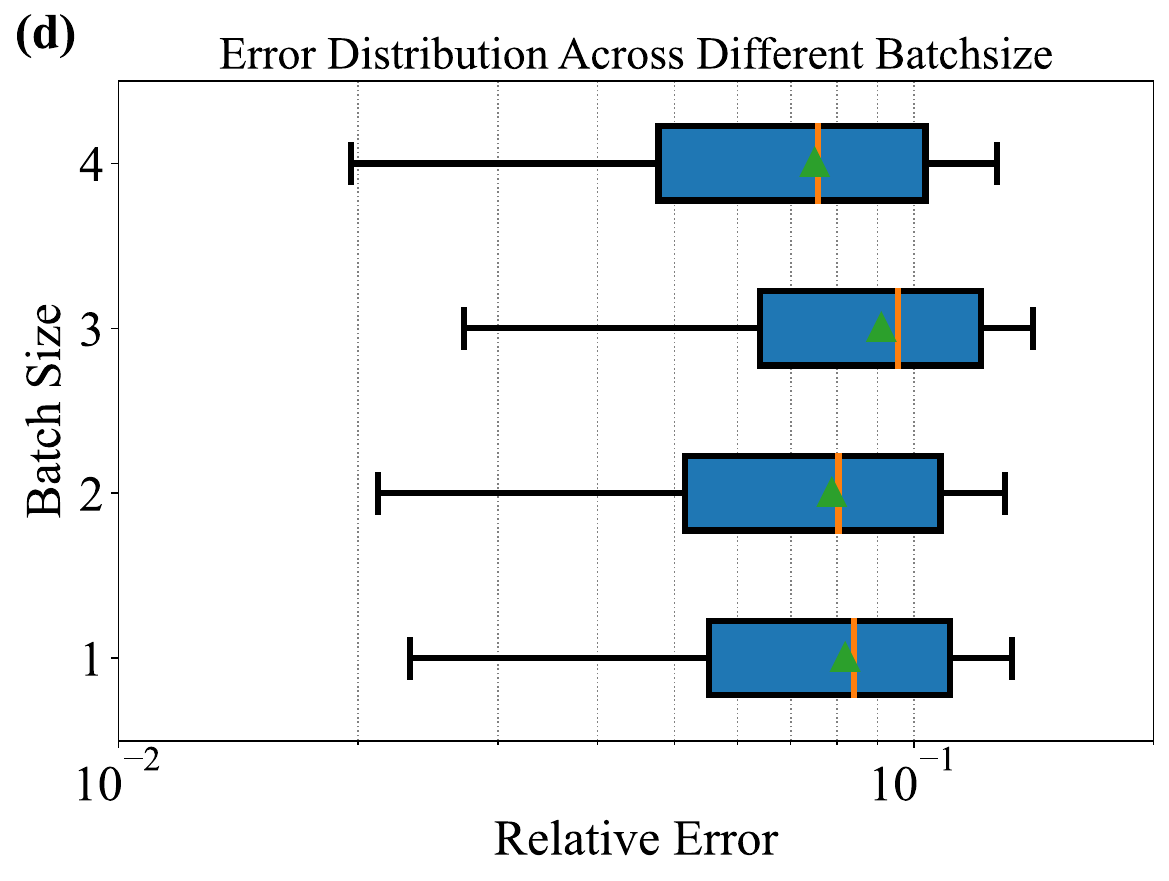}}
\end{minipage}
\vspace{-1pt}
\begin{minipage}{0.45\linewidth}
\centerline{\includegraphics[width=\textwidth]{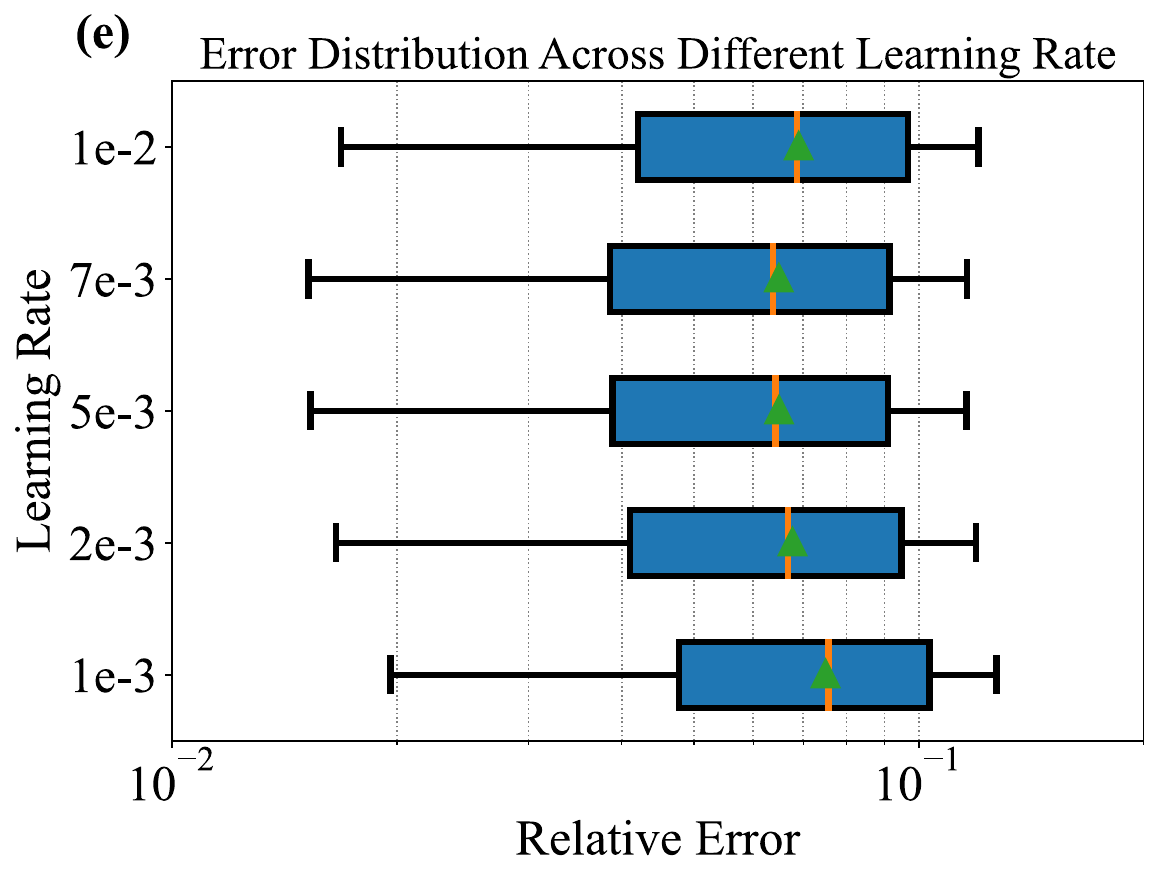}}
\end{minipage}
\hspace{1pt}
\begin{minipage}{0.45\linewidth}
\centerline{\includegraphics[width=\textwidth]{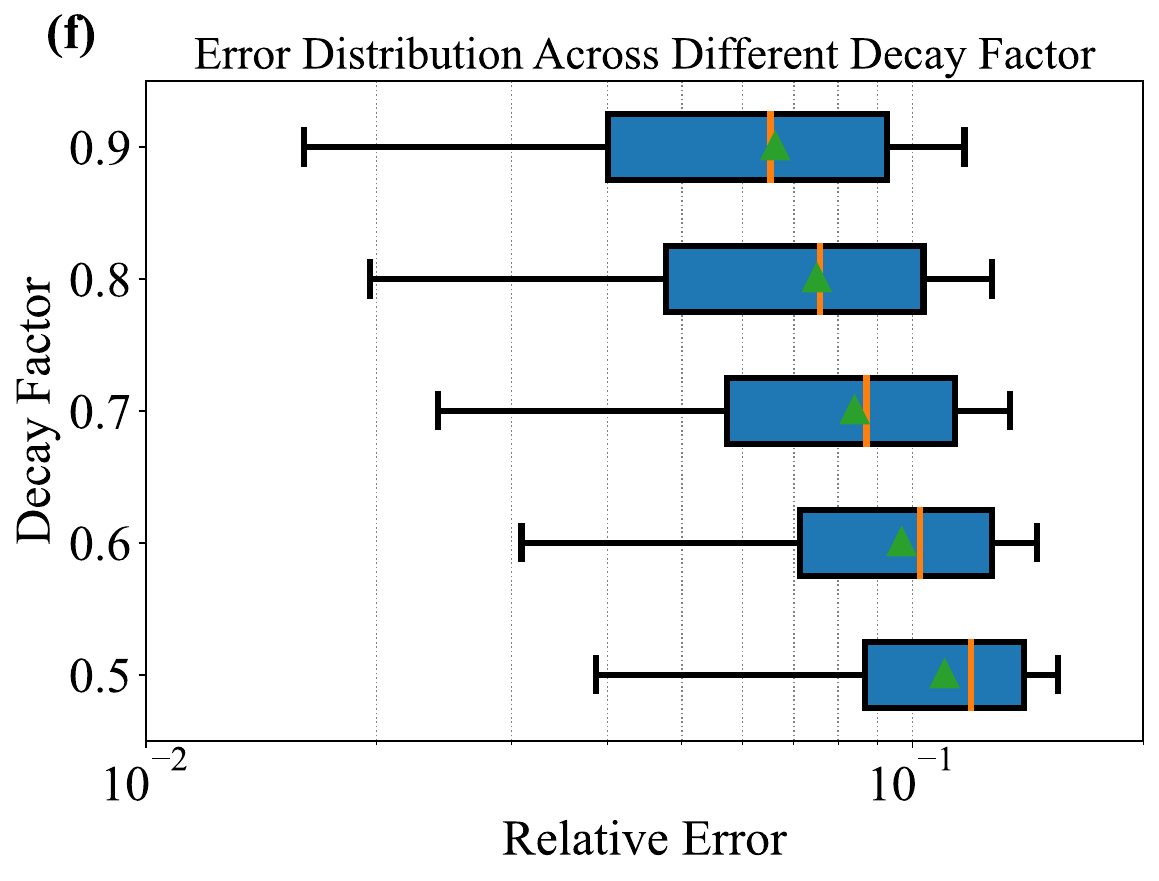}}
\end{minipage}
\caption{Hyper-parameters impact on test dataset error.}
\label{fig:hyperparameter-effects}
\end{figure}

\afterpage{\clearpage}

\begin{itemize}
\item \textbf{Fourier layers}: The number of network layers critically influences the model's capacity to capture deep, hierarchical features in the input data. The box plot of the Fourier layers shown in Fig. \ref{fig:hyperparameter-effects}(a) indicates that deeper networks (i.e., models with more layers) generally yield lower relative errors, and that once the depth reaches four layers, further increasing the number of layers brings only limited improvements in accuracy.

\item \textbf{Fourier width}: The Fourier width denotes the number of neurons contained in each individual Fourier layer. The experimental results (Fig. \ref{fig:hyperparameter-effects}(b)) indicate that increasing the network width can significantly enhance the prediction accuracy of the model. However, due to computational resource limitations, the width was limited to 80 in our experiments.

\item \textbf{Fourier modes}: In neural networks, the number of Fourier modes reflects the amount of frequency information represented by the model. Using more modes enables the model to capture a broader range of flow features and thereby learn the mapping of the flow field more effectively. However, an excessive number of modes increases the computational cost, and it should not exceed half of the physical spatial resolution. Fig. \ref{fig:hyperparameter-effects}(c) presents the performance of the LESnets model under different numbers of Fourier modes. In our experiments, the Fourier layer with 32 modes exhibits a clear performance advantage in our experiment.

\item \textbf{Batch size}: The batch size refers to the number of data samples processed by the neural network in a single training iteration. The box plot in Fig. \ref{fig:hyperparameter-effects}(d) illustrates the relative errors obtained by the model for four different batch sizes. The results show that larger batch sizes generally lead to smaller relative errors.

\item \textbf{Learning rate}: The learning rate is used in the neural network optimization process to control the magnitude of parameter updates during each training iteration. A lower learning rate leads to more stable convergence but requires more training epochs to reach the minimum error. A higher learning rate converges faster, but may cause larger fluctuations in error before the final solution is reached. As shown in Fig. \ref{fig:hyperparameter-effects}(e), the model performs best when the learning rate is set to 0.005.

\item \textbf{Decay factor}: The decay factor controls the rate at which the learning rate is reduced during training. The box plot in Fig. \ref{fig:hyperparameter-effects}(f) shows that a higher learning decay rate yields a lower relative error, possibly because a smoother learning-rate decrease provides a better optimization path for the model.
\end{itemize}

\clearpage

\section{Nomenclature}
\label{app:nomenclature}

\renewcommand{\thetable}{F\arabic{table}}
\renewcommand{\thefigure}{F\arabic{figure}}
\renewcommand{\theequation}{F\arabic{equation}}

\captionsetup{
  font=small,
  labelfont=bf,
  width=\textwidth
}

\setlength{\abovecaptionskip}{0pt}

\begingroup
\renewcommand{\arraystretch}{0.9}

\begin{longtable}{
  @{}
  >{\centering\arraybackslash}m{5cm}|
  m{11cm}
  @{}
}

\caption{Notation used throughout the paper.}
\label{tab:nomenclature}
\\

\toprule
\textbf{\large Symbol}
&
\textbf{\large Description}
\\
\midrule


$\Omega_c$ denotes the set of constrained grid points
 & \textbf{ Physical Variables} \\
 $x,y,z,t$  & \small Spatial and time coordinates \\
 $u,v,w,\bar{u}$  & \small Velocity components and filter velocity \\
 $\rho,p$  & \small Density and pressure \\
 $Re_h,\nu$  & \small Reynolds number and kinematic viscosity\\
 $Re_{\tau},u_{\tau}$  & \small Friction Reynolds number and friction velocity\\
 $\delta,\delta_{\nu},\Delta$  & \small Channel half-width, viscous length scale, and filter width \\
 $\delta_{ij},k$  & \small  Kronecker symbol and wavenumber\\
 $G,M,D,\Omega_c$  & \small Filter kernel, physical domain, bounded domain and the set of constrained grid points\\
 $y^+,\Delta y_{w}^+$ & \small Normalized wall-normal distance and distance of the first grid point off the wall\\
 $L_x,L_y,L_z$ & \small Computational domain in three directions\\
 $N_x,N_y,N_z,N_d,$ & \small Computational grids and number of the velocity components and pressure\\
 $\mathcal{N}_l,n_{\mathrm{pif}}$ & \small Interpolation node set and the number of neighboring fluid grid points\\
 $\Delta t,\Delta T$ & \small Numerical time step and non-dimensional units\\
\midrule
 & \textbf{ Neural Networks} \\
 $\Delta t_{train},\Delta t_{inf},\Delta t_{fDNS}$  & \small Training, inference, and supplement fDNS dataset time interval \\
 $T_{train},T_{inf},T_{output},T_{fDNS}$  & \small Training, inference, output and supplement fDNS dataset time steps \\
 $\mathcal{R},\mathcal{B},\mathcal{I}$  & \small Nonlinear differential operator, boundary condition, and initial condition\\
 $a^\dagger,u^\dagger,\mathcal{G}^{\dagger}$  & \small Instance, true solution, and solution operator, \\
 ${\theta},u_{\theta},\mathcal{G}_{\theta}$  & \small Network parameters, network output, and neural operator output \\
 ${P},{Q}$  & \small Point-wise operators \\
 $\mathcal{K},\mathcal{W},\sigma$  & \small Kernel integral operator, point-wise linear operator, and non-linear activation function\\
\midrule
  & \textbf{ Loss Functions} \\
 $\alpha,\beta,\gamma,\lambda$  & \small Loss weight hyper-parameters \\
 $\mathcal{L}_{\mathrm{ic}},\mathcal{L}_{\mathrm{bc}},\mathcal{L}_{\mathrm{pde}},\mathcal{L}_{\mathrm{data}},\mathcal{L}_{\mathrm{SGS}},\mathcal{L}(\theta)$  & \small Initial, boundary, PDE, data, SGS and physics-informed losses \\
\midrule
  & \textbf{ Optimization} \\
 $E,\eta$  & \small Number of epochs and learning rate\\
 $N_\mathrm{batch}$  & \small Batch size \\
 $\theta_{0},\bm{\Theta}$   & \small Initial networks parameters and finite-dimensional parameter space \\
 $\beta_1,\beta_2$ & \small Parameters of SOAP \\
  $T_{iter}$ & \small Number of iterations \\
\midrule
  & \textbf{ Turbulence Diagnostics} \\
 $U^+$ & \small Normalized mean streamwise velocity (wall units)\\
 $\tau_{w}$  & \small Wall shear stress \\
 ${\mathcal{F}_i},\bar{\mathcal{F}}_i$ & \small Forcing \\
 $\bar{S}_{ij},\bar{\Omega}$ & \small Filtered strain rate and antisymmetric rate of rotation\\
 $C_w,C_w(\theta)$ & \small Constant and learnable coefficient of WALE \\
 $u',v',w'$ & \small Root-mean-square (RMS) fluctuating velocities (wall units) \\
 $\langle u'v'\rangle, \langle u'v'\rangle^+$ & \small Shear Reynolds stress and normalized Reynolds shear stress (wall units) \\
\midrule
  & \textbf{Sampling} \\
 $N, N_s$  & \small Number of samples for training datasets and supplement dataset \\
 $\mathcal{C}_{train}^{180},\mathcal{C}_{train}^{590},\mathcal{C}_{train}^{1000},\mathcal{C}_{train}^{2000},\mathcal{H}_{train}^{5600},\mathcal{C}_{fDNS}^{180}$  & \small Training and supplement datasets \\
\midrule
\end{longtable}
\endgroup

\bibliographystyle{elsarticle-num} 
\bibliography{bibtxt}

\end{document}